\documentclass[%
superscriptaddress,
preprint,
 amsmath,amssymb,
 aps,
prb,
floatfix,
]{revtex4-2}

\usepackage{graphicx}% Include figure files
\usepackage{dcolumn}% Align table columns on decimal point
\usepackage{bm}% bold math
\usepackage{tikz}

\usepackage[pdftex,plainpages=false,colorlinks=true,linkcolor=blue, citecolor=blue, urlcolor=blue]{hyperref}

\usepackage{comment}

\begin{document}

%\preprint{Draft/8}

\title{Anisotropic bond-current susceptibilities and real-space current topology in correlated electron systems}

\author{Worapon Phatcharasirinawakun}
 \affiliation{Department of Condensed Matter Physics, Graduate School of Science, Hokkaido University, Sapporo 060-0810, Japan}
 \affiliation{%
 Research Center for Materials Nanoarchitectonics (MANA), National Institute for Materials Science (NIMS), Tsukuba 305-0047, Japan
}%
\author{Hiroyuki Yamase}%
 \email{YAMASE.Hiroyuki@nims.go.jp}
\affiliation{%
 Research Center for Materials Nanoarchitectonics (MANA), National Institute for Materials Science (NIMS), Tsukuba 305-0047, Japan
}
 \affiliation{Department of Condensed Matter Physics, Graduate School of Science, Hokkaido University, Sapporo 060-0810, Japan}

\date{\today}

\begin{abstract}
We derive all symmetry-allowed bond-current form factors generated by nearest-neighbor bond-charge interactions on square, triangular, and kagome lattices. We find that the anisotropy of the bond-current susceptibility systematically favors ordering wave vectors that support closed loop-current states, whereas symmetry-related wave vectors with weaker susceptibility generate noncirculating current textures. Near van Hove filling, this correspondence is robust across all lattice geometries considered, producing staggered flux phases on the square lattice, diamond-shaped current patterns on the triangular lattice and both chiral and nonchiral loop-current states on the kagome lattice. Our results establish a direct link between bond-current form factors, susceptibility anisotropy, and real-space current topology, providing a general framework for identifying loop-current orders and interpreting signatures of time-reversal symmetry breaking in correlated electron materials.
\end{abstract}


\maketitle

%\tableofcontents

\section{\label{sec.Introduction}Introduction}

Electronic ordering phenomena are a hallmark of strongly correlated materials. Depending on the interplay among electron interactions, lattice geometry, and band structure, correlated electron systems exhibit a variety of ordered phases, including charge-density waves (CDWs) \cite{tranquada95, kivelson98, yamase15b,li17,yamase26b}, nematic order \cite{yamase00a,yamase00b,metzner00,gallais13,fernandes14,sato17,auvray19,yamase21,nakata21}, unconventional superconductivity \cite{lee06,klein18,scalapino12}, and loop-current states \cite{varma97,varma06a,varma14,tazai23,friedlan26,schultz26}. Beyond conventional charge order, correlated electrons can develop bond orders residing on nearest-neighbor links. Such bond-order states may break translational, rotational, and time-reversal symmetries, giving rise to unconventional charge and current distributions.

Experimental interest in bond order has been stimulated by observations of intertwined electronic phases in cuprate and kagome materials. In cuprate superconductors \cite{keimer15}, charge order, nematicity, and signatures of time-reversal symmetry breaking have been reported throughout the pseudogap regime, suggesting the presence of hidden orders beyond simple density modulation \cite{shekhter09,varma14}. More recently, kagome metals AV$_3$Sb$_5$ have emerged as a platform for studying competing electronic orders, exhibiting charge ordering, superconductivity, anomalous Hall responses, and evidence for time-reversal symmetry breaking \cite{ortiz19,ortiz20}. These observations have renewed interest in bond-current and loop-current states as possible mechanisms underlying unconventional ordered phases.

A prototypical bond-current state is the flux phase proposed by Affleck and Marston \cite{affleck88a}, in which staggered circulating currents generate alternating magnetic fluxes through neighboring plaquettes. Because its order parameter possesses $d_{x^2-y^2}$ symmetry, the state is often refered to as a $d$-density-wave (DDW) or $d$-wave charge-density-wave ($d$CDW) phase \cite{cappelluti99,chakravarty01,bejas12}. Related loop-current states have subsequently been proposed for square \cite{shekhter09,varma14,varma97,varma06a,wang06}, triangular \cite{wang05,wang06}, and kagome \cite{dong23,zhan26,feng21,feng21a} lattices, suggesting that current-carrying order may be a generic order of correlated electron systems. 

Microscopically, bond-current order naturally emerges from bond-charge interactions, which generate momentum-dependent form factors that determine the order's symmetry. Previous studies have investigated a variety of bond-order \cite{zafur24,bejas12,bejas17,sachdev13,yamase21c,kiesel13}, however, the connection between bond-current form factors, susceptibility anisotropy, and the resulting real-space current topology remains largely unexplored. In particular, it is unclear whether the ordering vectors selected by bond-current fluctuations are intrinsically linked to the formation of closed loop-current states. The question is especially relevant for multisublattice systems such as the kagome lattice, where both single-$\mathbf{Q}$ ($1\mathbf{Q}$) and triple-$\mathbf{Q}$ ($3\mathbf{Q}$) current orders have been proposed  \cite{park21,denner21,zhan26,friedlan26,fu25,christensen22}. The kagome lattice is also of particular experimental interest because AV$_3$Sb$_5$ compounds exhibit a $2\times2$ charge-density-wave order with ordering vectors at the Brillouin-zone $M$ points \cite{li22,jiang21,nie22}, together with chiral magnetic responses and signatures of time-reversal symmetry breaking \cite{mielke22,khasanov22,guo22,xu22,guguchia23,yu21}. These observations have motivated numerous proposals for chiral ordered states enabled by the multisublattice structure of the kagome lattice. Understanding how such states emerge microscopically from bond-current interactions therefore remains an important open problem.

In this work, we investigate bond-current orders induced by nearest-neighbor bond-charge interactions on square, triangular, and kagome lattices. We derive all symmetry-allowed odd-parity bond-current form factors and evaluate their static susceptibilities. By reconstructing the corresponding real-space current configurations, we directly relate momentum-space fluctuations to current topology. We find that, near van Hove filling, the dominant susceptibility maxima preferentially occur at ordering vectors that generate closed-loop current states, whereas symmetry-related wave vectors with weaker susceptibility typically produce noncirculating current textures. This behavior originates from the pronounced susceptibility anisotropies induced by the odd-parity bond-current form factors, which break the rotational symmetry in momentum space, a nematic state,  and encode information about the topology of the resulting ordered state. In the kagome lattice, both $1\mathbf{Q}$ and $3\mathbf{Q}$ orderings support multiple loop-current phases, including a chiral flux state. These results establish a unified relationship between bond-current form factors, susceptibility anisotropy, and current topology across distinct lattice geometries.

This paper is organized as follows. Section~\ref{sec.Theory} introduces the bond-charge interaction formalism, derives the symmetry-allowed bond-current form factors and the real-space representation of the bond-current orders for the square, triangular, and kagome lattices. Section~\ref{sec.Results} presents the susceptibility analysis together with the corresponding real-space current configurations. Section~\ref{sec.discussion} discusses the physical implications of our results, and Section~\ref{sec.conclusions} summarizes the main conclusions. The appendices provide the complete set of bond-current form factors, discuss the periodicity of the bond-current susceptibility in the extended Brillouin zone, and present results away from van Hove filling.

\section{\label{sec.Theory}Formalism}
Bond-current orders are investigated within the bond-charge model \cite{zafur24}
\begin{eqnarray}
    H =-\sum_{i,j,\sigma} t_{ij}c^\dagger_{i,\sigma}c_{j,\sigma} + V_b \sum_{i\tau}\chi^\dagger_{i\tau}\chi_{i\tau},
\end{eqnarray}
where $t_{ij}=t~(t')$ denotes the hopping integral between first- (second-) nearest-neighbor sites, and
$c^{\dagger}_{i,\sigma}$ ($c_{i,\sigma}$) is the electron creation (annihilation) operator with spin $\sigma$ at site $i$.
The bond-charge interaction strength satisfies $V_b<0$, and the bond operator is defined as
\begin{equation}
    \chi_{i\tau}
    =
    \sum_{\sigma}
    c^{\dagger}_{i,\sigma}
    c_{i+\tau,\sigma},
\end{equation}
where $\tau$ denotes a nearest-neighbor bond direction. 

After performing the Fourier transformation, the bond-charge interaction term may become
\begin{align}
    H_b
    =
    \frac{V_b}{N}
    \sum_{\mathbf{k},\mathbf{k}',\mathbf{q}}
    \sum_{\sigma,\sigma'}
    \gamma_{\mathbf{k}}
    \gamma_{\mathbf{k}'}
    \,
    c^{\dagger}_{\mathbf{k}-\mathbf{q}/2,\sigma}
    c_{\mathbf{k}+\mathbf{q}/2,\sigma}
    \nonumber \\
    \times
    c^{\dagger}_{\mathbf{k}'+\mathbf{q}/2,\sigma'}
    c_{\mathbf{k}'-\mathbf{q}/2,\sigma'},
\end{align}
where $N$ is the number of lattice sites. The momentum dependence of the interaction is encoded in the form factor $\gamma_{\mathbf{k}}$, which satisfies
\begin{equation}
  \gamma_{\mathbf{k}}\gamma_{\mathbf{k}'} = \sum_{\boldsymbol{\tau}}  e^{i(\mathbf{k}\cdot\boldsymbol{\tau}-\mathbf{k}'\cdot\boldsymbol{\tau}) }.
  \label{eq.BondFormFacG}
\end{equation}
The symmetry properties of $\gamma_{\mathbf{k}}$ determine the nature of the resulting bond order: even-parity form factors correspond to bond-charge channels, whereas odd-parity form factors describe bond-current channels. Unlike approaches that assume a particular real-space ordering pattern \cite{tazai23}, Eq.~\eqref{eq.BondFormFacG} derives the symmetry-allowed ordering channels directly from the microscopic interaction, providing a systematic classification of interaction-driven bond orders.

The interaction term naturally introduces the bond-order operator at ordering wave vector $\mathbf{Q}$,
\begin{equation}
    \Delta_\gamma(\mathbf{Q}) = \sum_{\mathbf{k},\sigma} \gamma_{\mathbf{k}} c^\dagger_{\mathbf{k}-\frac{\mathbf{Q}}{2},\sigma} c_{\mathbf{k}+\frac{\mathbf{Q}}{2},\sigma},
    \label{eq.BondOperator}
\end{equation}
which serves as the order parameter for the corresponding bond-charge or bond-current order. In this work, we focus on the odd-parity channels, for which the sine-like momentum dependence is the momentum-space representation of the antisymmetric bond-current operator,
$
c_i^\dagger c_j-c_j^\dagger c_i
$, 
obtained by Fourier transformation \cite{chen25}. 
%The complete set of symmetry-allowed form factors, including the even-parity channels, is presented in Appendix~\ref{app.FormFac}.

The tendency toward bond-current order is studied by the random-phase-approximation (RPA) susceptibility,   
\begin{equation}
    \chi (\mathbf{Q}) = \dfrac{\chi_0 (\mathbf{Q} )}{1+ V_b \chi_0 (\mathbf{Q})},
    \label{eq.xRPA}
\end{equation}
 where the corresponding bare susceptibility is
\begin{equation}
    \chi_0 (\mathbf{Q} ) = -\dfrac{2}{N} \sum_{\mathbf{k}} \gamma^2_\mathbf{k} \dfrac{f(\varepsilon_{\mathbf{k} + \mathbf{Q}/2}) - f(\varepsilon_{\mathbf{k} - \mathbf{Q}/2} )}{  \varepsilon_{\mathbf{k} + \mathbf{Q}/2} - \varepsilon_{\mathbf{k} - \mathbf{Q}/2}}.
    \label{eq.x0}
\end{equation}
Here, $f(x)$ is the Fermi-Dirac distribution function and $\varepsilon_{\mathbf{k}}$ is the electronic energy dispersion.
Since $V_b$ is taken to be momentum-independent, the momentum structure of the RPA susceptibility is governed entirely by $\chi_0(\mathbf{Q})$. Consequently, the maxima of the static bare susceptibility identify the leading ordering wave vectors and provide the primary diagnostics of bond-current orders.

In the following subsections, this formalism is applied to square, triangular, and kagome lattices, establishing a systematic correspondence between bond-current susceptibilities and the resulting real-space current configurations across different lattice geometries.

\subsection{Square lattice}
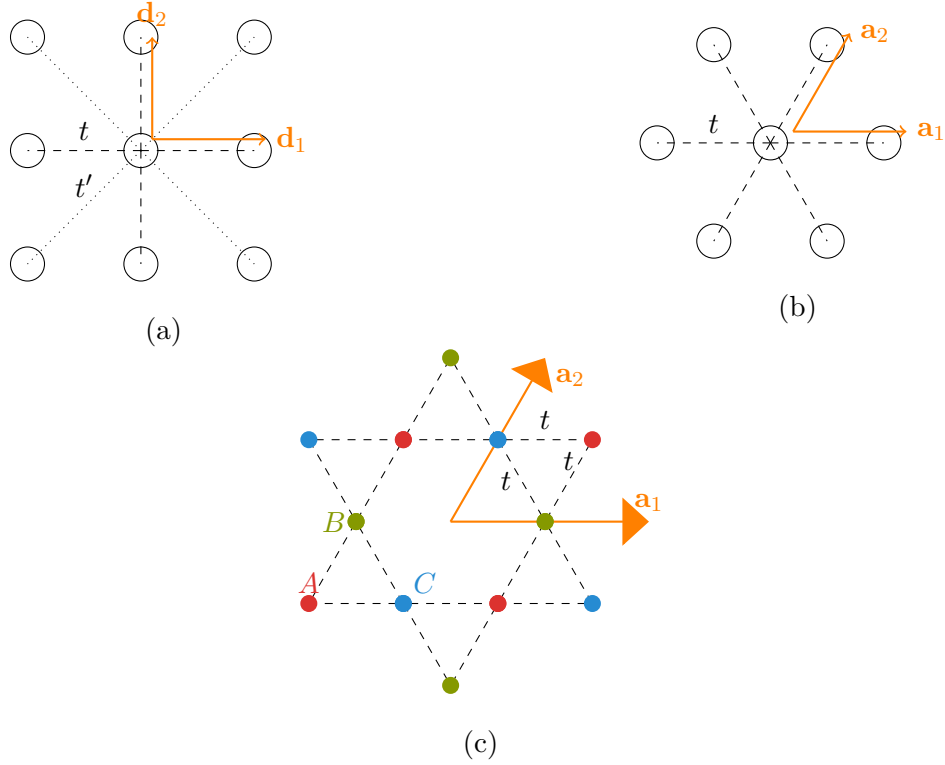
\begin{figure}[t]
    \centering
      \begin{minipage}{0.49\linewidth}
        \centering
              \begin{tikzpicture}[scale = 1.5]
    % central site
    \draw (0,0) circle (0.15);

    % nearest neighbors
    \draw (1,0) circle (0.15);
    \draw (-1,0) circle (0.15);
    \draw (0,1) circle (0.15);
    \draw (0,-1) circle (0.15);

    % next-nearest neighbors
    \draw (1,1) circle (0.15);
    \draw (-1,1) circle (0.15);
    \draw (1,-1) circle (0.15);
    \draw (-1,-1) circle (0.15);

    % nearest-neighbor bonds
    \draw[dashed] (0,0)--node[above]{$t$}(-1,0);
    \draw[dashed] (0,0)--(0,-1);

    \draw[dashed] (0,0)--(1,0);
    \draw[dashed] (0,0)--(0,1);

    \draw[thick,->,orange] (0.1,0.1)--(1.1,0.1)node[right]{$\mathbf{d}_1$};
    \draw[thick,->,orange] (0.1,0.1)--(0.1,1)node[above]{$\mathbf{d}_2$};

    % next-nearest-neighbor bonds
    \draw[dotted] (0,0)--(-1,1);
    \draw[dotted] (0,0)--node[above]{$t'$}(-1,-1);

    \draw[dotted] (0,0)--(1,1);
    \draw[dotted] (0,0)--(1,-1);
\end{tikzpicture}
\\
        (a)
    \end{minipage}
    \hfill
    \begin{minipage}{0.49\linewidth}
        \centering
\begin{tikzpicture}[scale = 1.5]
    \draw (0,0) circle (0.15);
    \draw (1,0) circle (0.15);
    \draw (-1,0) circle (0.15);
    \draw (0.5,1.732/2) circle (0.15);
    \draw (0.5,-1.732/2) circle (0.15);
    \draw (-0.5,1.732/2) circle (0.15);
    \draw (-0.5,-1.732/2) circle (0.15);
    \draw[dashed](0,0)--node[above]{$t$}(-1,0) ;
    \draw[dashed](0,0)--(-0.5,1.732/2) ;
    \draw[dashed](0,0)--(-0.5,-1.732/2) ;
    \draw[dashed](0,0)--(1,0) ;
    \draw[dashed](0,0)--(0.5,1.732/2) ;
    \draw[dashed](0,0)--(0.5,-1.732/2) ;
    \draw[thick,->,orange](0.2,0.1)--(1.2,0.1)node[right]{$\mathbf{a}_1$} ;
    \draw[thick,->,orange](0.2,0.1)--(0.7,{1.732/2 +0.1})node[right]{$\mathbf{a}_2$} ;
   % \draw[thick,->](0,0)--(0.5,-1.732/2)node[ right]{$\mathbf{a}_3$} ;
\end{tikzpicture}
\\
        (b)
    \end{minipage}

    \begin{minipage}{0.49\linewidth}
        \centering
\begin{tikzpicture}[scale=1.25]

\usetikzlibrary{calc}
\usetikzlibrary{arrows.meta}
%================================================
% colors
%================================================
\definecolor{Acol}{RGB}{220,50,47}
\definecolor{Bcol}{RGB}{38,139,210}
\definecolor{Ccol}{RGB}{133,153,0}

%================================================
% lattice vectors
%================================================
\coordinate (a1) at (2,0);
\coordinate (a2) at (1,{sqrt(3)});

%=======================================
\draw[-{Triangle[width=18pt,length=10pt]},thick,orange] (1.5,{sqrt(3)/2})--(3.6,{sqrt(3)/2})node[above]{$\mathbf{a}_1$};
\draw[-{Triangle[width=18pt,length=10pt]},thick,orange] ({0+1.5},{sqrt(3)/2})--({1.0+1.5},{sqrt(3)+ sqrt(3)/2})node[below right]{$\mathbf{a}_2$};
%================================================
%================================================
% reusable star unit
%================================================
\newcommand{\starunit}[1]{

    %--------------------------------------------
    % coordinates
    %--------------------------------------------
    \coordinate (A)  at ($#1 +(0,0)$);
    \coordinate (B)  at ($#1 +(1,0)$);
    \coordinate (C)  at ($#1 +(0.5,{sqrt(3)/2})$);

    \coordinate (A2) at ($#1 +(0,0)$);
    \coordinate (B2) at ($#1 +(-1,0)$);
    \coordinate (C2) at ($#1 +(-0.5,-{sqrt(3)/2})$);

    %--------------------------------------------
    % bonds
    %--------------------------------------------
    \draw[dashed] (A)--(B)--(C)--cycle;
    \draw[dashed] (A2)--(B2)--(C2)--cycle;

    %--------------------------------------------
    % sites
    %--------------------------------------------
    \fill[Acol] (A)  circle (0.090);
    \fill[Acol] (A2) circle (0.090);

    \fill[Bcol] (B)  circle (0.090);
    \fill[Bcol] (B2) circle (0.090);

    \fill[Ccol] (C)  circle (0.090);
    \fill[Ccol] (C2) circle (0.090);
}

\newcommand{\starunitone}[1]{

    %--------------------------------------------
    % coordinates
    %--------------------------------------------
    \coordinate (A)  at ($#1 +(0,0)$);
    \coordinate (B)  at ($#1 +(1,0)$);
    \coordinate (C)  at ($#1 +(0.5,{sqrt(3)/2})$);

    %--------------------------------------------
    % bonds
    %--------------------------------------------
    \draw[dashed] (A)--(B)--(C)--cycle;

    %--------------------------------------------
    % sites
    %--------------------------------------------
    \fill[Acol] (A)  circle (0.090);
    \fill[Bcol] (B)  circle (0.090);
    \fill[Ccol] (C)  circle (0.090);
}
\newcommand{\starunittwo}[1]{

    %--------------------------------------------
    % coordinates
    %--------------------------------------------

    \coordinate (A2) at ($#1 +(0,0)$);
    \coordinate (B2) at ($#1 +(-1,0)$);
    \coordinate (C2) at ($#1 +(-0.5,-{sqrt(3)/2})$);

    %--------------------------------------------
    % bonds
    %--------------------------------------------
    \draw[dashed] (A2)--node[above]{$t$}(B2)--node[left]{$t$}(C2)--node[above]{$t$}cycle;

    %--------------------------------------------
    % sites
    %--------------------------------------------
    \fill[Acol] (A2) circle (0.090);
    \fill[Bcol] (B2) circle (0.090);
    \fill[Ccol] (C2) circle (0.090);
}
%================================================
% draw units
%================================================
\starunitone{(0,0)}
\starunit{(a1)}
\starunit{(a2)}
\starunittwo{(a1)+(a2)}

% labels
%================================================
\node[Acol,above]  at (0,0) {$A$};
\node[Bcol,above right] at (1,0) {$C$};
\node[Ccol,left] at (0.5,{sqrt(3)/2}) {$B$};
\end{tikzpicture}
\\
        (c)
    \end{minipage}

    \caption{   Lattice geometries considered in this work:
(a) square,
(b) triangular, and
(c) kagome lattices.
Dashed and dotted bonds denote nearest- and next-nearest-neighbor hopping paths, respectively. The primitive lattice vectors are defined by arrows. 
   }
    \label{fig:squareTrian}
\end{figure}

We first consider the square lattice shown in Fig.~\ref{fig:squareTrian}(a). The primitive lattice vectors are
\begin{align}
    \mathbf{d}_1 = [1,0],\quad \mathbf{d}_2 = [0,1].
\end{align}

The corresponding electronic dispersion is
\begin{equation}
  \varepsilon_\mathbf{k} = -2t (\cos k_x + \cos k_y ) - 4t'\cos k_x \cos k_y - \mu,  
\end{equation}
where $\mu$ denotes the chemical potential. 

Solving Eq.~\eqref{eq.BondFormFacG} for the nearest-neighbor bonds yields two independent odd-parity form factors,
\begin{align}
    p^\pm(\mathbf{k}) = \sin k_x \pm \sin k_y, \label{eq.pFormfactors}
\end{align}
which describe the bond-current channels of the square lattice. In particular, at the commensurate ordering vector, $\mathbf{Q}_p=(\pi,\pi)$, the translation
\begin{equation}
    p^{-}\!\left(\mathbf{k}+\frac{\mathbf{Q}_p}{2}\right) =\cos k_x-\cos k_y
\end{equation}
maps the odd-parity form factor onto the $d_{x^2-y^2}$ bond-order form. This is a reason why $p^{-}$ is called as the $d$CDW (or flux-phase) \cite{affleck88a,cappelluti99,chakravarty01}. By contrast, $p^{+}$ represents a channel of symmetry-distinct bond-current. The even-parity bond-charge channels are summarized in Appendix~\ref{app.FormFac}.

By Fourier transforming the bond-current operator in Eq.~\eqref{eq.BondOperator}, the order parameters can be expressed in terms of currents flowing on nearest-neighbor bonds. The two odd-parity form factors $p^{+}$ and $p^{-}$ give rise to the bond-current order parameters,
\begin{align}
    \Delta_{p^\pm}(\mathbf{Q}) =& \dfrac{-i}{2}\sum_{i,\sigma} 
        [ \Delta_{ix}(\mathbf{Q})
    \pm \Delta_{iy}(\mathbf{Q})
     ] ,
    \label{eq.Flux_order_square}
\end{align}
where
\begin{align}
\Delta_{ix}(\mathbf{Q}) =   \langle c^\dagger_{i,\sigma}c_{i+x,\sigma} - c^\dagger_{i+x,\sigma}c_{i,\sigma} \rangle e^{-i(R_i+(\frac{1}{2},0))\cdot\mathbf{Q}}, \\
\Delta_{iy}(\mathbf{Q}) =   \langle c^\dagger_{i,\sigma}c_{i+y,\sigma}  - c^\dagger_{i+y,\sigma}c_{i,\sigma}\rangle e^{-i(R_i+(0,\frac{1}{2}))\cdot\mathbf{Q}}.
\end{align}
Here, $\Delta_{ix}$ and $\Delta_{iy}$ denote bond-current operators on nearest-neighbor bonds oriented along the $x$ and $y$ directions, respectively. The phase factors encode the spatial modulation associated with the ordering wave vector $\mathbf{Q}$, and
\begin{eqnarray}
    R_i = n_x \mathbf{d}_1 + n_y \mathbf{d}_2,
\end{eqnarray}
with integers $n_x$ and $n_y$ labeling lattice sites.

\subsection{Triangular lattice}
The triangular-lattice structure shown in Fig.~\ref{fig:squareTrian}(b) is defined by the primitive vectors
\begin{align}
    \mathbf{a}_1 &= [1,0], \\
    \mathbf{a}_2 &= \bigg[\dfrac{1}{2},\dfrac{\sqrt{3}}{2} \bigg],
\end{align}
together with the third bond direction
\begin{equation}
    \mathbf{a}_3 = \mathbf{a}_1 - \mathbf{a}_2 = \bigg[\dfrac{1}{2}, - \dfrac{\sqrt{3}}{2} \bigg].
\end{equation}

For isotropic nearest-neighbor hopping $t$, the electronic dispersion is
\begin{equation}
    \varepsilon_\mathbf{k} = -2t(\cos \mathbf{k}_1 + \cos \mathbf{k}_2 + \cos \mathbf{k}_3 ) - \mu,
\end{equation}
where
\begin{align}
    \mathbf{k}_1 &= k_x, \label{eq.k1}\\
    \mathbf{k}_2 &=
    \frac{k_x}{2}
    +
    \frac{\sqrt{3}k_y}{2}, \label{eq.k2}\\
    \mathbf{k}_3 &=
    \frac{k_x}{2}
    -
    \frac{\sqrt{3}k_y}{2}. \label{eq.k3}
\end{align}
The corresponding band structure is shown by the dashed-red line in Fig.~\ref{fig:EnergyDispersion}.

Applying Eq.~\eqref{eq.BondFormFacG} to the three nearest-neighbor bonds yields four independent odd-parity form factors,
    \begin{align}
    p^{++}_\mathbf{k} = \sin \mathbf{k}_1 + \sin \mathbf{k}_2 + \sin \mathbf{k}_3 , \label{eq.TriPxFormfac}\\
    p^{--}_\mathbf{k} = \sin \mathbf{k}_1 - \sin \mathbf{k}_2 - \sin \mathbf{k}_3 ,\label{eq.TriFFormfac}\\
    p^{+-}_\mathbf{k} = \sin \mathbf{k}_1 + \sin \mathbf{k}_2 - \sin \mathbf{k}_3 ,\label{eq.TriPpmFormfac}\\
    p^{-+}_\mathbf{k} = \sin \mathbf{k}_1 - \sin \mathbf{k}_2 + \sin \mathbf{k}_3 .\label{eq.TrimpFormfac}
\end{align}
These form factors constitute the bond-current channels of the triangular lattice and differ by the relative phases among the three bond directions. The complete set of symmetry-allowed form factors, including the even-parity bond-charge channels, is given in Appendix~\ref{app.FormFac}.

Substituting these form factors into Eq.~\eqref{eq.BondOperator} yields the corresponding bond-current order parameters
\begin{align}
    \Delta_{p^{++}}(\mathbf{Q}) &= \Delta_{b,1}(\mathbf{Q}) + \Delta_{b,2}(\mathbf{Q}) +\Delta_{b,3}(\mathbf{Q}), \label{eq.Delta_p1}\\
    \Delta_{p^{--}}(\mathbf{Q}) &= \Delta_{b,1}(\mathbf{Q}) - \Delta_{b,2}(\mathbf{Q}) -\Delta_{b,3}(\mathbf{Q}),\\
    \Delta_{p^{+-}}(\mathbf{Q}) &= \Delta_{b,1}(\mathbf{Q}) + \Delta_{b,2}(\mathbf{Q}) -\Delta_{b,3}(\mathbf{Q}),\\
    \Delta_{p^{-+}}(\mathbf{Q}) &= \Delta_{b,1}(\mathbf{Q}) - \Delta_{b,2}(\mathbf{Q}) +\Delta_{b,3}(\mathbf{Q}),\label{eq.Delta_p4}
\end{align}
where $\Delta_{b,\mu}$ ($\mu=1,2,3$) denotes the bond-current component associated with the nearest-neighbor bond direction $\mathbf{a}_\mu$:

Performing the Fourier transformation gives the real-space representation
\begin{widetext}
\begin{align}
    \Delta_{b,1}(\mathbf{Q}) &= \dfrac{-i}{2} \sum_{i,\sigma} \langle c^\dagger_{i,\sigma} c_{i+\mathbf{a}_1,\sigma} - c^\dagger_{i+\mathbf{a}_1,\sigma} c_{i,\sigma}\rangle e^{-i( r_i + \left( \frac{1}{2},0\right)  )\cdot \mathbf{Q}}, \label{eq.D1} \\
    \Delta_{b,2}(\mathbf{Q}) &= \dfrac{-i}{2}\sum_{i,\sigma} \langle c^\dagger_{i,\sigma}c_{i+ \mathbf{a}_2,\sigma}  - c^\dagger_{i+\mathbf{a}_2,\sigma}c_{i,\sigma} \rangle e^{-i( r_i + \left(\frac{1}{4},\frac{\sqrt{3}}{4}\right)  )\cdot \mathbf{Q}}, \label{eq.D2} \\
    \Delta_{b,3}(\mathbf{Q}) &= \dfrac{-i}{2}\sum_{i,\sigma}  \langle c^\dagger_{i,\sigma}c_{i+ \mathbf{a}_3,\sigma}  - c^\dagger_{i+\mathbf{a}_3,\sigma}c_{i,\sigma} \rangle e^{-i( r_i + \left(\frac{1}{4},-\frac{\sqrt{3}}{4}\right)  )\cdot \mathbf{Q}}, \label{eq.D3}
\end{align}
\end{widetext}
with lattice positions
\begin{eqnarray}
    r_i = n \mathbf{a}_1 + m \mathbf{a}_2,
\end{eqnarray}
where $n$ and $m$ are integers labeling the lattice sites.

\subsection{Kagome lattice}
The kagome lattice structure, is shown in Fig.~\ref{fig:squareTrian}(c), consists of three sublattices ($A$, $B$, and $C$) on a triangular Bravais lattice with primitive vectors $\mathbf{a}_1$ and $\mathbf{a}_2$. Nearest-neighbor bonds connect different sublattices and are therefore associated with bond vectors equal to one-half of the primitive lattice vectors.

\begin{figure}[t]
    \centering
    \includegraphics[width=0.75\linewidth]{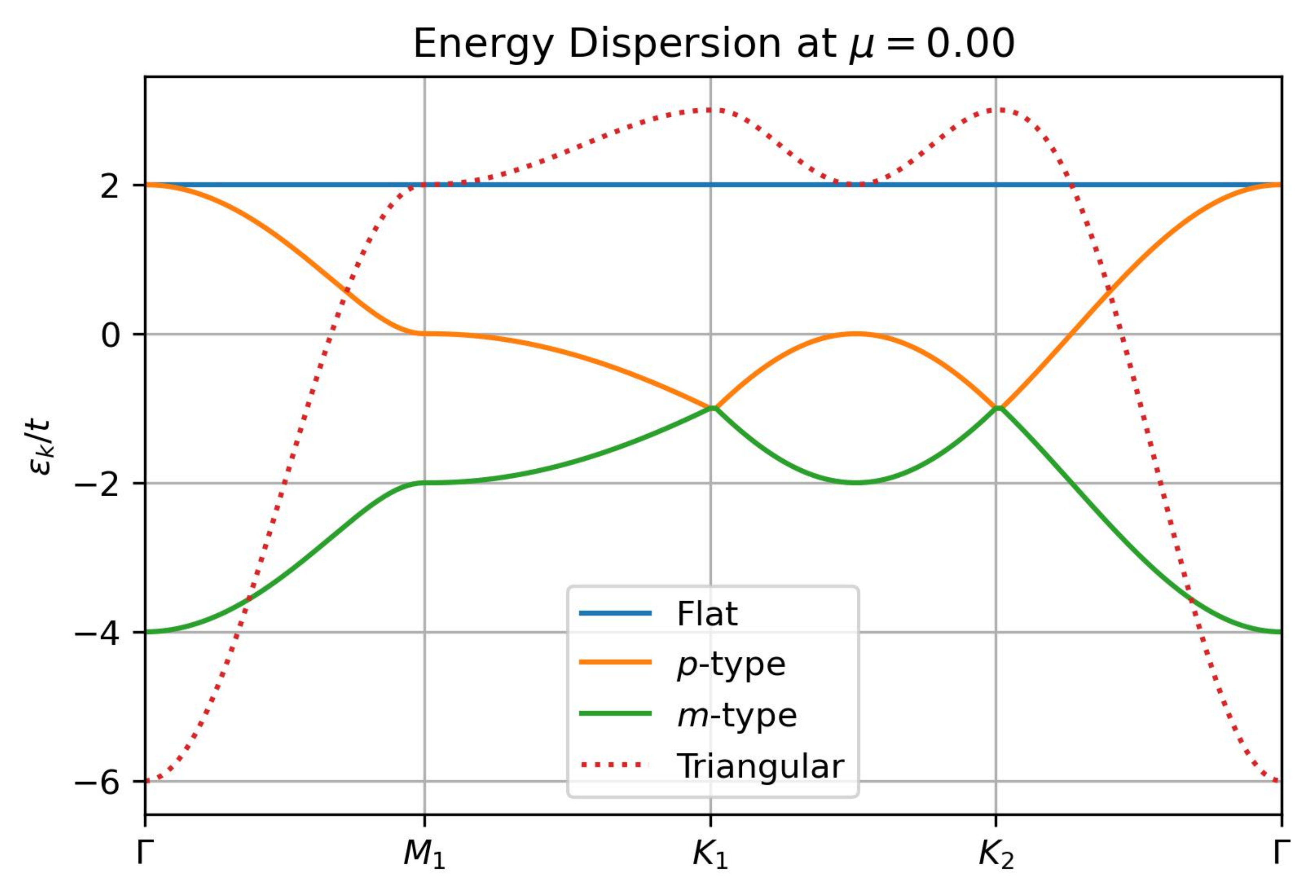}
\\
     \begin{minipage}{\linewidth}
        \centering
  \begin{tikzpicture}[scale = 0.5]
    \draw[] (2*3.14/3,2*3.14/1.73)--(4*3.14/3,0) -- (2*3.14/3,-2*3.14/1.73)--(-2*3.14/3,-2*3.14/1.73) -- (-4*3.14/3,0)  --(-2*3.14/3,2*3.14/1.73)--(2*3.14/3,2*3.14/1.73);
    \filldraw[red] (2*3.14/3,2*3.14/1.73) circle (4pt) node[anchor=west]{$K_1$};
    \filldraw[red] (4*3.14/3,0) circle (4pt) node[anchor=west]{$K_2 = \bigg( \frac{4\pi}{3} ,0 \bigg)$};
    \filldraw[red] (2*3.14/3,-2*3.14/1.73) circle (4pt) node[anchor=west]{$K_3$};
    \filldraw[red] (-2*3.14/3,-2*3.14/1.73) circle (4pt) node[anchor=east]{$K_4$};
    \filldraw[red] (-4*3.14/3,0) circle (4pt) node[anchor=east]{$K_5$};
    \filldraw[red] (-2*3.14/3,2*3.14/1.73) circle (4pt) node[anchor=east]{$K_6$};
    \filldraw[blue] (0,2*3.14/1.73) circle (4pt) node[anchor=south]{$M_1 = \bigg( 0,\frac{2\pi}{\sqrt{3}} \bigg)$};
    \filldraw[blue] (3.14,3.14/1.73) circle (4pt) node[anchor=west]{$M_2$};
    \filldraw[blue] (3.14,-3.14/1.73) circle (4pt) node[anchor=west]{$M_3$};
    \filldraw[blue] (0,-2*3.14/1.73) circle (4pt) node[anchor=north]{$M_4$};
    \filldraw[blue] (-3.14,-3.14/1.73) circle (4pt) node[anchor=east]{$M_5$};
    \filldraw[blue] (-3.14,3.14/1.73) circle (4pt) node[anchor=east]{$M_6$};
    \filldraw[black] (0,0) circle (4pt) node[anchor=north]{$\Gamma = (0,0)$};
\end{tikzpicture}
    \end{minipage}
    \caption{ Electronic band structures of the triangular (dashed line) and kagome (solid lines) lattices at $\mu=0$.
The hexagon below indicates the high-symmetry points of the first Brillouin zone in a hexagonal lattice.
}
    \label{fig:EnergyDispersion}
\end{figure}
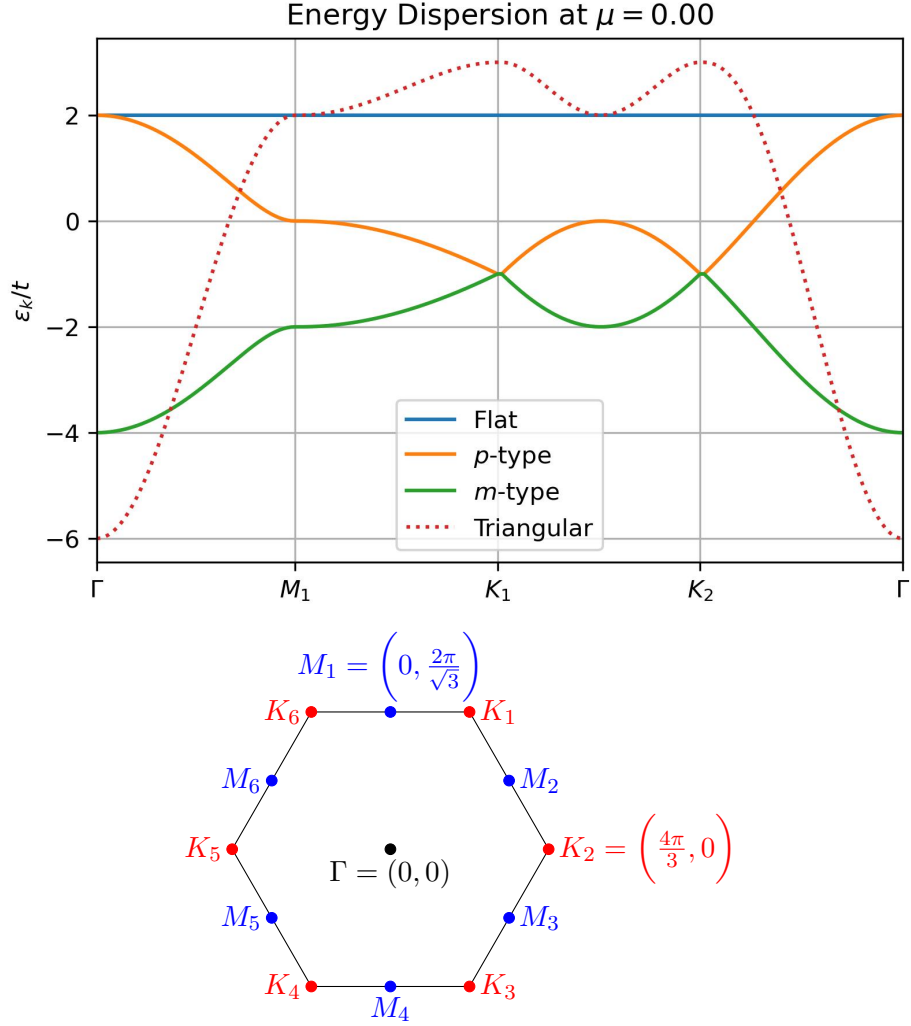

The nearest-neighbor tight-binding model yields three bands \cite{guo09,wu21}:
    \begin{align}
        E_1 &= 2t -\mu, \\
        E_2 &= t(-1 + \sqrt{4A_\mathbf{k}  - 3}) -\mu ,\label{eq.p_band} \\
        E_3 &= t(-1 - \sqrt{4A_\mathbf{k}  - 3}) -\mu,
    \end{align}
where $A_\mathbf{k} = \cos^2(\frac{k_x}{2}) + \cos^2(\frac{k_x}{4} + \frac{\sqrt{3}k_y}{4})  + \cos^2(\frac{k_x}{4} - \frac{\sqrt{3}k_y}{4})$. The flat band $E_1$ is separated from the dispersive upper ($p$-type) and lower ($m$-type) bands, $E_2$ and $E_3$, respectively. The corresponding band structure is shown in Fig.~\ref{fig:EnergyDispersion}.

Applying Eq.~\eqref{eq.BondFormFacG} to the three nearest-neighbor bonds yields the odd-parity form factors
\begin{align}
    p_{AC}(\mathbf{k}) &= \sin \bigg(\frac{k_x}{2}\bigg), \label{eq.PAC} \\
    p_{AB}(\mathbf{k}) &= \sin \bigg(\frac{k_x}{4} + \frac{\sqrt{3}k_y}{4}\bigg), \label{eq.PAB}\\
    p_{BC}(\mathbf{k}) &=  \sin \bigg(\frac{k_x}{4} - \frac{\sqrt{3}k_y}{4}\bigg), \label{eq.PBC}
\end{align}
associated with currents on the $AC$, $AB$, and $BC$ bonds, respectively. The even-parity form factors are discussed in Appendix~\ref{app.FormFac}.

The ordering structure of kagome bond-order states is under active discussions \cite{park21,denner21,zhan26,friedlan26,fu25,christensen22}. In the most general case, the three bond components may condense at different ordering vectors, leading to a $3\mathbf{Q}$ state described by
\begin{eqnarray}
    \Delta_b(\mathbf{Q}_1,\mathbf{Q}_2,\mathbf{Q}_3) =(\Delta_{AC}(\mathbf{Q}_1),\Delta_{AB}(\mathbf{Q}_2),\Delta_{BC}(\mathbf{Q}_3)), \label{eq.DeltaBKagome3Q}    
\end{eqnarray}
where $\Delta_{\alpha\beta}(\mathbf{Q}_i)$ denotes the bond-current order parameter on bond $\alpha\beta$.

In addition, an alternative ordering pattern frequently discussed in kagome systems is the $1\mathbf{Q}$ state. In this work, we also consider the case in which all three bond-current components are modulated by the same ordering wave vector,
\begin{equation}
\mathbf{Q}_1=\mathbf{Q}_2=\mathbf{Q}_3 = \mathbf{Q}.
\end{equation}
Throughout this paper, we refer to this configuration as a $1\mathbf{Q}$ state because the ordered phase is characterized by a single modulation wave vector. We emphasize that this terminology differs from that adopted in some previous studies \cite{park21,zhan26,fu25,christensen22}, where a $1\mathbf{Q}$ state denotes the condensation of only a single bond-current component (or a single ordering channel), while the remaining components vanish. In contrast, our $1\mathbf{Q}$ state retains all three bond-current components, which remain finite but share the same ordering vector.

The corresponding bond-current form factors are still given by Eqs.~\eqref{eq.PAC}--\eqref{eq.PBC}. However, because all three bond-current components share the same ordering vector, it is convenient to combine them into a set of linear combinations. Motivated by the construction for the triangular lattice,  a set of the $1\mathbf{Q}$ form factors is defined by
\begin{align}
    K^{++}_\mathbf{k} &=  p_{AC}(\mathbf{k}) + p_{AB}(\mathbf{k}) + p_{BC}(\mathbf{k}), \label{eq.Kx}\\
    K^{--}_\mathbf{k} &=  p_{AC}(\mathbf{k}) - p_{AB}(\mathbf{k}) - p_{BC}(\mathbf{k}), \label{eq.Kf}\\
    K^{+-}_\mathbf{k} &=  p_{AC}(\mathbf{k}) + p_{AB}(\mathbf{k}) - p_{BC}(\mathbf{k}), \label{eq.Kpm}\\
    K^{-+}_\mathbf{k} &= p_{AC}(\mathbf{k}) - p_{AB}(\mathbf{k}) + p_{BC}(\mathbf{k}). \label{eq.Kmp}
\end{align}
This basis provides a more convenient description of the $1\mathbf{Q}$ bond-current states than treating the three bond-current components independently. As shown in the following, it simplifies both the susceptibility analysis and the classification of the corresponding real-space current patterns.

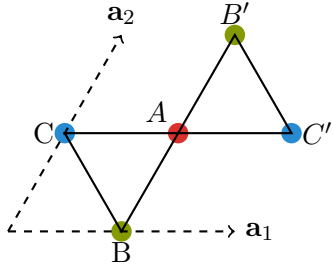
\begin{figure}[t]
    \centering
\begin{tikzpicture}[scale=1.5]
\definecolor{Acol}{RGB}{220,50,47}
\definecolor{Bcol}{RGB}{133,153,0}
\definecolor{Ccol}{RGB}{38,139,210}
    \fill[Acol] (0,0) circle (0.09);
    \fill[Bcol] (-0.5,-1.732/2) circle (0.09);
    \fill[Ccol] (-1,0) circle (0.09);
        \fill[Bcol] (0.5,1.732/2) circle (0.09);
    \fill[Ccol] (1,0) circle (0.09);
\draw[thick](0,0)node[above left]{$A$}--(1,0)node[right]{$C'$}--(0.5,1.732/2)node[above]{$B'$}--(0,0) ;
    \draw[thick](0,0)--(-1,0)node[left]{C}--(-0.5,-1.732/2)node[below]{B}--(0,0) ;
\draw[->,thick,dashed](-0.75*2,-1.732*2/4) -- (0.25*2,-1.732*2/4)node[right]{$\mathbf{a}_1$};
\draw[->,thick,dashed](-0.75*2,-1.732*2/4) -- (-0.25*2,1.732*2/4)node[above]{$\mathbf{a}_2$};
\end{tikzpicture}
\caption{
Unit-cell convention and sublattice coordinates used in the real-space construction of kagome-lattice bond-current operators. Sublattice $A$ is chosen as the unit-cell center, while $B$, $C$, $B'$, and $C'$ denote nearest-neighbor sites related by primitive lattice translations.
}
    \label{fig:UnitCell}
\end{figure}

To reconstruct the real-space bond-current operators on the kagome lattice, it is necessary to specify the sublattice embedding, since the order parameters couple different sublattices. Different choices of the unit-cell origin correspond to different phase conventions for the Bloch states and therefore lead to different real-space representations of the same momentum-space order parameter. These representations are, however, physically equivalent and describe the same bond-current state.
%, as demonstrated in Appendix~\ref{app:gauge}
Throughout this work, we adopt the convention shown in Fig.~\ref{fig:UnitCell}, with sublattice $A$ located at the center of the unit cell. The corresponding sublattice coordinates are
\begin{align}
    \tau_A &= \left(\frac{3}{4},\frac{\sqrt{3}}{4}\right), \\
    \tau_B &= \left(\frac{1}{2},0\right) , \quad
    &&\tau_{B'} = \left(1,\frac{\sqrt{3}}{2}\right), \\
    \tau_C &= \left(\frac{1}{4},\frac{\sqrt{3}}{4}\right), \quad
    &&\tau_{C'} = \left(\frac{5}{4},\frac{\sqrt{3}}{4}\right).
\end{align}

Using the bond-current form factors $p_{AC}$, $p_{AB}$, and $p_{BC}$ defined in Eqs.~\eqref{eq.PAC}--\eqref{eq.PBC}, the momentum-space bond-current operator between sublattices $\alpha$ and $\beta$ is
\begin{widetext}
\begin{align}
    \Delta_{\alpha\beta}(\mathbf{Q}) 
    = \sum_{\mathbf{k},\sigma} p_{\alpha\beta}\bigg( {\mathbf{k}+\frac{\mathbf{Q}}{2}}\bigg) \langle c^\dagger_{\mathbf{k},\sigma;\alpha} c_{\mathbf{k}+\mathbf{Q},\sigma;\beta} +  c^\dagger_{\mathbf{k},\sigma;\beta} c_{\mathbf{k}+\mathbf{Q},\sigma;\alpha} \rangle . \label{eq.DeltaAlphaBeta}
\end{align}

Fourier transformation yields the real-space bond-current operators
\begin{align}
        \Delta_{AC}(\mathbf{Q}) &= \dfrac{-i}{2}\sum_{j,\sigma} (\Delta^j_{AC'}(\mathbf{Q}) + \Delta_{AC}^j(\mathbf{Q}) ), \label{eq.DAC}
\\
             \Delta_{AB}(\mathbf{Q}) &= \dfrac{-i}{2}\sum_{j,\sigma} ( \Delta_{AB'}^j(\mathbf{Q})  + \Delta_{AB}^j(\mathbf{Q}) ), \label{eq.DAB}
\\
                  \Delta_{BC}(\mathbf{Q}) &= \dfrac{-i}{2}\sum_{j,\sigma} ( \Delta_{B'C'}^j(\mathbf{Q})  + \Delta_{BC}^j(\mathbf{Q})), \label{eq.DBC}
\end{align}
where the bond operators $\Delta^j_{\alpha\beta}$ are defined below.
\begin{align}
  &\Delta^j_{AC'}(\mathbf{Q}) =   \langle c^\dagger_{j,\sigma;A}  c_{j,\sigma;C'} - c^\dagger_{j,\sigma;C'}c_{j,\sigma;A} \rangle e^{-i ( r_j +  (1,\frac{\sqrt{3}}{4})) \cdot\mathbf{Q}  } , \label{eq.DjAC1} \\
  &\Delta_{AC}^j(\mathbf{Q}) = \langle c^\dagger_{j,\sigma;C}  c_{j,\sigma;A} - c^\dagger_{j,\sigma;A}  c_{j,\sigma;C}\rangle e^{-i ( r_j + (\frac{1}{2},\frac{\sqrt{3}}{4}))\cdot\mathbf{Q}  } ,\\
  &\Delta_{AB'}^j(\mathbf{Q}) = \langle  c^\dagger_{j,\sigma;A}  c_{j,\sigma;B'} - c^\dagger_{j,\sigma;B'} c_{j,\sigma;A}\rangle e^{-i ( r_j +  (\frac{7}{8},\frac{3\sqrt{3}}{8}))\cdot\mathbf{Q}  } , \\
 &\Delta_{AB}^j(\mathbf{Q}) = \langle c^\dagger_{j,\sigma;B}  c_{j,\sigma;A} - c^\dagger_{j,\sigma;A}c_{j,\sigma;B} \rangle e^{-i ( r_j +  (\frac{5}{8},\frac{\sqrt{3}}{8}))\cdot\mathbf{Q}  } ,\\
 &\Delta_{B'C'}^j(\mathbf{Q}) = \langle  c^\dagger_{j,\sigma;B'}  c_{j,\sigma;C'} - c^\dagger_{j,\sigma;C'} c_{j,\sigma;B'}\rangle e^{-i ( r_j +  (\frac{9}{8},\frac{3\sqrt{3}}{8}))\cdot\mathbf{Q}  }, \\
 &\Delta_{BC}^j(\mathbf{Q}) =\langle c^\dagger_{j,\sigma;C}  c_{j,\sigma;B} - c^\dagger_{j,\sigma;B}c_{j,\sigma;C} \rangle e^{-i ( r_j +  (\frac{3}{8},\frac{\sqrt{3}}{8}))\cdot\mathbf{Q}  } . \label{eq.DjBC2}
\end{align}
\end{widetext}
The three operators $\Delta_{AC}$, $\Delta_{AB}$, and $\Delta_{BC}$ form a complete basis for nearest-neighbor bond-current order on the kagome lattice and naturally describe the $3\mathbf{Q}$ ordering discussed in Sec.~\ref{sec.Results}.

%Unlike the square and triangular lattices, the kagome lattice admits an alternative description of bond-current order in terms of even-parity form factors. As discussed in Appendix~\ref{app.FormFac}, these even-parity channels are related to the odd-parity representation through reciprocal-lattice translations and therefore provide an equivalent description of the same bond-current order. Nevertheless, throughout this work we employ the odd-parity basis, which provides the most direct description of bond-current orders.

For the $1\mathbf{Q}$ states, all nearest-neighbor bond-current components are modulated by the same ordering vector. The corresponding real-space bond-current operators are therefore conveniently expressed in the symmetry-adapted basis associated with the momentum-space form factors in Eqs.~\eqref{eq.Kx}--\eqref{eq.Kmp},
\begin{align}
    \Delta_{K^{++}}(\mathbf{Q}) &=  \Delta_{AC}(\mathbf{Q}) + \Delta_{AB}(\mathbf{Q}) + \Delta_{BC}(\mathbf{Q}), \label{eq.DKx}\\
    \Delta_{K^{--}}(\mathbf{Q}) &=  \Delta_{AC}(\mathbf{Q}) - \Delta_{AB}(\mathbf{Q}) - \Delta_{BC}(\mathbf{Q}), \label{eq.DKf}\\
    \Delta_{K^{+-}}(\mathbf{Q}) &=  \Delta_{AC}(\mathbf{Q}) + \Delta_{AB}(\mathbf{Q}) - \Delta_{BC}(\mathbf{Q}), \label{eq.DKpm}\\
   \Delta_{K^{-+}}(\mathbf{Q}) &= \Delta_{AC}(\mathbf{Q}) - \Delta_{AB}(\mathbf{Q}) + \Delta_{BC}(\mathbf{Q}). \label{eq.DKmp} 
\end{align}

\subsection{Real-space bond-current order}

 For the square and triangular lattices, each real-space bond-current operator introduced in  Eqs.~\eqref{eq.Flux_order_square} and \eqref{eq.Delta_p1}--\eqref{eq.Delta_p4} can be written in the general form as
\begin{equation}
  \Delta_\ell(\mathbf{Q}) =   -i \sum_{i} v_\ell e^{-i\mathbf{Q}\cdot(\mathbf{R}_i+r_\ell)}
\end{equation}
where $v_\ell= \frac{1}{2}\sum_{\sigma}\langle c_{i,\sigma}^\dagger c_{i+\delta_\ell,\sigma}-c_{i+\delta_\ell,\sigma}^\dagger c_{i,\sigma}\rangle$ is the bond-current expectation value on the nearest-neighbor bond $\delta_\ell$, and $r_\ell=\delta_\ell/2$ denotes the corresponding bond-center position. 

Because the position-dependent phase factor $e^{-i\mathbf{Q}\cdot(\mathbf{R}_i+r_\ell)}$ is generally complex, the order parameter $\Delta_\ell(\mathbf{Q})$ is itself a complex momentum-space order parameter rather than the physical current on an individual bond.  Only for special commensurate ordering wave vectors, corresponding to conventional flux phases, does $\Delta_\ell(\mathbf{Q})$ become purely imaginary. Unlike the bond-current order parameters commonly discussed in previous studies~\cite{tazai23,fu25}, the present expression explicitly retains the spatial phase factor.
%required to reconstruct the ordered current distribution associated with a finite ordering wave vector. 

The physical current is therefore identified with the imaginary part of the reconstructed order parameter,
\begin{equation}
J_\ell(\mathbf{R}_i,\mathbf{Q})
=
\mathrm{Im}
\left[
\Delta_\ell(\mathbf{Q})
\right],
\label{eq:realcurrent}
\end{equation}
which gives the current flowing through the bond centered at $\mathbf{R}_i+r_\ell$. This definition enables the real-space current network to be reconstructed directly from the momentum-space order parameter, allowing loop-current, source-sink, and other current topologies associated with a given ordering wave vector. %to be identified.

For the kagome lattice, each inequivalent bond in Eqs.\eqref{eq.DAC}--\eqref{eq.DBC} consists of two symmetry-related bond directions. The bond-current order parameter can therefore be written as
\begin{equation}
\Delta_{\alpha\beta}(\mathbf{Q})
=
-i
\sum_i
\left(
v_{\alpha\beta}
e^{-i\mathbf{Q}\cdot(\mathbf{R}_i+r_{\alpha\beta})}
+
v_{\alpha'\beta'}
e^{-i\mathbf{Q}\cdot(\mathbf{R}_i+r_{\alpha'\beta'})}
\right),
\end{equation}
where $v_{\alpha\beta}$ and $v_{\alpha'\beta'}$ denote the bond-current expectation values on the two symmetry-related bonds.
As in the square and triangular lattices, the physical current is obtained from the imaginary part of the bond-current order parameter,
\begin{equation}
\mathrm{Im}\!\left[\Delta_{\alpha\beta}(\mathbf{Q})\right]
=
J_{\alpha\beta}(\mathbf{R}_i,\mathbf{Q})
+
J_{\alpha'\beta'}(\mathbf{R}_i,\mathbf{Q}), 
\label{eq.Jalphabetaalphabeta}
\end{equation}
with the current on each bond defined by
\begin{equation}
    J_{\alpha\beta}(\mathbf{R}_i,\mathbf{Q})
=
\mathrm{Im}
\left[
-i
v_{\alpha\beta}
e^{-i\mathbf{Q}\cdot(\mathbf{R}_i+r_{\alpha\beta})}
\right].
\end{equation}

\section{\label{sec.Results}Results}
We evaluate the static bond-current susceptibility, Eq.~\eqref{eq.x0}, for the symmetry-allowed odd-parity form factors on the square, triangular, and kagome lattices. Calculations are performed at van Hove filling, where the Fermi surface intersects the saddle points of the electronic dispersion and the bond-current susceptibility is maximally enhanced. Decreasing the temperature increases the susceptibility and sharpens the peak structure, but does not qualitatively alter the momentum-space distribution of the fluctuations. We therefore fix the temperature to $T=0.01t$ throughout the susceptibility analysis.

To connect the momentum-space susceptibility to the corresponding ordered state, we reconstruct the real-space bond-current distribution from the bond-current order parameter. For the purpose of visualizing the current topology, we fix the normalization $v_\ell=i$ throughout this section, thereby fixing the overall phase and magnitude of the order parameter while preserving the current topology. Under this convention, the bond current in Eq.~\eqref{eq:realcurrent} is determined solely by the phase modulation,
\begin{equation}
    J_\ell(\mathbf{R}_i,\mathbf{Q})
=
\mathrm{Im}
\!\left[
e^{-i\mathbf{Q}\cdot(\mathbf{R}_i+r_\ell)}
\right].
\end{equation}

For kagome systems, we also adopt the normalization
$v_{\alpha\beta}=v_{\alpha'\beta'}=i$ in Eq.~\eqref{eq.Jalphabetaalphabeta},
so that the current pattern is determined solely by the position-dependent phase factor, 
\begin{equation}
J_{\alpha\beta}(\mathbf{R}_i,\mathbf{Q}) = \mathrm{Im}[e^{-i\mathbf{Q}\cdot(\mathbf{R}_i+r_{\alpha\beta})}]. 
\label{eq.realCurrentKagome}
\end{equation}

\subsection{Square lattice}
\begin{figure}[t]
    \centering
\includegraphics[width=0.9\linewidth]{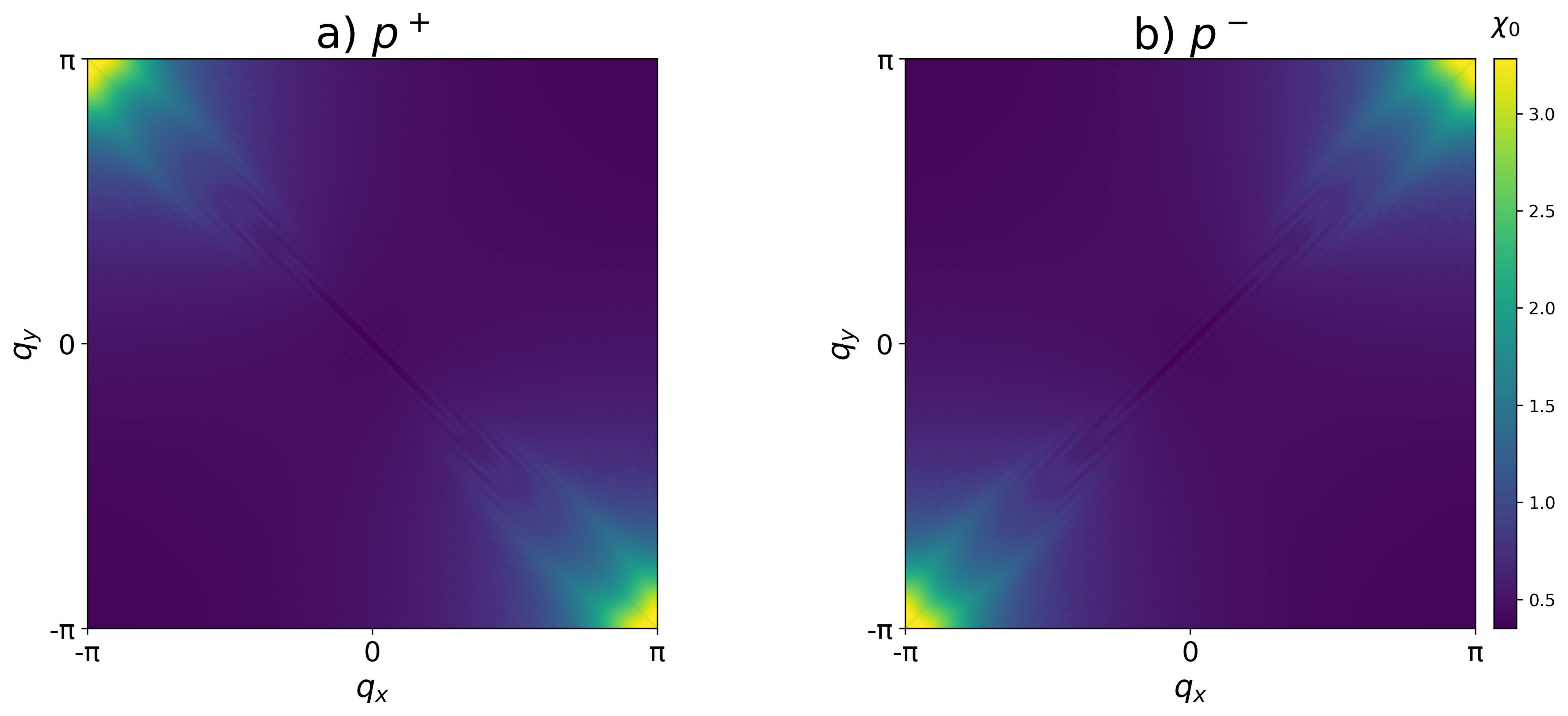}
    \caption{
Momentum-space bond-current susceptibilities of the square lattice at van Hove filling ($\mu=-0.8t$) and $T=0.01t$. The $p^{+}$ channel (a) is maximized at $\mathbf{Q}=(\pm\pi,\mp\pi)$, whereas the $p^{-}$ channel (b) exhibits maxima at $\mathbf{Q}=\pm(\pi,\pi)$.
    }
    \label{fig:Sussquare}
\end{figure}

For the square lattice, we adopt $t'=-0.2t$, for which the van Hove singularity occurs at $\mu=-0.8t$. The momentum-dependent bond-current susceptibilities associated with the form factors $p^{+}$ and $p^{-}$ are shown in Fig.~\ref{fig:Sussquare}.

The $p^{+}$ susceptibility exhibits pronounced maxima at $\mathbf{Q}=(\pm\pi,\mp\pi)$, whereas the $p^{-}$ channel is maximized at $\mathbf{Q}=\pm(\pi,\pi)$, thereby identifying the leading ordering vectors of the two bond-current channels. Although the square lattice possesses $C_4$ rotational symmetry, the susceptibility distribution is strongly anisotropic, with spectral weight concentrated at only two of the four Brillouin-zone corners. This nematicity originates from the odd-parity form factors.

\begin{figure*}
    \centering
        \begin{minipage}{0.49\linewidth}
        \centering
    \includegraphics[width=0.95\linewidth]{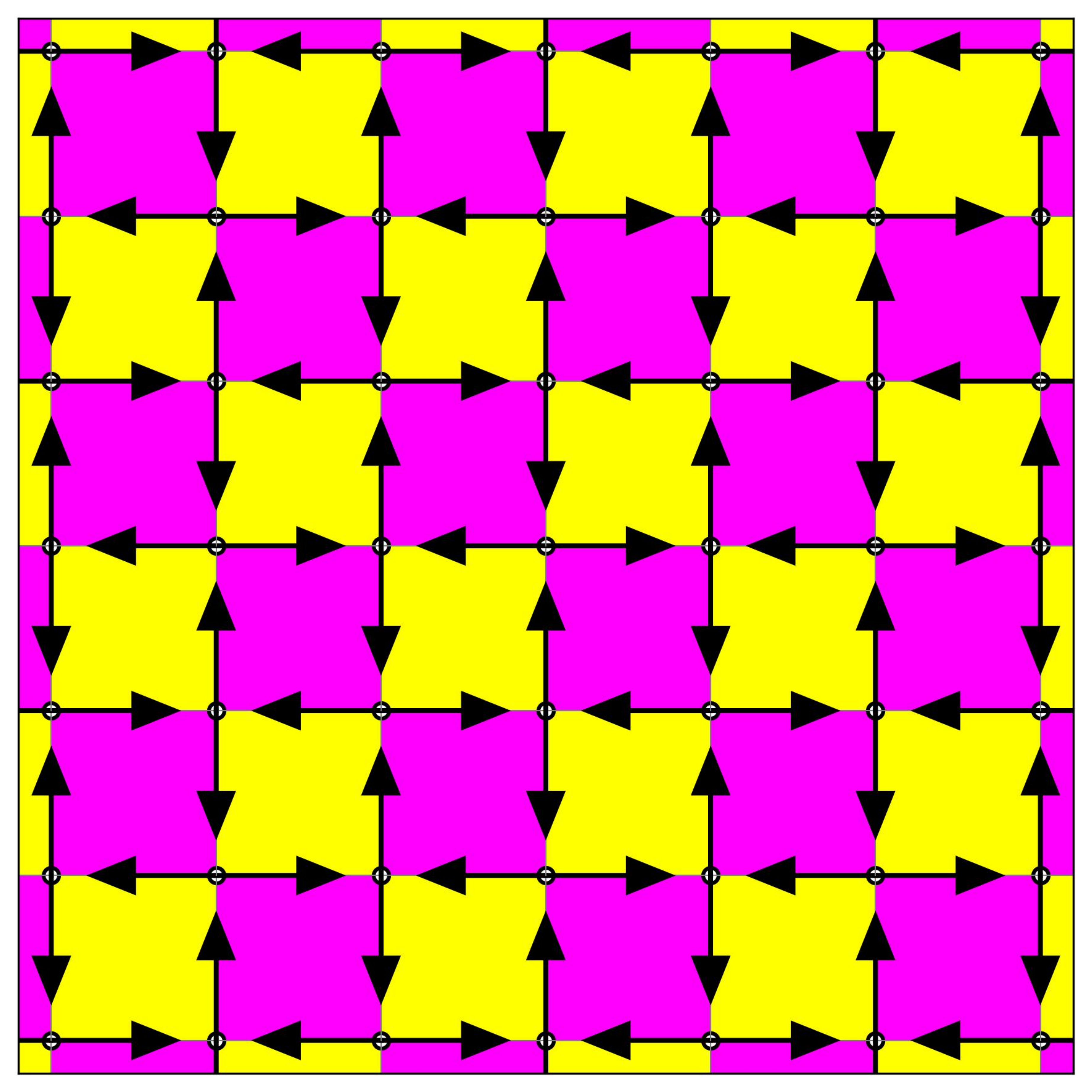}
\\
        (a)
    \end{minipage}
    \hfill
    \begin{minipage}{0.49\linewidth}
        \centering
    \includegraphics[width=0.95\linewidth]{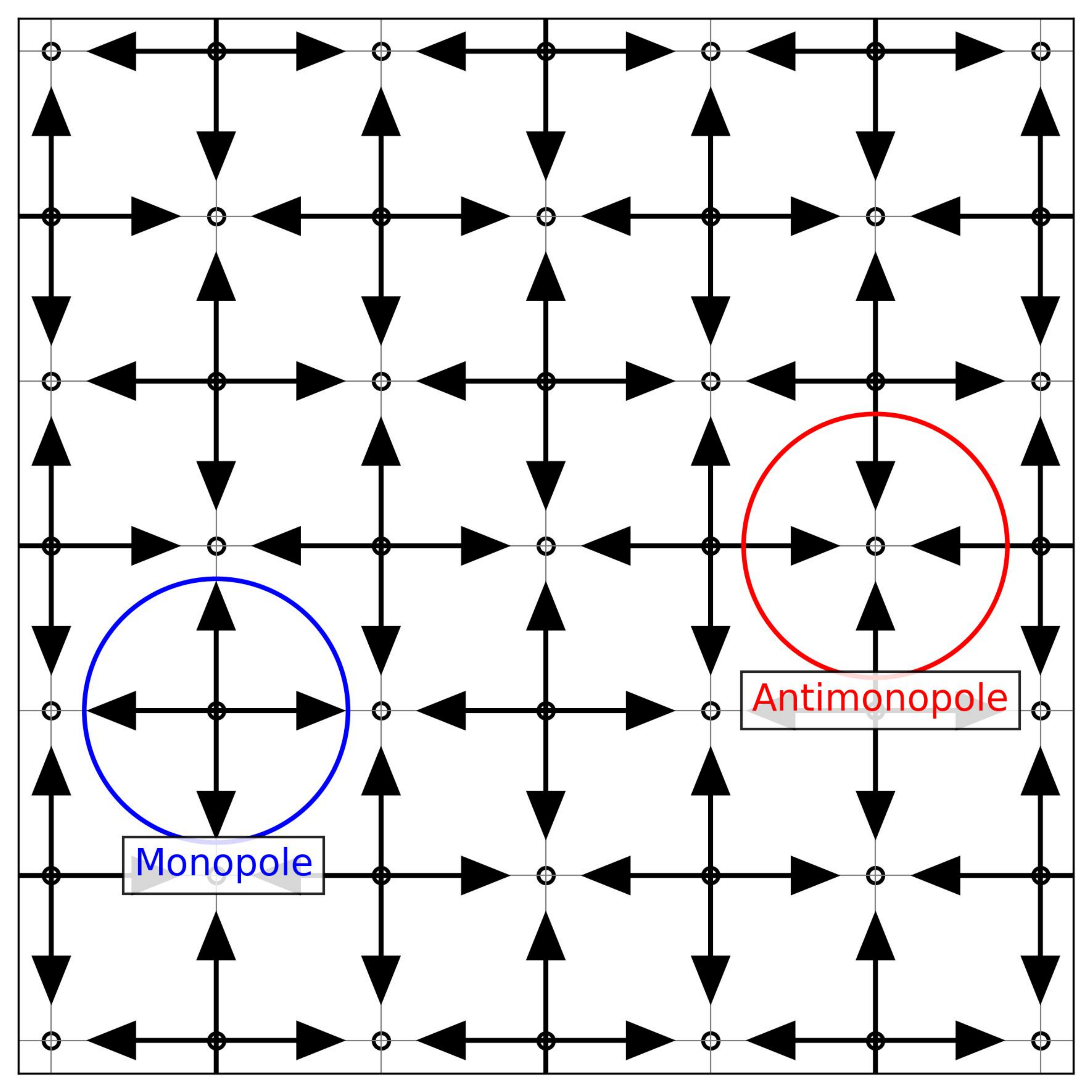}
\\
      (b) 
    \end{minipage}
    \caption{
    Real-space bond-current configurations on the square lattice generated by the $p^{+}$ and $p^{-}$ form factors. Arrows indicate the current direction. (a) Staggered loop-current state obtained from the susceptibility maxima, $\mathbf{Q}=(\pi,-\pi),(-\pi,\pi)$ for $p^{+}$ and $\mathbf{Q}=(\pi,\pi),(-\pi,-\pi)$ for $p^{-}$. Pink (yellow) plaquettes denote clockwise (counterclockwise) circulating currents. (b) Current configuration generated by $\mathbf{Q}=(\pi,\pi),(-\pi,-\pi)$ for $p^{+}$ and $\mathbf{Q}=(\pi,-\pi),(-\pi,\pi)$ for $p^{-}$. The resulting pattern lacks closed loops and exhibits alternating monopole-antimonopole structures.
    }
    \label{fig:RealSquare}
\end{figure*}

The real-space current configurations corresponding to these ordering vectors are shown in Fig.~\ref{fig:RealSquare}. For the dominant susceptibility maxima [Fig.~\ref{fig:RealSquare}(a)], the currents form staggered closed loops around neighboring plaquettes, generating alternating magnetic fluxes and breaking both translational and time-reversal symmetries. By contrast, the symmetry-related wave vectors associated with weaker susceptibility, $\mathbf{Q}=(\pi,\pi),(-\pi,-\pi)$ for $p^{+}$ and $\mathbf{Q}=(\pi,-\pi),(-\pi,\pi)$ for $p^{-}$, produce noncirculating current patterns consisting of alternating source- and sink-like, namely monopole-antimonopole textures, as shown in Fig.~\ref{fig:RealSquare}(b). These results establish a direct correspondence between the momentum-space susceptibility and the topology of the ordered current state: the strongest susceptibility maxima systematically select ordering vectors that support closed-loop currents, whereas the symmetry-related wavevector with weaker susceptibility favors noncirculating monopole-antimonopole current textures. The susceptibility anisotropy, therefore, acts as a momentum-space indicator for loop-current formation.

\subsection{Triangular lattice}
\begin{figure}[t]
    \centering
    \includegraphics[width=0.45\linewidth]{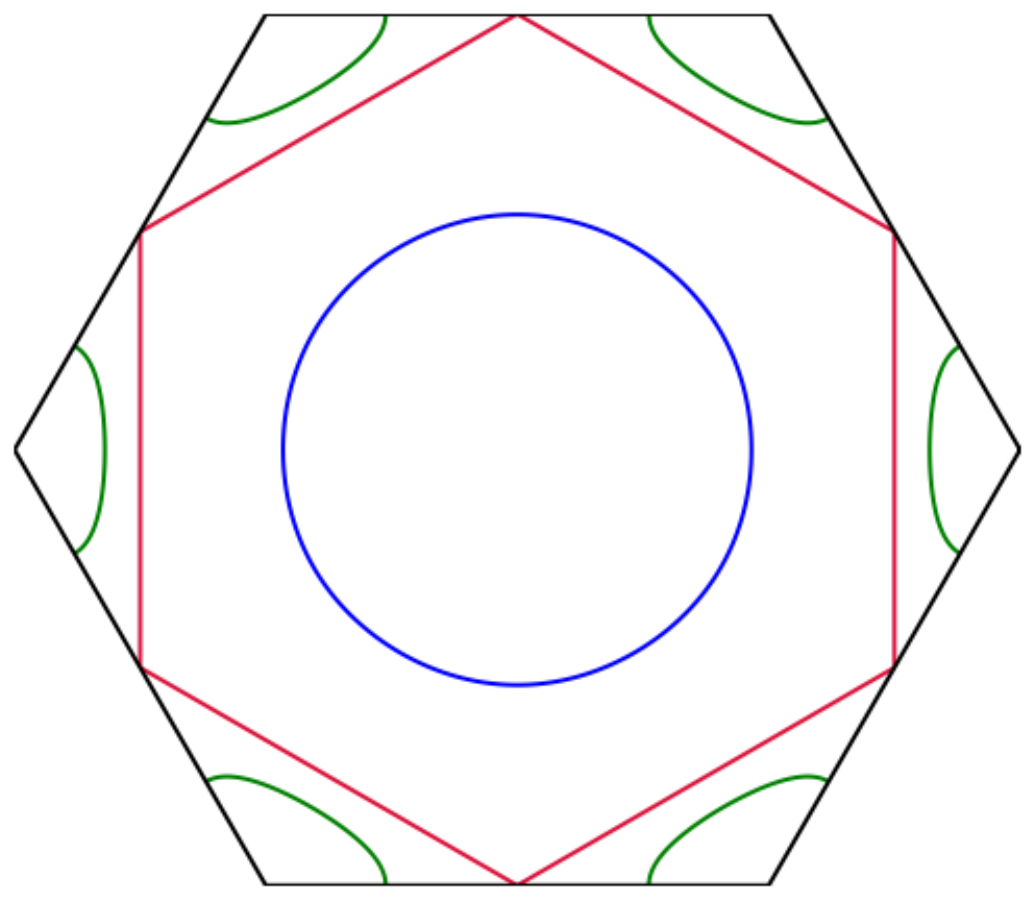}
   \caption{
Fermi surfaces of the triangular lattice at three representative fillings. The red curve corresponds to van Hove filling ($\mu=2.0t$), where the Fermi surface passes through the $M$ points of the Brillouin zone. The blue and green curves correspond to fillings below ($\mu=-1.5t$) and above ($\mu=2.5t$) van Hove filling, respectively. High-symmetry points are defined in Fig.~\ref{fig:EnergyDispersion}.
}
    \label{fig:FSTri}
\end{figure}
We next examine the bond-current susceptibilities associated with the four form factors defined in Eqs.~\eqref{eq.TriPxFormfac}--\eqref{eq.TrimpFormfac}. The calculations are performed at van Hove filling $\mu=2t$, where the density of states is maximal and the Fermi surface passes through the saddle points at the Brillouin-zone $M$ points (Fig.~\ref{fig:FSTri}). The corresponding susceptibilities are shown in Fig.~\ref{fig:SusTri}; their evolution with filling is presented in Appendix~\ref{app.ChangeOfSUSonMu}.

\begin{figure}[t]
    \centering
        \includegraphics[width=.7\linewidth]{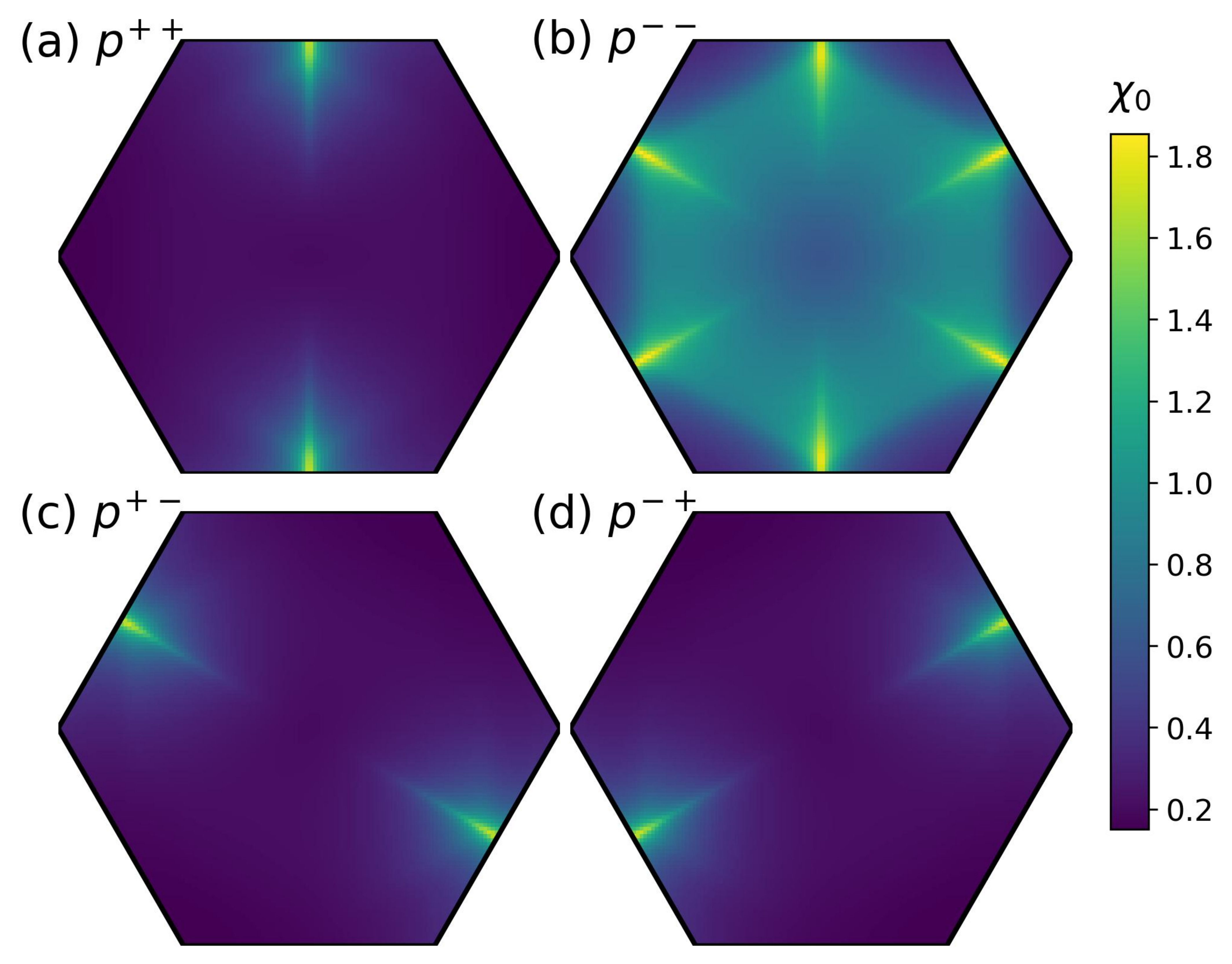}
    \caption{
   Momentum-space bond-current susceptibilities of the triangular lattice at van Hove filling ($\mu=2t$) and $T=0.01t$ for 
   (a) $p^{++}$, 
   (b) $p^{--}$, 
   (c) $p^{+-}$, and 
   (d) $p^{-+}$, in the first Brillouin zone, (labelling is shown in Fig.~\ref{fig:EnergyDispersion}).
   The dominant susceptibility peaks occur at the $M$ points on the zone boundary. }
    \label{fig:SusTri}
\end{figure}

In all channels, the dominant susceptibility maxima occur at the $M_i$ points on the Brillouin-zone boundary. The $p^{--}$ form factor preserves the sixfold rotational symmetry of the lattice and exhibits equivalent maxima at all six $M$ points. In contrast, the remaining channels display pronounced anisotropy: $p^{++}$ is enhanced at $(M_1,M_4)$, $p^{-+}$ at $(M_2,M_5)$, and $p^{+-}$ at $(M_3,M_6)$. In each case, the susceptibility selects only one pair of symmetry-related $M$ points, reducing the sixfold rotational symmetry of the susceptibility landscape to a twofold pattern. This is a $Z_3$ nematic state in momentum space, where one of three equivalent orientations is spontaneously selected despite the underlying lattice retaining its $C_6$ symmetry. The observed anisotropy therefore reflects the directional character of the underlying bond-current form factors.

\begin{figure*}
    \centering
    \includegraphics[width = \linewidth]{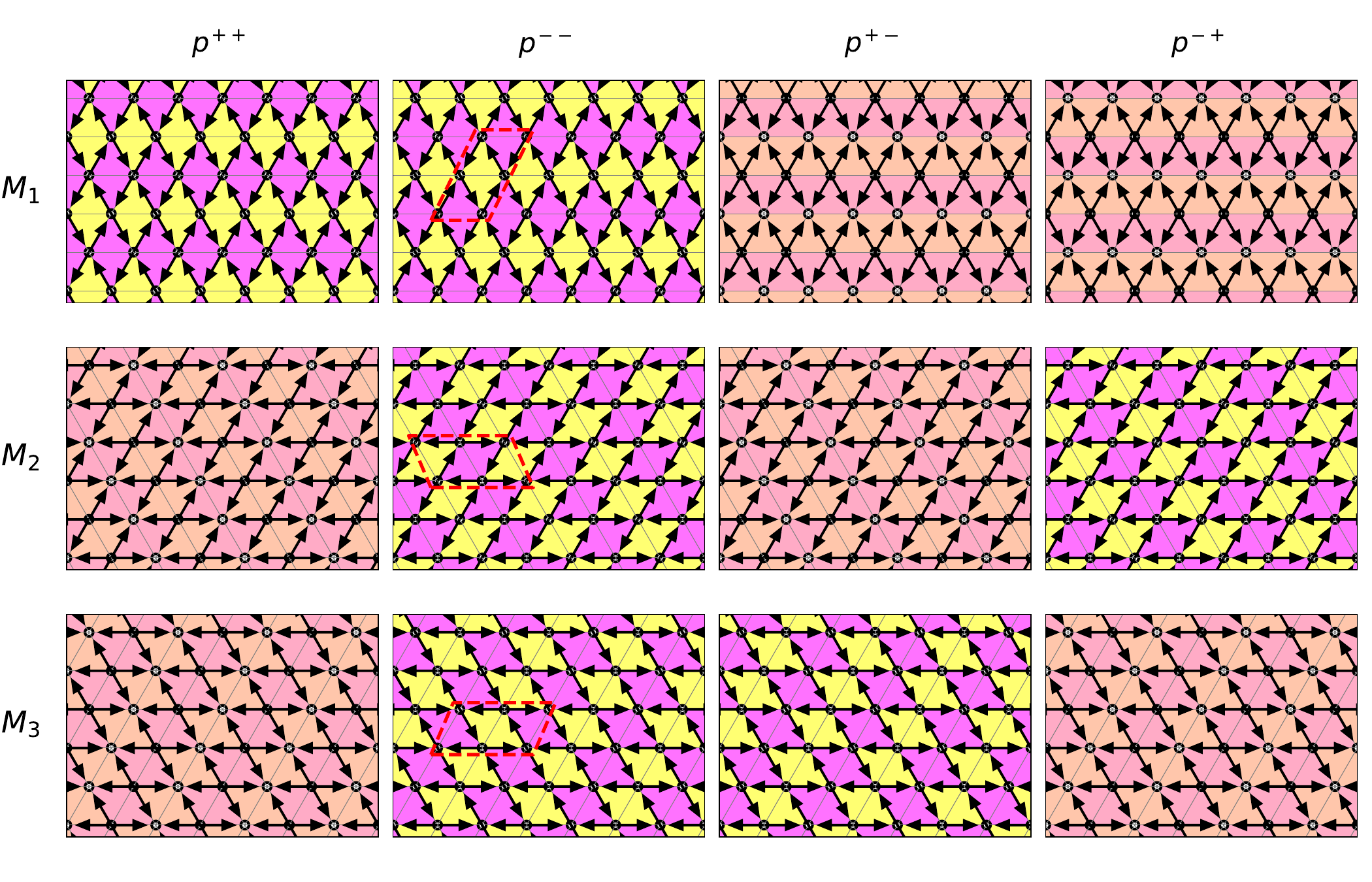}
    \caption{ Real-space bond-current configurations on the triangular lattice for the ordering vectors $M_1$ (top), $M_2$ (middle), and $M_3$ (bottom). Columns correspond to the form factors $p^{++}$, $p^{--}$, $p^{+-}$, and $p^{-+}$. Arrows indicate the bond-current direction, while pink (yellow) plaquettes represent clockwise (counterclockwise) circulating currents. The dashed-red lines describe the enlarged $1\times 2$ unit cells.
  }
    \label{fig:M1}
\end{figure*}

To determine the associated ordered states, we reconstruct the real-space current textures generated by the $M$-point ordering vectors. Since the current texture at $M_1$ and $M_4$, similary $M_2$ and $M_5$, and $M_3$ and $M_6$, differ only by an overall reversal of the current direction, representative configurations are shown in Fig.~\ref{fig:M1}. In all cases, the susceptibility maxima produce loop-current states analogous to the square-lattice flux phase. The circulating currents form diamond-shaped loops \cite{wang06}, with the inactive bond direction determined by the ordering vector: currents along $\mathbf{a}_1$, $\mathbf{a}_3$, and $\mathbf{a}_2$ vanish for $\mathbf{Q}=M_1(M_4)$, $\mathbf{Q}=M_2(M_5)$, and $\mathbf{Q}=M_3(M_6)$, respectively.

These loop-current states break time-reversal symmetry and enlarge the translational unit cell to a $1\times 2$ superstructure. In contrast, symmetry-related vectors located in regions of suppressed susceptibility generate noncirculating monopole-antimonopole current textures. As in the square lattice, a clear correspondence emerges between susceptibility anisotropy and current topology: susceptibility maxima favor loop-current formation, whereas weak-susceptibility regions are associated with noncirculating monopole--antimonopole patterns.

\subsection{Kagome lattice}
\begin{figure}[t]
    \centering
    \includegraphics[width=0.45\linewidth]{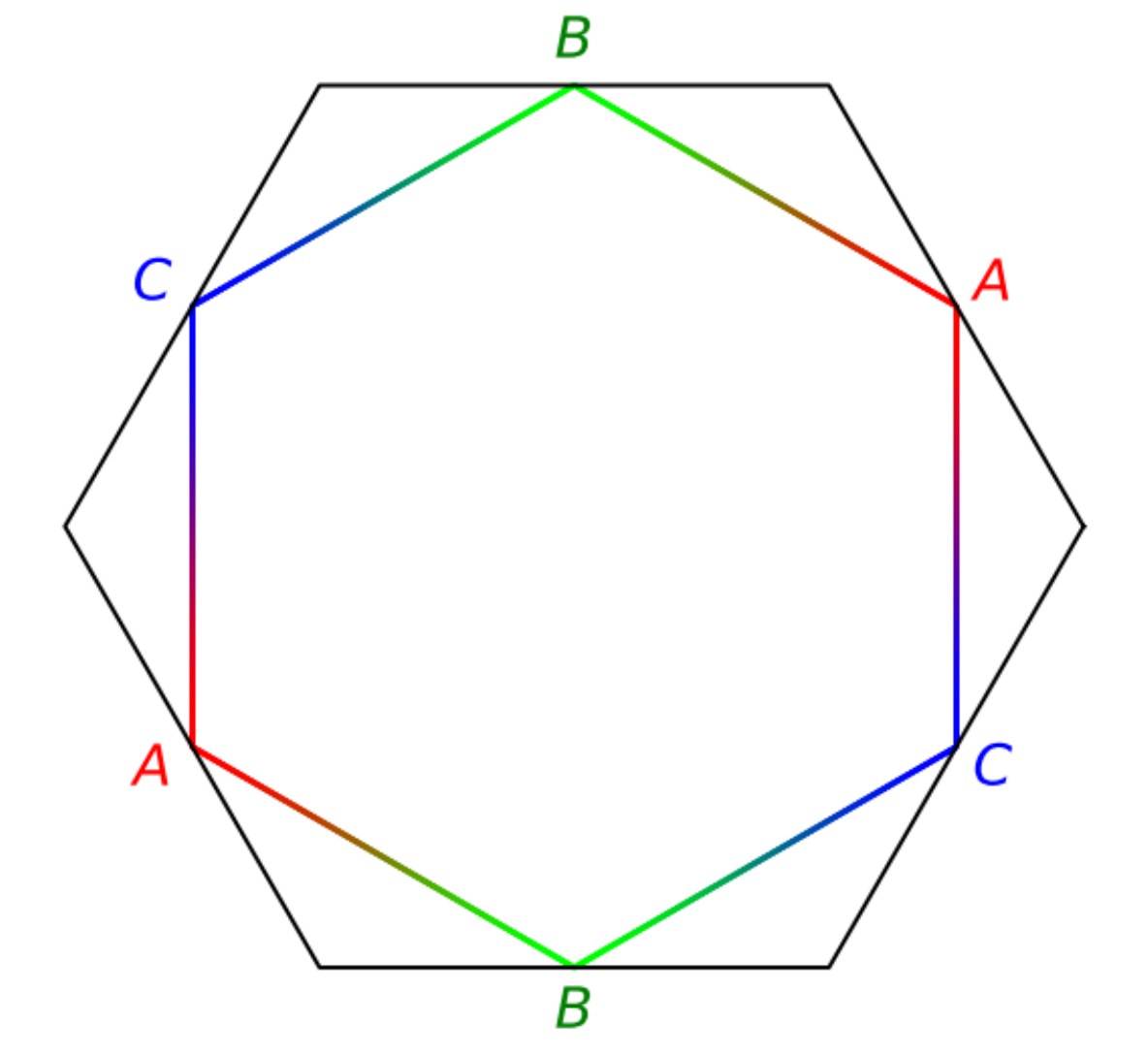}
    \caption{Fermi surface of the kagome lattice at van Hove filling of the dispersive $p$ band. The sublattice weights are indicated by red ($A$), green ($B$), and blue ($C$). High-symmetry points are defined in Fig.~\ref{fig:EnergyDispersion}.}
    \label{fig:FSKagome}
\end{figure}
The kagome lattice exhibits van Hove singularities at $\mu=0$, and $-2t$ for the upper dispersive ($p$), and lower dispersive ($m$) bands, respectively (Fig.~\ref{fig:EnergyDispersion}). In this work, we focus on van Hove filling of the upper dispersive band ($\mu=0$), where the saddle points coincide with the Brillouin-zone $M$ points. At this filling, the Fermi surface touches the Brillouin-zone boundary, enhancing particle-hole scattering at the $M$-point wave vectors and thereby maximizing the bond-current susceptibility. The corresponding Fermi surface, together with its sublattice weight, is shown in Fig.~\ref{fig:FSKagome}.

As discussed in Sec.~\ref{sec.Theory}, bond-current order on the kagome lattice can emerge either as a $3\mathbf{Q}$ state, characterized by independent ordering vectors on the three inequivalent bonds, or as a $1\mathbf{Q}$ state constructed from symmetry-adapted combinations of these bond-current channels. In the following, we examine both scenarios and compare their susceptibility landscapes and associated real-space current configurations.

\subsubsection{\texorpdfstring{$3\mathbf{Q}$}{3Q}-ordering}
\begin{figure}[t]
    \centering
   \includegraphics[width=0.65\linewidth,trim= 50 40 40 50,clip]{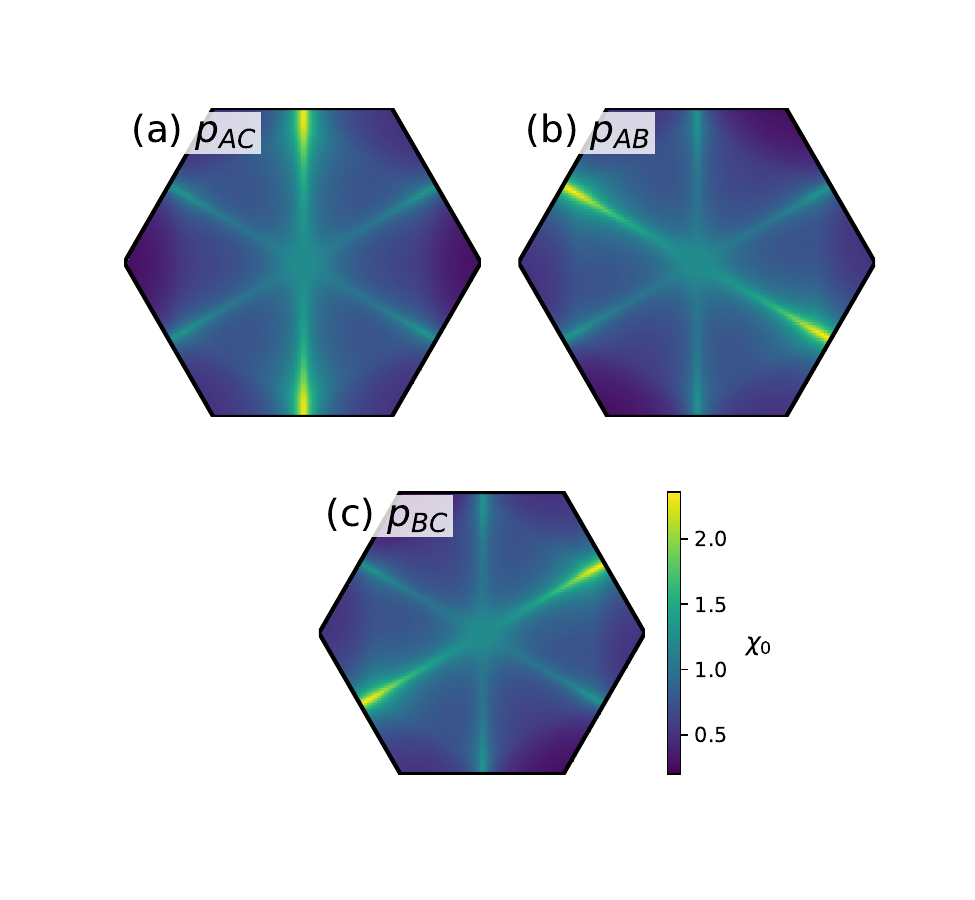}
    \caption{
Momentum-space bond-current susceptibilities of the kagome lattice at the $p$-band van Hove filling ($\mu=0$) and $T=0.01t$ for 
(a) $p_{AC}$, 
(b) $p_{AB}$, and 
(c) $p_{BC}$. 
Each bond-current channel exhibits maxima at a distinct pair of $M$ points.
    }
    \label{fig:SusKagome}
\end{figure}

We first consider the bond-current fluctuations associated with the form factors $p_{AC}$, $p_{AB}$, and $p_{BC}$, Eqs.~\eqref{eq.PAC}--\eqref{eq.PBC}. The corresponding susceptibilities at van Hove filling of the $p$-band are shown in Fig.~\ref{fig:SusKagome}.

For each form factor, the susceptibility is maximized at a distinct pair of symmetry-related $M$ points:
\begin{align}
    \mathbf{Q}_{AC} &= M_1,M_4= \pm(0,\frac{2\pi}{\sqrt{3}})  ,\\
    \mathbf{Q}_{AB} &= M_3, M_6 = \pm(\pi,-\frac{\pi}{\sqrt{3}}) , \\
    \mathbf{Q}_{BC} &= M_2, M_5= \pm(\pi,\frac{\pi}{\sqrt{3}}).
\end{align}
Each bond-current channel therefore selects a different pair of ordering vectors, reflecting the strong momentum-space anisotropy imposed by the corresponding bond-current form factor. Similar to the triangular lattice, different bond-current channels preferentially enhance fluctuations at different $M$ points. The susceptibility landscape therefore exhibits a momentum-space $Z_3$ nematicity, despite the underlying $C_6$ symmetry of the kagome lattice.

The resulting susceptibility profiles closely resemble the bond-order susceptibilities reported by Tazai \textit{et al.}~\cite{tazai23}. However, an important distinction is that the bond-current form factors considered here are derived directly from the symmetry decomposition of the nearest-neighbor bond-charge interaction, providing a systematic classification of all symmetry-allowed bond-current channels rather than focusing on a specific bond-order pattern in each site of the lattice.

To construct the corresponding ordered state, we introduce the bond-current order-parameter vector
\begin{equation}
    \Delta_b = (\Delta_{AC}(\mathbf{Q}_{AC}),\Delta_{AB}(\mathbf{Q}_{AB}),\Delta_{BC}(\mathbf{Q}_{BC})),
    \label{eq.DeltaChiralFluxPhase}
\end{equation}
and consider a time-reversal-symmetry-breaking solution in which all components are purely imaginary and share a common phase,
$\Delta_{\alpha\beta}(\mathbf{Q}_{\alpha\beta})=i$.
This equal-phase ansatz is motivated by the symmetry equivalence of the three bond-current channels and by their nearly identical susceptibility enhancements at the corresponding ordering vectors, making a symmetric superposition the natural candidate for the leading $3\mathbf{Q}$ order. The resulting ordered state is a chiral flux phase discussed below. 

Since the three bond-current components correspond to independent ordering vectors, other equal-magnitude sign combinations, such as $(i,i,-i), (i,-i,i)$, and $(i,-i,-i)$, are also symmetry-allowed. These choices modify the current directions on individual bonds but preserve the overall topology of the ordered state, yielding symmetry-related realizations of the same chiral flux phase.  Determining the energetically preferred phase relation, however, requires a microscopic free-energy analysis, which beyond the scope of the present work.

\begin{figure*}
    \centering
    \begin{minipage}{0.9\linewidth}
    (a) $(i,i,i)$ \\
        \centering
        \includegraphics[width=0.8\linewidth]{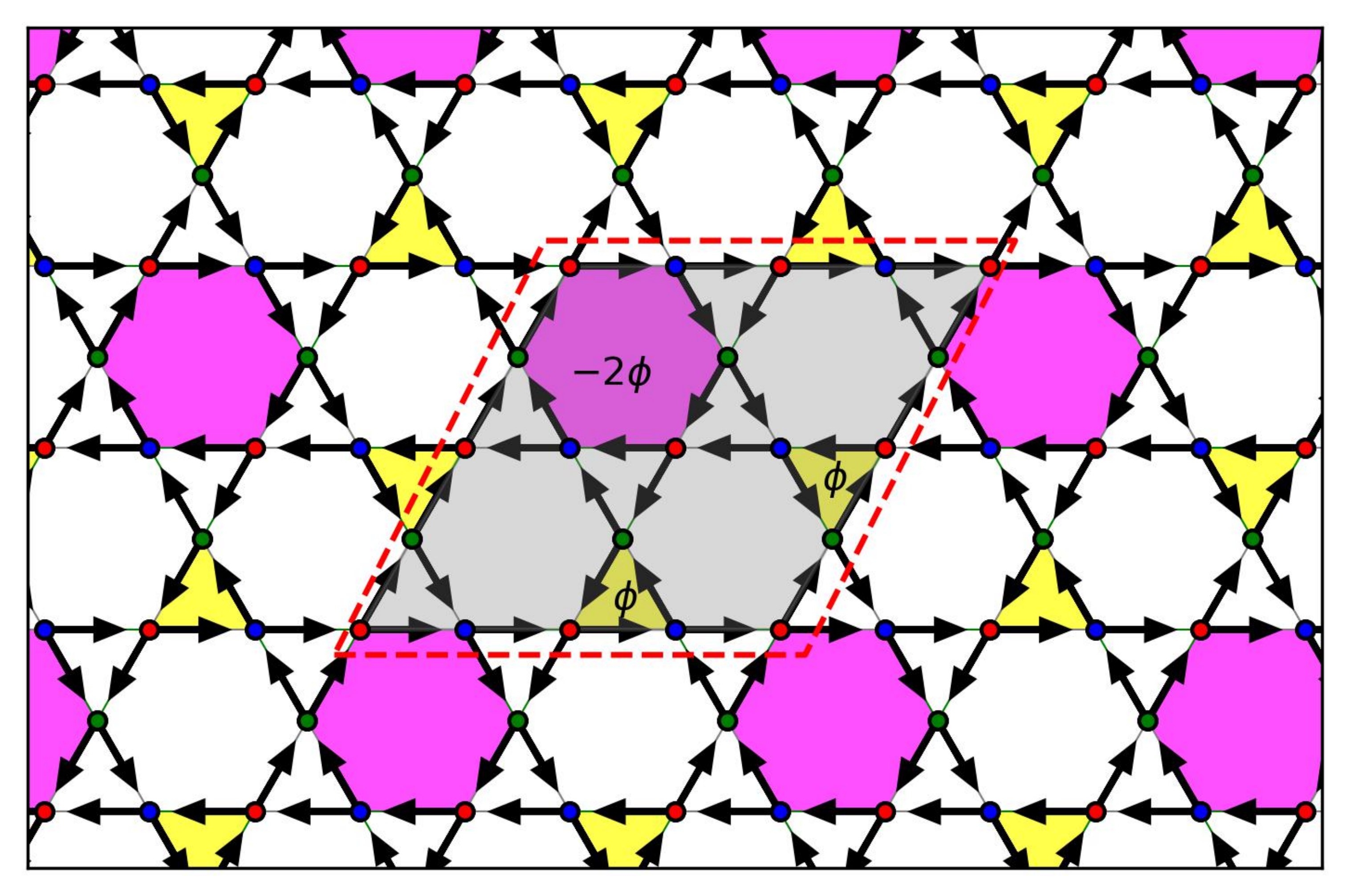} 
    \end{minipage}
        \begin{minipage}{0.35\linewidth}
        (b) $(i,-i,-i)$\\
        \centering
        \includegraphics[width=\linewidth]{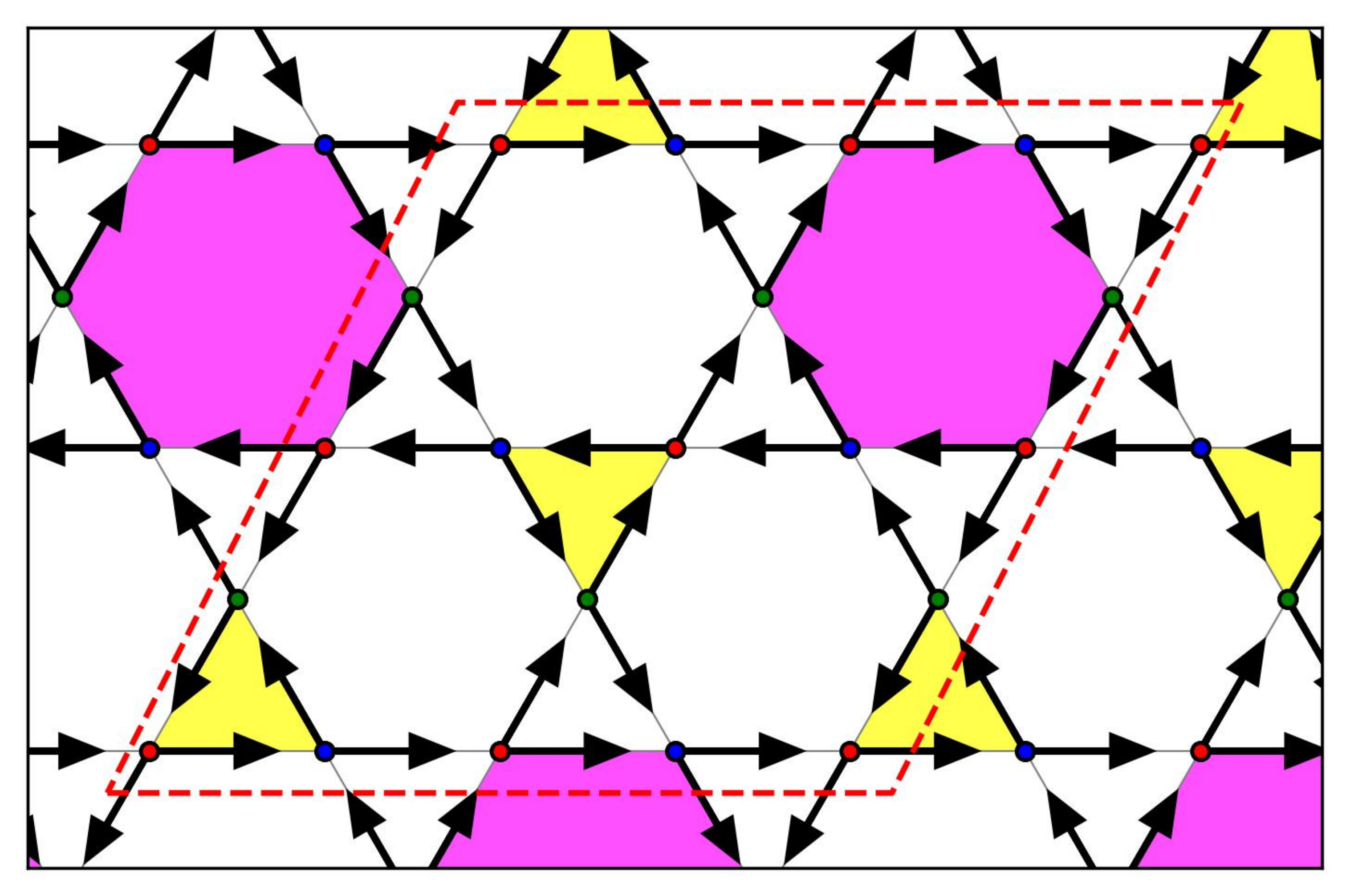}
        \end{minipage}
        \begin{minipage}{0.35\linewidth}
        (c) $(-i,i,-i)$\\
        \centering
        \includegraphics[width=\linewidth]{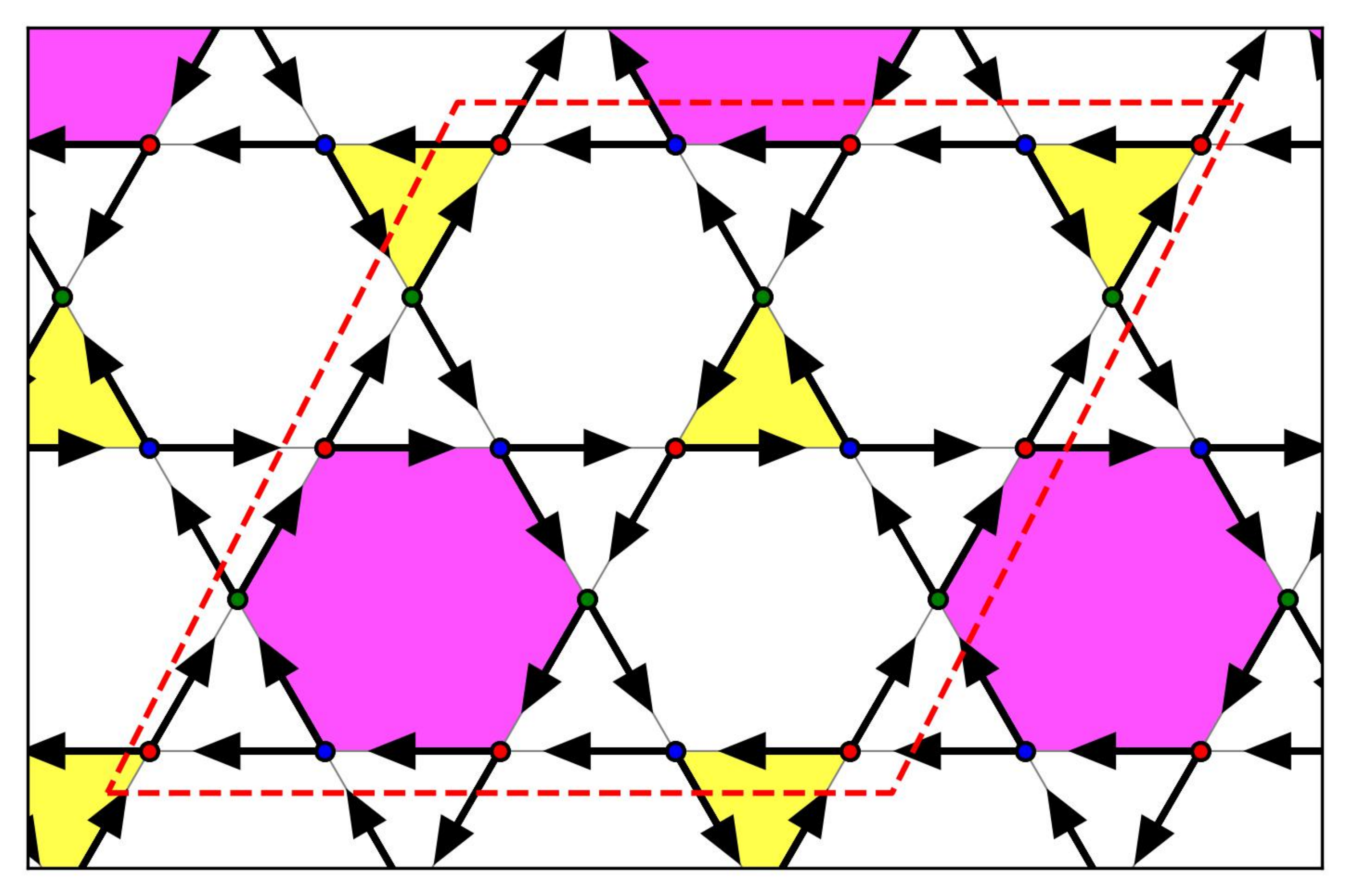}
        \end{minipage}

        \begin{minipage}{0.35\linewidth}
        (d) $(-i,-i,i)$\\
        \centering
        \includegraphics[width=\linewidth]{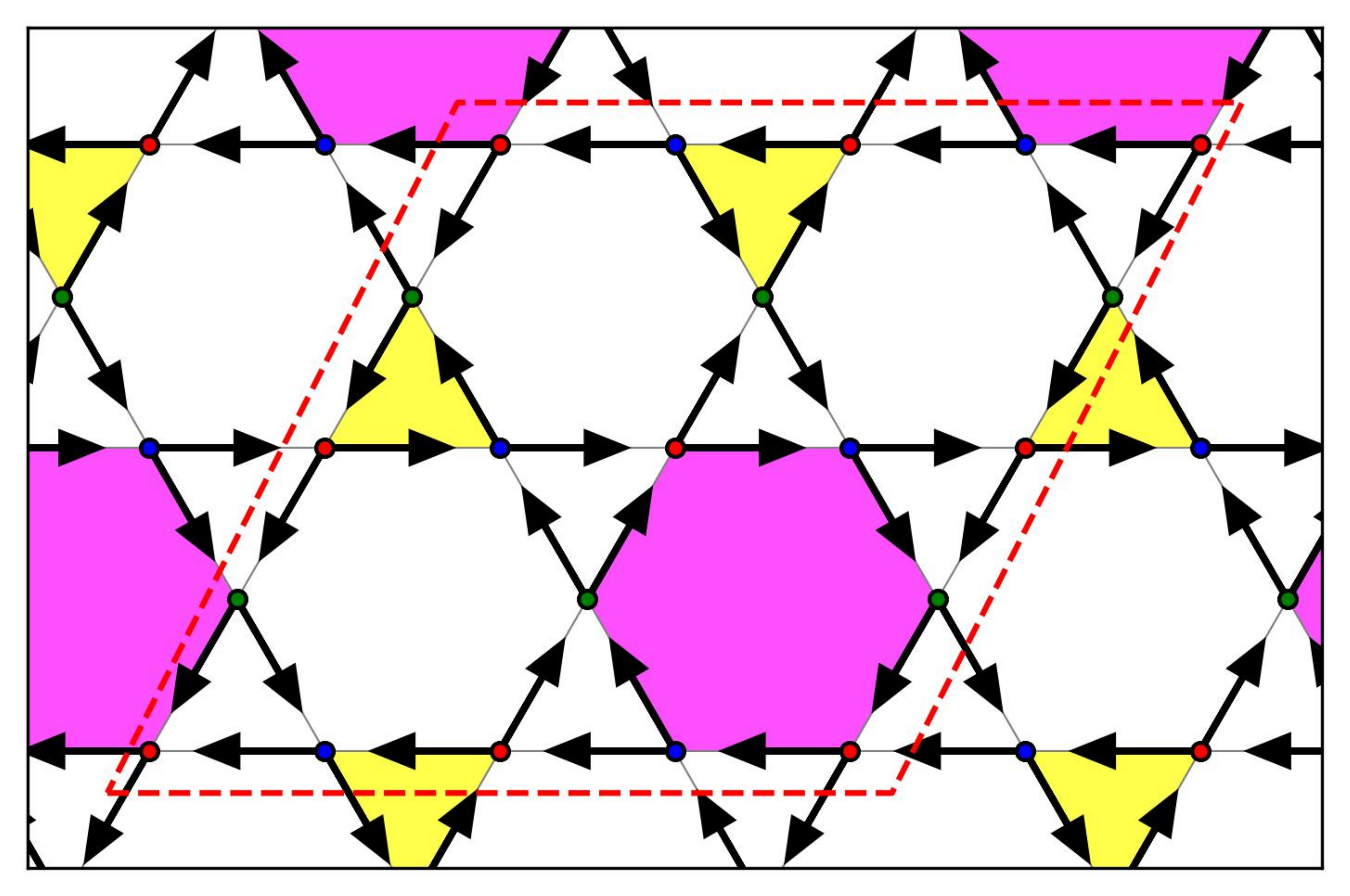}
        \end{minipage}
        \begin{minipage}{0.35\linewidth}
        (e) $(-i,-i,-i)$\\
        \centering
        \includegraphics[width=\linewidth]{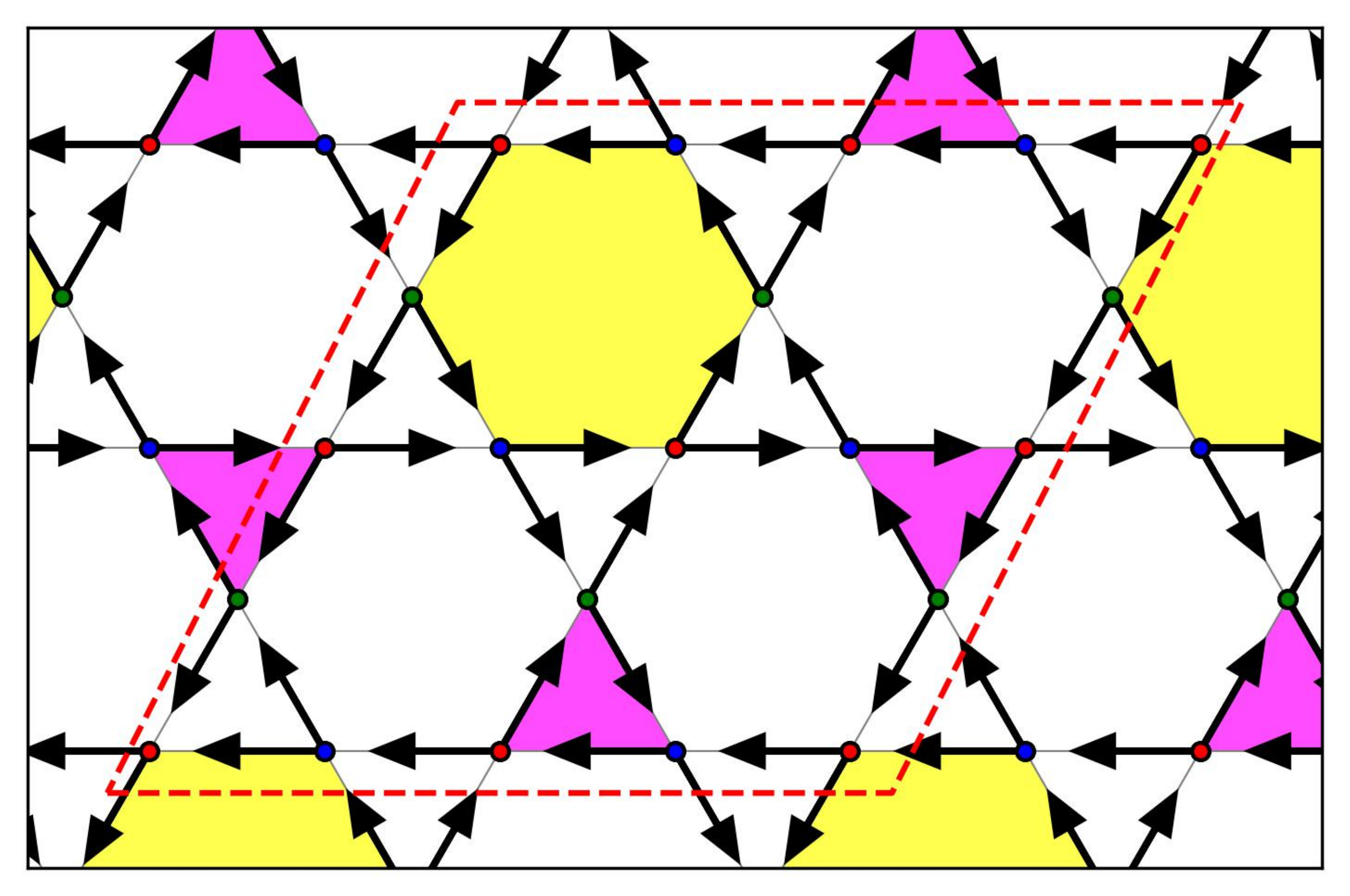}
    \end{minipage}
\caption{
Real-space bond-current configurations on the kagome lattice generated by the ordering-vector set $(\mathbf{Q}_{AC},\mathbf{Q}_{AB},\mathbf{Q}_{BC})$. Arrows indicate the current direction on each bond, while pink (yellow) plaquettes denote clockwise (counterclockwise) circulating currents. Panel (a) shows a representative chiral flux phase with a $2\times2$ supercell (red dashed line) containing one hexagonal and two triangular current loops. Panels (b)--(e) illustrate symmetry-equivalent realizations obtained from different relative signs of the bond-current expectation values $(\Delta_{AC},\Delta_{AB},\Delta_{BC})$ and equivalent choices of the ordering vectors at the $M$ points. Although the current directions on individual bonds are rearranged, all configurations exhibit the same loop-current topology and therefore belong to the same chiral flux phase.
}
    \label{fig:ChiralFlux}
\end{figure*}
The resulting real-space current patterns, shown in Fig.~\ref{fig:ChiralFlux}, corresponds to the chiral flux phase \cite{feng21,feng21a}. 
This state breaks both translational and time-reversal symmetries and enlarges the unit cell to a $2\times2$ structure. Within the enlarged unit cell, the currents form one hexagonal and two triangular closed loops. The circulation around the hexagon is opposite to that of the triangular plaquettes, generating a staggered-flux pattern with fluxes $-2\phi$ and $\phi$, respectively.

For a given $2\times2$ supercell, the chiral flux phase has eight symmetry-equivalent realizations. Four of them exhibit clockwise circulation around the hexagonal loop accompanied by counterclockwise circulation around the two triangular loops, while the remaining four are obtained by reversing all bond currents through time-reversal symmetry. Within each group of four, the hexagonal loop occupies one of the four symmetry-equivalent positions in the enlarged unit cell, corresponding to different choices of the symmetry-related ordering vectors $\mathbf{Q}_{\alpha\beta}$ at the $M$ points, or different
relative signs of the bond-current expectation value. Representative examples are shown in Fig.~\ref{fig:ChiralFlux}(a)–(e): panels (a)–(d) illustrate the four translational variants with the same circulation sense, whereas panel (e) shows the corresponding time-reversed configuration. Although the current directions on individual bonds differ among these configurations, they share the same loop-current topology and staggered-flux pattern. They therefore represent symmetry-equivalent realizations of the same chiral flux phase rather than distinct ordered states.

The coexistence of inequivalent triangular and hexagonal fluxes gives rise to a chiral magnetic response and spontaneous time-reversal-symmetry breaking. Such chiral loop-current states have been proposed as candidate ordered phases in kagome metals and may be relevant to the experimentally observed time-reversal-symmetry breaking in AV$_3$Sb$_5$ compounds \cite{jiang21}.

%Among the possible $3\mathbf{Q}$ bond-current configurations, this phase follows directly from the dominant susceptibility maxima at van Hove filling and therefore represents the leading bond-current order within the present analysis. Alternative current states can be obtained from different combinations of ordering vectors and may become favorable away from van Hove filling; one such example is discussed in Appendix~\ref{app.ChangeOfSUSonMu}.

\subsubsection{\texorpdfstring{$1\mathbf{Q}$}{1Q}-ordering}
\begin{figure}[t]
    \centering
  \includegraphics[width = 0.7\linewidth]{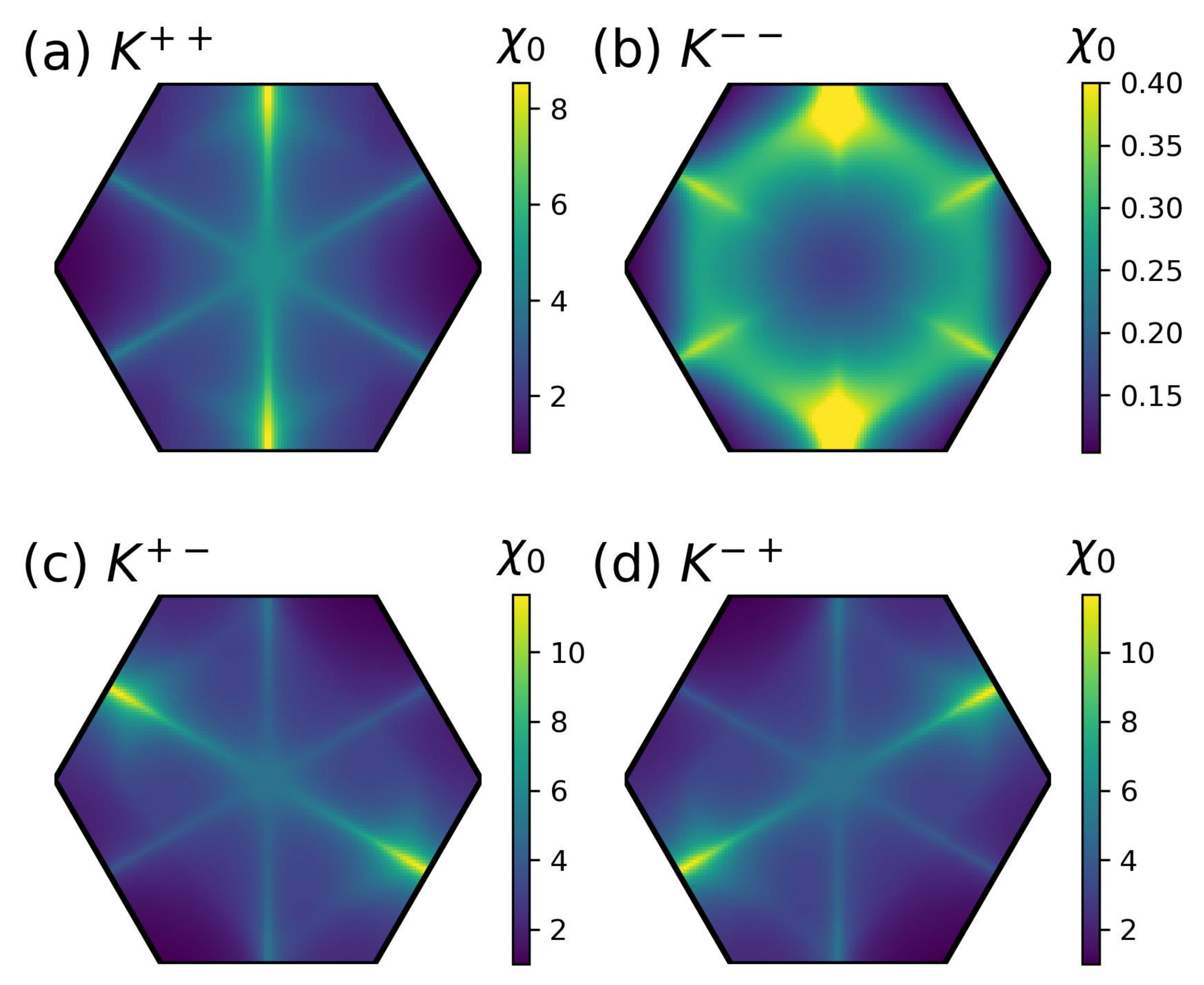}
   \caption{Momentum-space bond-current susceptibilities of the kagome lattice at the $p$-band van Hove filling ($\mu=0$) and $T=0.01t$ in the $1\mathbf{Q}$ ordering scenario. Panels (a)--(d) correspond to the form factors $K^{++}$, $K^{--}$, $K^{+-}$, and $K^{-+}$, respectively. While the $K^{--}$ channel exhibits enhanced susceptibility at all six $M$ points, the remaining channels display pronounced anisotropy, with susceptibility enhancement restricted to a single pair of $M$ points.}
    \label{fig:SusKagome1Q}
\end{figure}

We next examine the $1\mathbf{Q}$ scenario, in which all bond-current components share a common ordering vector. The corresponding symmetry-adapted form factors, Eqs.~\eqref{eq.Kx}--\eqref{eq.Kmp}, yield the susceptibilities shown in Fig.~\ref{fig:SusKagome1Q}.

As in the triangular lattice, the dominant susceptibility maxima are located at the Brillouin-zone $M_i$ points. The $K^{--}$ channel exhibits  enhanced susceptibility at all six symmetry-related $M$ points and maxima at $(M_1,M_4)$. In contrast, the $K^{++}$, $K^{+-}$, and $K^{-+}$ channels display pronounced anisotropy, with dominant peaks at $(M_1,M_4)$, $(M_3,M_6)$, and $(M_2,M_5)$, respectively. By selecting only one of the three symmetry-equivalent pairs of $M$ points, these channels exhibit a momentum-space $Z_3$ nematicity of the kagome lattice.

\begin{figure*}
    \centering
    \begin{minipage}{\linewidth}
    (a) $\mathbf{Q}=M_1,M_4$ \\
        \centering
        \includegraphics[width=0.82\linewidth]{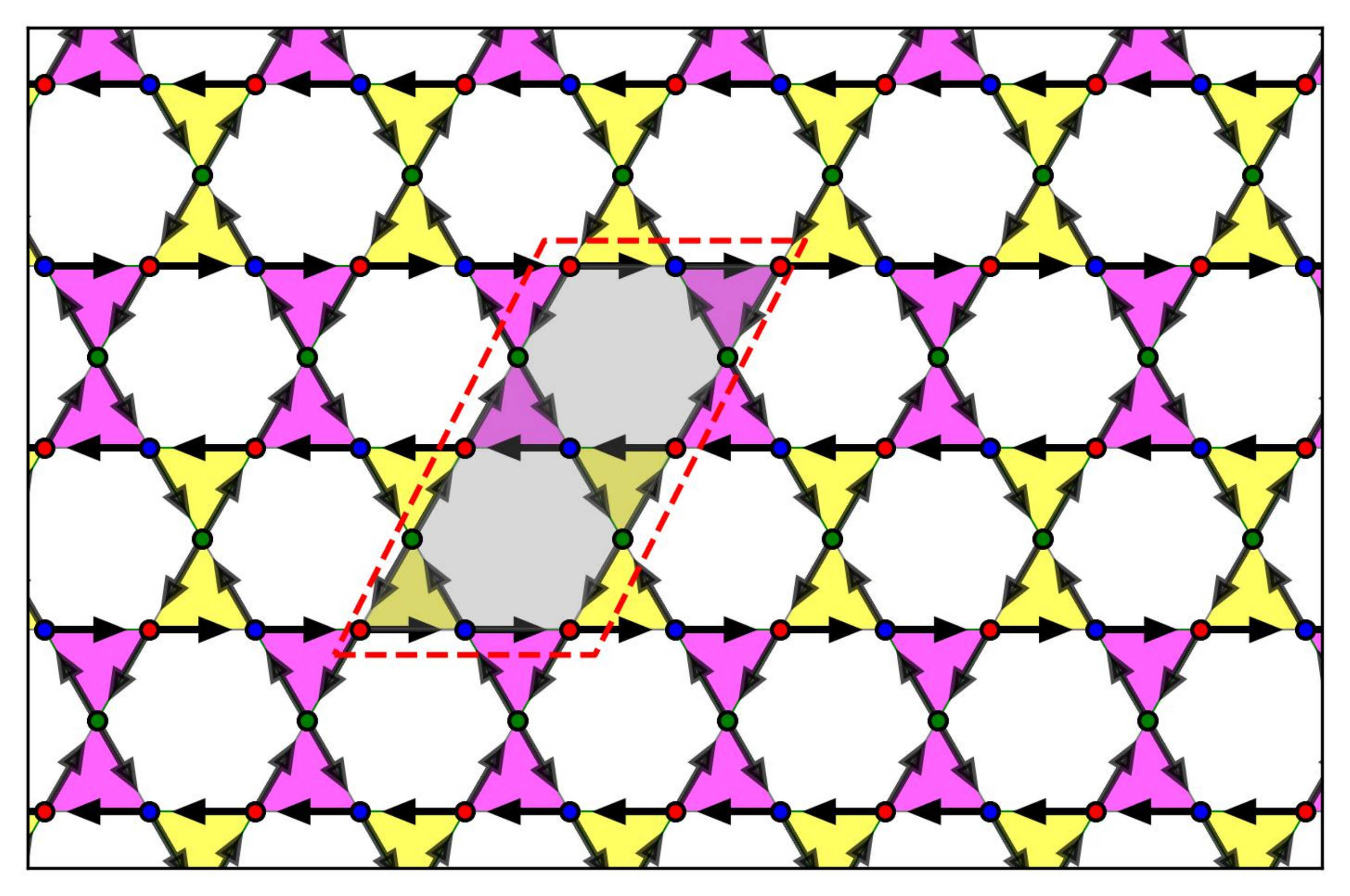} %0.6
    \end{minipage}

        \begin{minipage}{0.4\linewidth}
        (b) $\mathbf{Q}=M_2,M_5$ \\
        \centering
        \includegraphics[width=\linewidth]{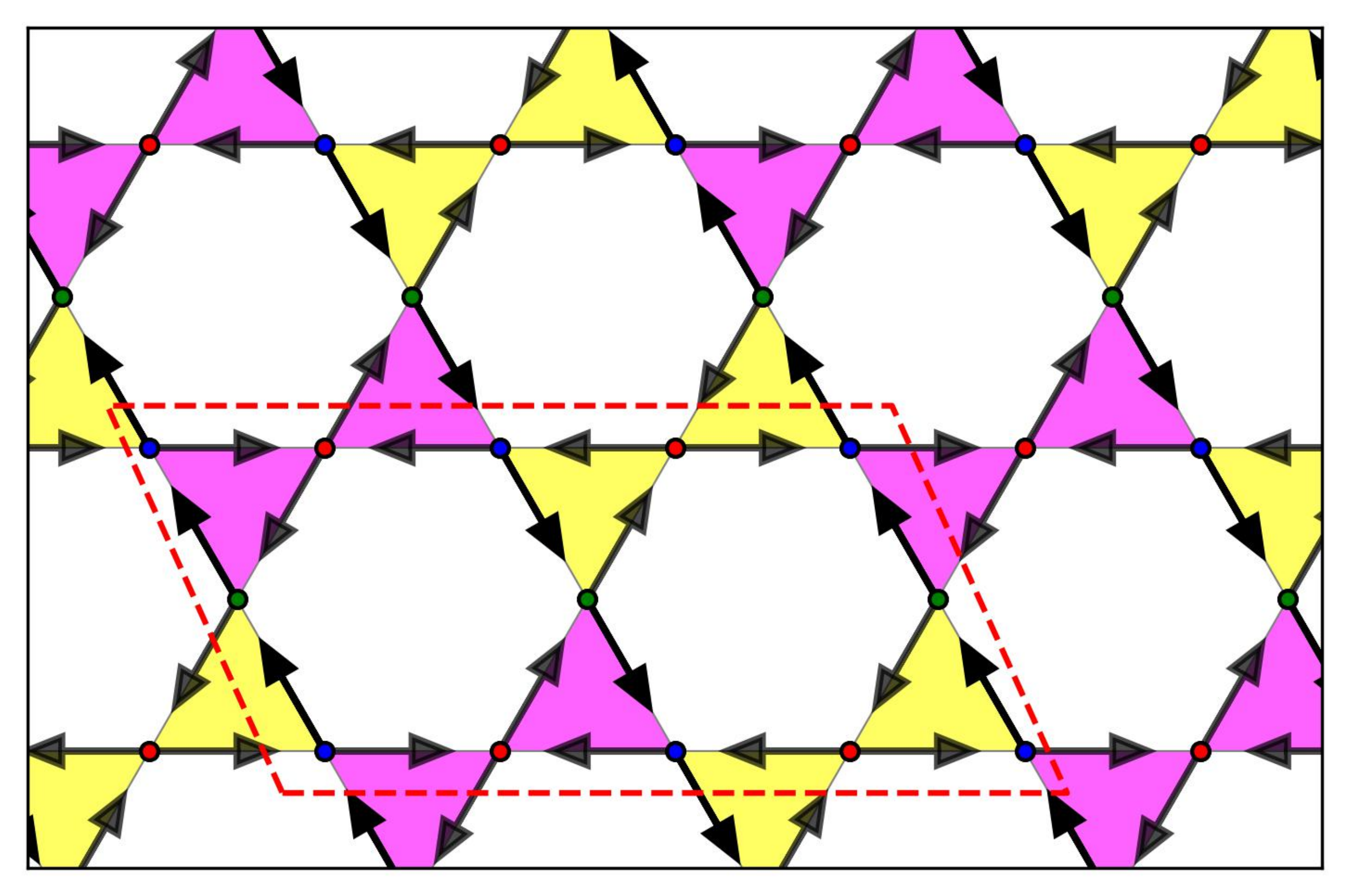}
            \end{minipage}
        \begin{minipage}{0.4\linewidth}
        (c) $\mathbf{Q}=M_3,M_6$ \\
        \centering
        \includegraphics[width=\linewidth]{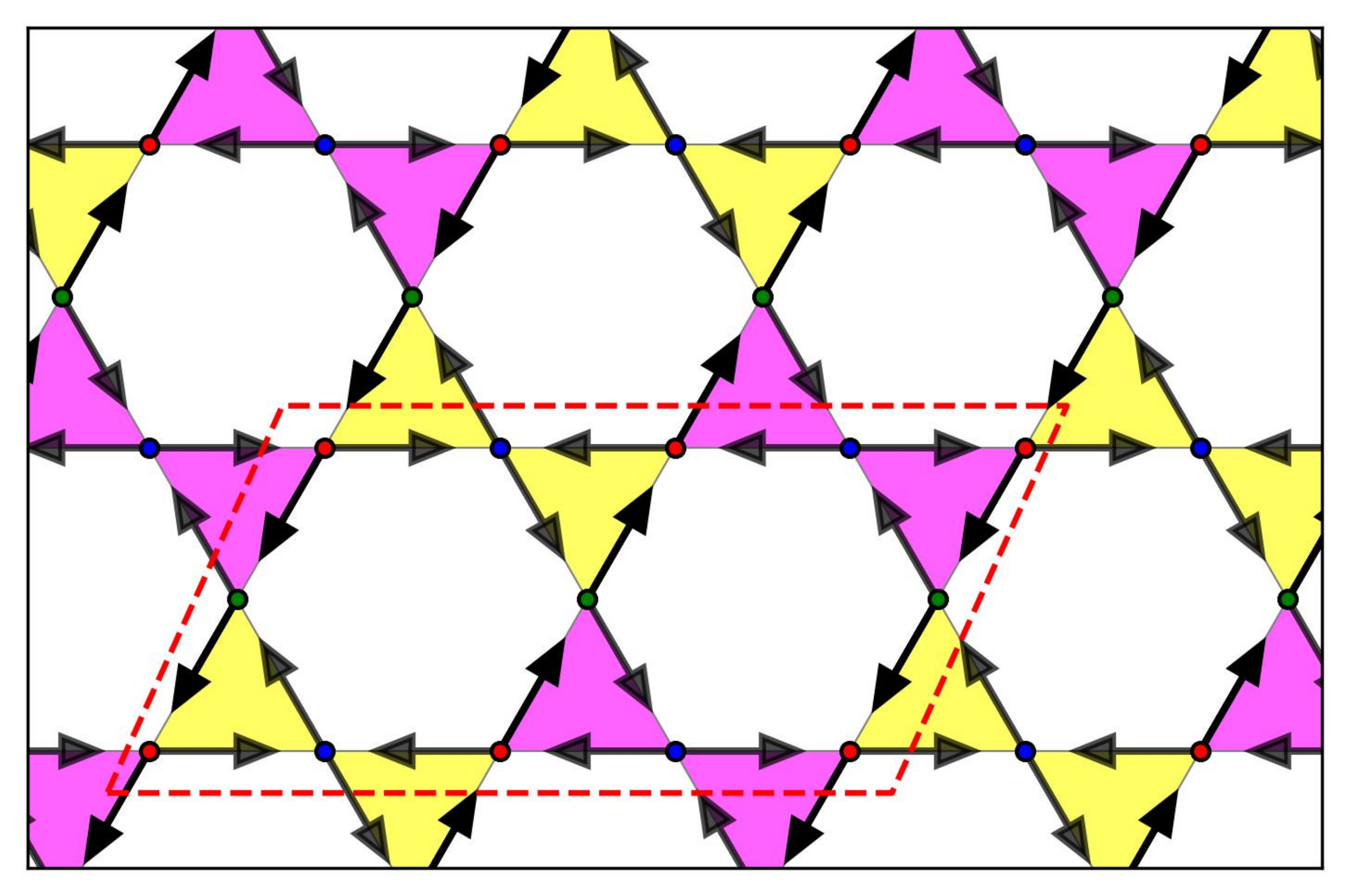}
    \end{minipage}
    \caption{
Real-space bond-current configurations of the kagome lattice in the $1\mathbf{Q}$ ordering state corresponding to the $D_{2a}$ phase generated by the $K^{--}$ form factor. Panels (a)–(c) show the current patterns for the symmetry-related ordering vectors $\mathbf{Q}=M_1,M_4$, $\mathbf{Q}=M_2,M_5$, and $\mathbf{Q}=M_3,M_6$, respectively. Arrows indicate the bond-current direction, while pink (yellow) plaquettes denote clockwise (counterclockwise) circulating currents. The dashed-red lines describe the enlarged $1\times2$ unit cell.
}
    \label{fig:LoopCurrent1QD2a}
\end{figure*}

\begin{figure*}
    \centering
    \begin{minipage}{\linewidth}
    (a) $K^{++}$ \\
        \centering
        \includegraphics[width=0.82\linewidth]{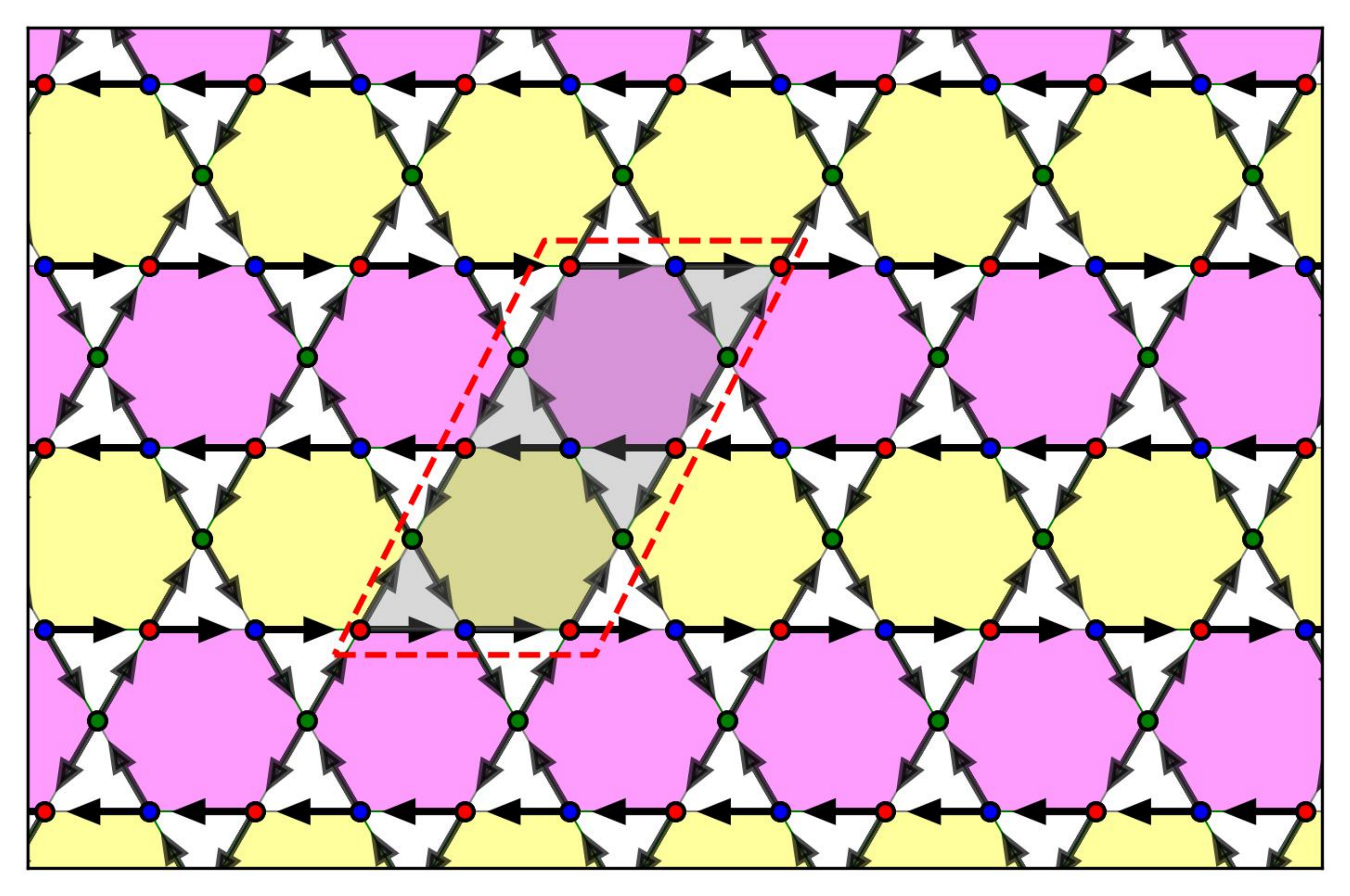}
    \end{minipage}
    
        \begin{minipage}{0.4\linewidth}
         (b) $K^{-+}$ \\
        \centering
        \includegraphics[width=\linewidth]{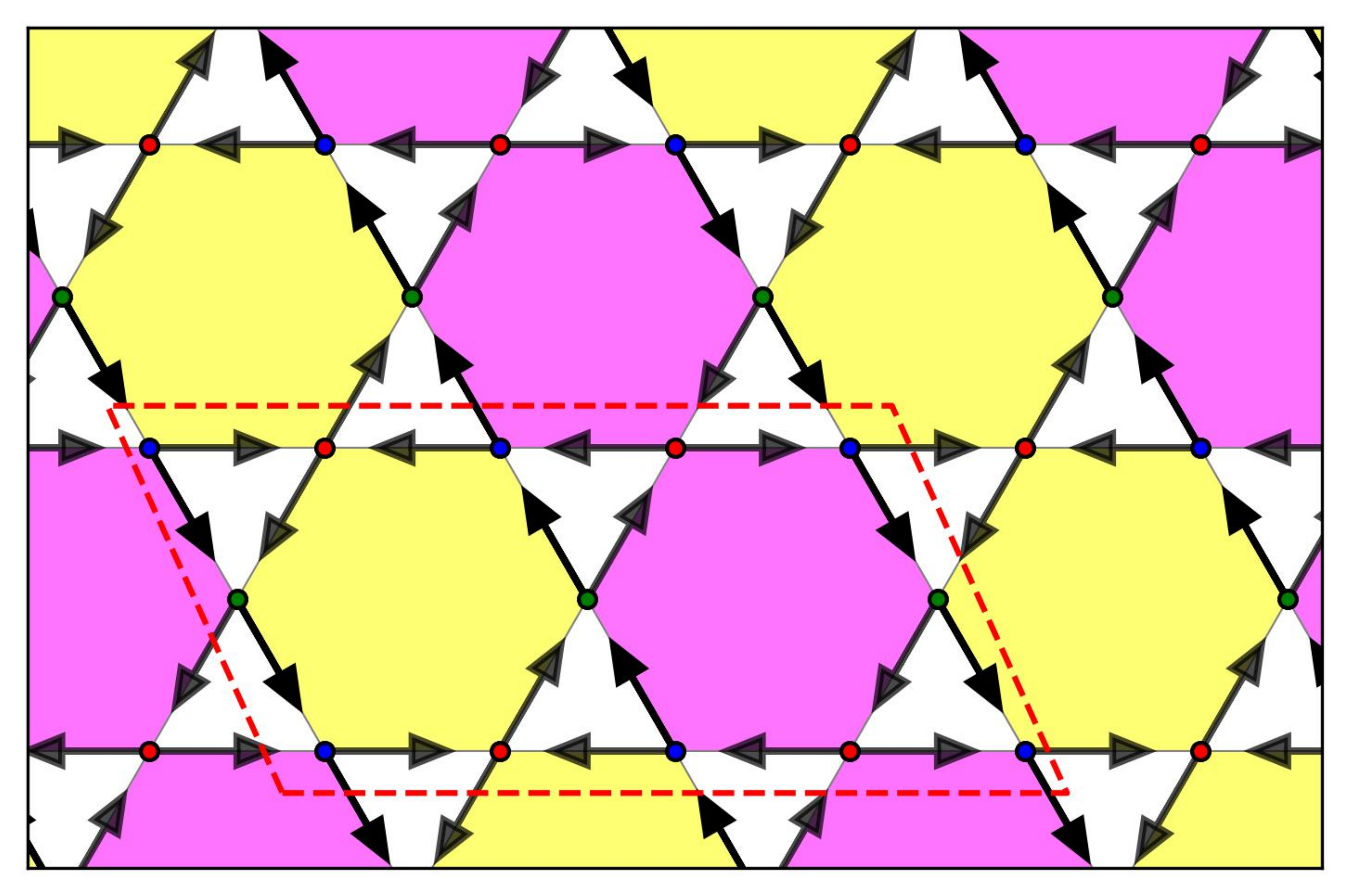}
       \end{minipage}
        \begin{minipage}{0.4\linewidth}
         (c) $K^{+-}$ \\
        \centering
        \includegraphics[width=\linewidth]{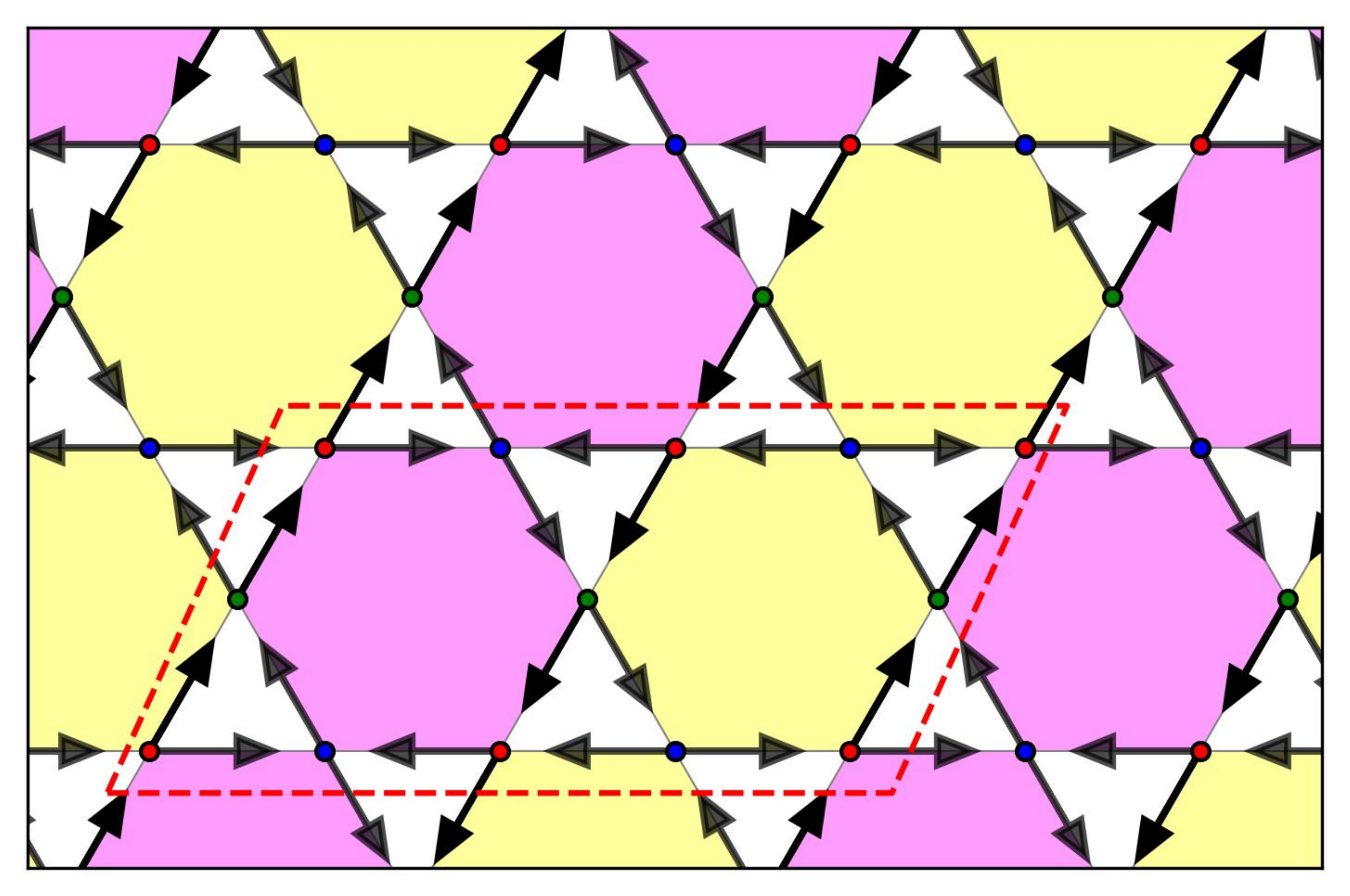}
    \end{minipage}
    \caption{
Real-space bond-current configurations of the kagome lattice in the $1\mathbf{Q}$ ordering state corresponding to the $D_{2b}$ phase. Panels (a)–(c) are generated by the $K^{++}$, $K^{-+}$, and $K^{+-}$ form factors, respectively. Arrows indicate the bond-current direction, while pink (yellow) plaquettes denote clockwise (counterclockwise) circulating currents. The dashed-red lines describe the enlarged $1\times2$ unit cell.
}
    \label{fig:LoopCurrent1QD2b}
\end{figure*}

The corresponding real-space current configurations, constructed from the dominant ordering vectors, are shown in Fig.~\ref{fig:LoopCurrent1QD2a} and \ref{fig:LoopCurrent1QD2b}. Following the symmetry classification of Ref.~\cite{feng21a}, two distinct loop-current phases are obtained: the $D_{2a}$ and $D_{2b}$ states, both characterized by a $1\times2$ enlarged unit cell. 

The $D_{2a}$ phase (Fig.~\ref{fig:LoopCurrent1QD2a}), generated by the $K^{--}$ form factor, is dominated by triangular current loops. Selecting different symmetry-related $M$-point ordering vectors rotates the current pattern but preserves its loop-current topology. In contrast, the $D_{2b}$ phase (Fig.~\ref{fig:LoopCurrent1QD2b}), generated by the $K^{++}$, $K^{+-}$, and $K^{-+}$ form factors, is characterized by predominantly hexagonal current loops. Although the current configurations differ by lattice rotations or symmetry-equivalent transformations, they exhibit the same loop-current topology and therefore belong to the same $D_{2b}$ symmetry class.

Both phases break translational and time-reversal symmetries, but differ in the topology of their circulating currents. Together with the $3\mathbf{Q}$ chiral flux phase discussed above, these results demonstrate that the kagome lattice naturally supports multiple symmetry-distinct loop-current states whose real-space structures are encoded directly in the momentum-space susceptibility.

\section{\label{sec.discussion}Discussions}

\begin{table*}[t]
\centering
\caption{
Summary of the dominant bond-current susceptibilities and the corresponding real-space current structures for the square, triangular, and kagome lattices near van Hove filling. The listed ordering vectors correspond to the susceptibility maxima in the first Brillouin zone. A check mark ($\checkmark$) indicates a pronounced susceptibility anisotropy that selects a subset of symmetry-related ordering vectors, whereas a dash ($-$) denotes isotropic susceptibility. The symbol ($\triangle$) indicates weak anisotropy, with the susceptibility remaining nearly isotropic despite slightly enhanced peaks at $M_1$ and $M_4$.
}
\label{tab:summary}
\begin{tabular}{lcccc}
\hline\hline
Lattice & Form factor & \hspace{5pt} Susceptibility anisotropy \hspace{5pt}& Dominant $\mathbf{Q}$ & Ordered state \\
\hline
Square
&
$p^{+}$
&
$\checkmark$
&
$(\pi,-\pi)$, $(-\pi,\pi)$
&
Staggered flux phase 
\\
 
&
$p^{-}$
&
$\checkmark$
&
$(\pi,\pi)$, $(-\pi,-\pi)$
&
Staggered flux phase 
\\

\hline

Triangular
&
$p^{++}$
&
$\checkmark$
&
$M_1$, $M_4$
&
Diamond flux phase
\\

&
$p^{+-}$
&
$\checkmark$
&
$M_3$, $M_6$
&
Diamond flux phase
\\

&
$p^{-+}$
&
$\checkmark$
&
$M_2$, $M_5$
&
Diamond flux phase
\\

&
$p^{--}$
&
$-$
&
All six $M$ points
&
Diamond flux phase
\\

\hline

Kagome ($3\mathbf{Q}$)
&
$p_{AC}$
&
$\checkmark$
&
$M_1$, $M_4$
&
Part of chiral flux phase
\\

&
$p_{AB}$
&
$\checkmark$
&
$M_3$, $M_6$
&
Part of chiral flux phase
\\

&
$p_{BC}$
&
$\checkmark$
&
$M_2$, $M_5$
&
Part of chiral flux phase
\\

\hline

Kagome ($1\mathbf{Q}$)
&
$K^{--}$
&
$\triangle$
&
All six $M$ points
&
$D_{2a}$ 
\\

&
$K^{++}$
&
$\checkmark$
&
$M_1$, $M_4$
&
$D_{2b}$ 
\\

&
$K^{+-}$
&
$\checkmark$
&
$M_3$, $M_6$
&
$D_{2b}$ 
\\

&
$K^{-+}$
&
$\checkmark$
&
$M_2$, $M_5$
&
$D_{2b}$ 
\\

\hline\hline
\end{tabular}
\end{table*}

\subsection{Choice of order-parameter expectation values}

In section~\ref{sec.Results}, the real-space current patterns are reconstructed from the bond-current order parameters in Eq.~\eqref{eq:realcurrent} and \eqref{eq.Jalphabetaalphabeta}.  To visualize the current configurations associated with the square lattice, and triangular lattice, we assign purely imaginary expectation values of equal magnitude, $v_\ell = i$, which fixes the overall phase and normalization while emphasizing the spatial topology and current directions.

For the square and triangular lattices, as well as the $1\mathbf{Q}$ states on the kagome lattice, all nearest-neighbor bond currents are modulated by the same ordering wave vector. In these cases, the relative signs of the bond-current expectation values in each bond are fixed by the symmetry of the chosen form factor. Consequently, changing the sign of one or more bond-current components does not simply correspond to a different realization of the same order, but instead selects a different symmetry channel. For example, assigning opposite signs to the expectation values on the $x$- and $y$-directed bonds in Eq.~\eqref{eq.Flux_order_square} transforms the current pattern associated with the $p^{+}$ form factor into that of the $p^{-}$ form factor, and vice versa.

The situation is different for the kagome $3\mathbf{Q}$ states, where the three inequivalent bond-current components are associated with different ordering vectors and therefore constitute independent order parameters. Different relative sign combinations, such as $(i,i,-i)$, $(i,-i,i)$, or $(-i,i,i)$, correspond to symmetry-equivalent realizations of the chiral flux phase described by Eq.~\eqref{eq.DeltaChiralFluxPhase}. Although these choices modify the direction of current on individual bonds, they preserve the overall loop-current topology and the chiral character of the ordered state.

More generally, the bond-current expectation values may be written as
\begin{equation}
v_\ell=\pm i|v_\ell|,
\end{equation}
where both the magnitudes and the relative signs are determined by the microscopic energetics. The susceptibility analysis presented here identifies the dominant ordering vectors but does not determine these equilibrium expectation values. Their values should be obtained from a microscopic theory, such as a self-consistent mean-field calculation or a Landau free-energy analysis.

\subsection{Relation between susceptibility anisotropy and real-space current structures}

A central result of this work is the systematic correspondence between susceptibility anisotropy and real-space current topology. Across the square, triangular, and kagome lattices, most odd-parity bond-current form factors generate strongly anisotropic susceptibility landscapes that spontaneously select only a subset of symmetry-related ordering wave vectors (Table~\ref{tab:summary}). In this sense, the susceptibility exhibits a momentum-space nematicity, despite the underlying lattice retaining its full rotational symmetry. The primary exceptions are the $p^{--}$ form factor on the triangular lattice and the $K^{--}$ form factor on the kagome lattice, whose susceptibilities remain nearly isotropic and display comparable enhancement at all symmetry-related $M$ points.

The real-space current configurations reveal the physical origin of this momentum-space selectivity. For every lattice considered, the ordering vectors associated with the susceptibility maxima generate staggered loop-current states with finite plaquette circulation and well-defined flux patterns. In contrast, symmetry-related wave vectors located in regions of suppressed susceptibility produce noncirculating monopole--antimonopole current textures. Although the corresponding bond-current order parameters remain finite, these configurations cannot sustain closed-current loops and therefore do not develop an associated flux structure.

These results demonstrate that the symmetry of an odd-parity bond-current form factor alone is insufficient to determine the nature of the ordered state. The ordering vector selected by the susceptibility plays an equally important role, effectively acting as a momentum-space filter that determines whether a given fluctuation channel evolves into a loop-current state or a noncirculating current texture. The monopole--antimonopole patterns identified here do not imply any violation of charge or current conservation; they merely represent local features of the current distribution, while the net current remains conserved throughout the lattice. Their suppression of the susceptibility therefore reflects geometric constraints on current circulation rather than a breakdown of conservation laws.

More broadly, the observed correspondence between susceptibility anisotropy and current topology suggests that momentum-space susceptibility profiles may serve as useful indicators of loop-current formation. Consequently, anisotropic bond-current fluctuations could provide a microscopic route to time-reversal symmetry breaking and unconventional charge ordering in correlated electron systems, particularly near van Hove fillings where the susceptibility enhancement is strongest.

\subsection{Physical realization of bond-current form factors}
The symmetry analysis developed in this work identifies the complete set of nearest-neighbor bond-current channels allowed by lattice symmetry. However, the existence of a symmetry-allowed form factor does not imply that the corresponding ordered state is realized. The stability of a bond-current phase depends on microscopic factors beyond symmetry, including the electronic dispersion, interaction strength, Fermi-surface geometry, and competition with other electronic orders. The susceptibilities calculated here therefore characterize candidate ordering channels and their preferred ordering vectors rather than the thermodynamic ground state.

This distinction is illustrated by previous large-$N$ studies of the $t$-$J$ model on the square lattice \cite{bejas12,bejas17,bejas14}. Although several bond-order channels are allowed by symmetry, the leading order was found in the $p^{-}$ channel, while the symmetry-related $p^{+}$ channel remained subdominant. Competing various bond-charge orders were also reported depending on the microscopic parameters. These results demonstrate that symmetry determines the possible ordering channels, whereas energetic considerations determine which channel ultimately condenses. 

%The framework developed here therefore provides a systematic foundation for studying bond-current orders. Determining whether these fluctuations condense into stable ordered phases requires self-consistent microscopic treatments, which are beyond the scope of the present work.

\subsection{Experimental implications and relevance to correlated materials}
The bond-current orders identified here have potential implications for experiments probing charge order and time-reversal symmetry breaking in correlated electron materials. For most odd-parity form factors, the bond-current susceptibility is strongly anisotropic, concentrating fluctuations at selected ordering wave vectors rather than distributing them uniformly among symmetry-equivalent momenta. 
If there is no domain structure in real materials,
such anisotropy could manifest as unequal scattering intensities in momentum-resolved probes, such as resonant x-ray scattering. Since the dominant ordering vectors are systematically associated with the formation of loop-current states, anisotropic momentum-space spectra may provide an experimental signature of bond-current order.

The kagome lattice provides an especially relevant realization of the present mechanism. Experiments on AV$_3$Sb$_5$ have revealed charge ordering at the Brillouin-zone $M$ points together with signatures of time-reversal symmetry breaking \cite{mielke22,khasanov22,jiang21,guo22,xu22,guguchia23,yu21,suetsugu26}. Within our framework, the leading bond-current susceptibilities are maximized at these same $M$ points, as those observed in the experiments, where the corresponding ordering vectors naturally reconstruct a $3\mathbf{Q}$ chiral flux phase consisting of coupled loop currents on the three inequivalent kagome bonds.  The resulting state simultaneously exhibits translational-symmetry breaking, circulating currents, and finite chirality, making it a possible candidate for the phenomena observed in kagome metals. Importantly, the ordering vectors $(\mathbf{Q}_{AC},\mathbf{Q}_{AB},\mathbf{Q}_{BC})$ are obtained directly as susceptibility maxima of the three bond-current channels rather than being imposed from symmetry arguments.

\section{\label{sec.conclusions}Conclusions}
We have systematically investigated bond-current orders driven by nearest-neighbor bond-charge interactions on square, triangular, and kagome lattices. By deriving the symmetry-allowed odd-parity form factors directly from the interaction, we analyzed their momentum-space susceptibilities and the corresponding real-space current configurations.

A central result is the emergence of a systematic correspondence between susceptibility anisotropy and current topology near van Hove filling. In this regime, the dominant susceptibility maxima across all lattice geometries considered occur at ordering wave vectors that generate closed-loop current states, whereas symmetry-related wave vectors located in regions of weak susceptibility produce noncirculating monopole--antimonopole current textures. The susceptibility therefore acts as a momentum-space indicator mechanism that strongly correlates with which current topologies are favored by a given fluctuation channel.

This mechanism stabilizes staggered flux phases on the square lattice, diamond-shaped loop-current phases on the triangular lattice, and a richer hierarchy of loop-current states on the kagome lattice, including a chiral flux phase in the $3\mathbf{Q}$ sector and the $D_{2a}$ and $D_{2b}$ states in the $1\mathbf{Q}$ sector. More generally, our results establish a direct connection between bond-current form factors, susceptibility anisotropy, and real-space current topology. The framework developed here provides a unified approach for identifying candidate bond-current orders and offers a useful perspective for understanding charge ordering, loop-current formation, and time-reversal-symmetry-breaking phenomena in correlated electron systems.

\begin{acknowledgments}
This work was supported by World Premier International Research Center Initiative (WPI), MEXT, Japan.
\end{acknowledgments}

\appendix%*

\section{\label{app.FormFac} Derivation of bond-charge form factors}
The bond-order form factors obtained from Eq.~\eqref{eq.BondFormFacG} can be rewritten as
\begin{align}
    \gamma_{\mathbf{k}}\gamma_{\mathbf{k}'} &=   \dfrac{1}{2}\sum_{\boldsymbol{\tau}}  e^{i(\mathbf{k}\cdot\boldsymbol{\tau}-\mathbf{k}'\cdot\boldsymbol{\tau})} + e^{i(\mathbf{k}\cdot(-\boldsymbol{\tau})-\mathbf{k}'\cdot(-\boldsymbol{\tau}))}, \nonumber \\
    &=  \sum_{\boldsymbol{\tau}}  \cos (\mathbf{k}\cdot\boldsymbol{\tau}-\mathbf{k}'\cdot\boldsymbol{\tau}) . \label{eq.cosGamma}
\end{align}
This expression can be decomposed into symmetry-resolved basis functions determined by the lattice geometry. The resulting basis functions form a complete set of symmetry-allowed nearest-neighbor bond-order channels and contain both even-parity (bond-charge) and odd-parity (bond-current) form factors. In the main text, we focus on the odd-parity channels relevant to bond-current order. Here we summarize the complete decomposition.

\subsection{Square lattice}
For the square lattice, Eq.~\eqref{eq.cosGamma} becomes \cite{yamase21c,zafur24}
\begin{equation}
    \gamma_{\mathbf{k}}\gamma_{\mathbf{k}'} = \cos(k_x - k'_x) + \cos (k_y - k'_y) ~,
\end{equation}
the interaction kernel can be decomposed into symmetry-resolved form factors as
\begin{equation}
    \gamma_{\mathbf{k}}\gamma_{\mathbf{k}'} = \dfrac{1}{2}(s_\mathbf{k} s_{\mathbf{k}'} 
    + d_\mathbf{k}d_{\mathbf{k}'} 
    + p^+_\mathbf{k}p^+_{\mathbf{k}'} 
    + p^-_\mathbf{k}p^-_{\mathbf{k}'} )  ,
\end{equation}
where 
\begin{align}
    s_\mathbf{k} &= \cos k_x +\cos k_y,\\
    d_\mathbf{k} &= \cos k_x -\cos k_y,
\end{align}
and the odd-parity form factors $p^{\pm}_{\mathbf{k}}$ are given in Eq.~\eqref{eq.pFormfactors}.

The form factors $s_{\mathbf{k}}$ and $d_{\mathbf{k}}$ are even under inversion; in contrast, $p^{\pm}_{\mathbf{k}}$ are odd-parity form factors associated with bond-current order. The present work focuses on these odd-parity channels, while the even-parity channels correspond to bond-order discussed extensively in previous studies \cite{yamase26b}.

\subsection{Triangular lattice}
For the isotropic triangular lattice, Eq.~\eqref{eq.cosGamma} becomes
\begin{equation}
    \gamma_{\mathbf{k}}\gamma_{\mathbf{k}'} = \cos(\mathbf{k}_1-\mathbf{k}'_1) + \cos(\mathbf{k}_2-\mathbf{k}'_2) + \cos(\mathbf{k}_3-\mathbf{k}'_3),
\end{equation}
where $\mathbf{k}_i$ are defined in Eqs.~\eqref{eq.k1}--\eqref{eq.k3}.  Expanding the cosine functions yields
\begin{widetext}
    \begin{equation}
   \gamma_{\mathbf{k}}\gamma_{\mathbf{k}'}  = \dfrac{1}{2}(s^t_\mathbf{k} s^t_{\mathbf{k}'} + d_{x^2 - y^2}(\mathbf{k})d_{x^2 - y^2}(\mathbf{k}') + d_{xy}(\mathbf{k})d_{xy}(\mathbf{k}'))  + \dfrac{1}{4} ( p_\mathbf{k}^{+-}p_{\mathbf{k}'}^{+-}  + p_\mathbf{k}^{-+}p_{\mathbf{k}'}^{-+} + p^{++}_\mathbf{k}p^{++}_{\mathbf{k}'} + p^{--}_\mathbf{k}p^{--}_{\mathbf{k}'} ) .
\end{equation}
\end{widetext}
The even-parity form factors are
\begin{align}
    s^t_\mathbf{k} &= \cos k_x + \sqrt{2}\cos(\frac{k_x}{2})\cos(\frac{\sqrt{3}k_y}{2}), \\
     d_{x^2 - y^2}(\mathbf{k}) &= \cos k_x - \sqrt{2}\cos(\frac{k_x}{2})\cos(\frac{\sqrt{3}k_y}{2}), \\
     d_{xy}(\mathbf{k}) &= 2\sin(\frac{k_x}{2})\sin(\frac{\sqrt{3}k_y}{2}),
\end{align}
while the odd-parity form factors $p^{++}_{\mathbf{k}}$, $p^{--}_{\mathbf{k}}$, $p^{+-}_{\mathbf{k}}$, and $p^{-+}_{\mathbf{k}}$ are given in Eqs.~\eqref{eq.TriPxFormfac}--\eqref{eq.TrimpFormfac}.

The triangular lattice therefore supports seven symmetry-resolved bond-order channels. The form factors $s^t_{\mathbf{k}}$, $d_{x^2-y^2}(\mathbf{k})$, and $d_{xy}(\mathbf{k})$ are even under inversion and describe bond-charge fluctuations, whereas the four $p$-channels are odd-parity bond-current modes. The even-parity channels were discussed previously in the context of charge orders on triangular lattices \cite{gneist22}.

\subsection{Kagome lattice}
Because nearest-neighbor hopping on the kagome lattice connects different sublattices, the interaction kernel decomposes into contributions from the three inequivalent bonds, $AC$, $AB$, and $BC$. Equation~\eqref{eq.cosGamma} therefore becomes
\begin{widetext}
    \begin{align}
    \gamma_{\mathbf{k}}\gamma_{\mathbf{k}'} \bigg|_{AC} &=  \cos \left( \frac{\mathbf{k}_1}{2} - \frac{\mathbf{k}'_1}{2} \right) = c_{AC}(\mathbf{k})c_{AC}(\mathbf{k}') + p_{AC}(\mathbf{k})p_{AC}(\mathbf{k}'),\\
    \gamma_{\mathbf{k}}\gamma_{\mathbf{k}'}\bigg|_{AB} &=  \cos \left( \frac{\mathbf{k}_2}{2} - \frac{\mathbf{k}'_2}{2} \right) = c_{AB}(\mathbf{k})c_{AB}(\mathbf{k}') + p_{AB}(\mathbf{k})p_{AB}(\mathbf{k}'),\\
    \gamma_{\mathbf{k}}\gamma_{\mathbf{k}'}\bigg|_{BC} &=  \cos \left( \frac{\mathbf{k}_3}{2} - \frac{\mathbf{k}'_3}{2} \right)= c_{BC}(\mathbf{k})c_{BC}(\mathbf{k}') + p_{BC}(\mathbf{k})p_{BC}(\mathbf{k}'),
\end{align}
\end{widetext}
where $\mathbf{k}_1$, $\mathbf{k}_2$, and $\mathbf{k}_3$ are defined in Eqs.~\eqref{eq.k1}--\eqref{eq.k3}.

The corresponding even-parity form factors are
\begin{align}
    c_{AC}(\mathbf{k}) &= \cos \bigg(\frac{k_x}{2}\bigg), \label{eq.cAC} \\
    c_{AB}(\mathbf{k}) &= \cos \bigg(\frac{k_x}{4} + \frac{\sqrt{3}k_y}{4}\bigg), \label{eq.cAB}\\
    c_{BC}(\mathbf{k}) &=  \cos \bigg(\frac{k_x}{4} - \frac{\sqrt{3}k_y}{4}\bigg) . \label{eq.cBC}
\end{align}
while the odd-parity form factors $p_{AC}(\mathbf{k})$, $p_{AB}(\mathbf{k})$, and $p_{BC}(\mathbf{k})$ are given in Eqs.~\eqref{eq.PAC}--\eqref{eq.PBC}.

Although the main text focuses on the odd-parity sine form factors because they naturally describe bond-current fluctuations, the cosine form factors become useful in the extended Brillouin-zone analysis presented in Appendix~\ref{app.ExtendedBZ}. On the kagome lattice, translations by reciprocal lattice vectors transform the sine form factors into equivalent cosine form factors while leaving the underlying bond-current order unchanged. Consequently, the same bond-current configuration can be described equivalently in either the sine- or cosine-form-factor representation. Thus, for the kagome lattice, the even-parity form factors also provide a valid framework for analyzing bond-current order.

To describe the bond-current channel using the cosine basis, the momentum-space operators must still be constructed from the antisymmetric combination of fermion bilinears \cite{fu25},
\begin{align}
    \Delta_{\alpha\beta}(\mathbf{Q}) 
    = \sum_{\mathbf{k},\sigma}  c_{\alpha\beta}\bigg( {\mathbf{k}+\frac{\mathbf{Q}}{2}}\bigg) \langle c^\dagger_{\mathbf{k},\sigma;\alpha} c_{\mathbf{k}+\mathbf{Q},\sigma;\beta} -  c^\dagger_{\mathbf{k},\sigma;\beta} c_{\mathbf{k}+\mathbf{Q},\sigma;\alpha}\rangle,
\end{align}
so that the operator remains odd under bond interchange and therefore represents a bond-current order parameter rather than a bond-charge order parameter.

The corresponding real-space expressions are
\begin{align}
        \Delta_{AC}(\mathbf{Q}) &= \dfrac{1}{2}\sum_{j,\sigma} ( \Delta^j_{AC'}(\mathbf{Q}) - \Delta_{AC}^j(\mathbf{Q}) ), \label{eq.DcAC}
\\
             \Delta_{AB}(\mathbf{Q}) &= \dfrac{1}{2}\sum_{j,\sigma} ( \Delta_{AB'}^j(\mathbf{Q})  - \Delta_{AB}^j(\mathbf{Q}) ) , \label{eq.DcAB}
\\
                  \Delta_{BC}(\mathbf{Q}) &= \dfrac{1}{2}\sum_{j,\sigma} ( \Delta_{B'C'}^j(\mathbf{Q})  - \Delta_{BC}^j(\mathbf{Q}) ), \label{eq.DcBC}
\end{align}
where the bond operators $\Delta^j_{\alpha\beta}$ are the same as Eqs.~\eqref{eq.DjAC1}--\eqref{eq.DjBC2} in the main text.
%Compared with the sine-type operators of Eqs.~\eqref{eq.DAC}--\eqref{eq.DBC}, the cosine channels differ only by the relative sign between symmetry-related bonds. 

\section{Form-factor periodicity in the extended Brillouin zone}\label{app.ExtendedBZ}
\begin{figure}[t]
    \centering
    \includegraphics[width=0.75\linewidth]{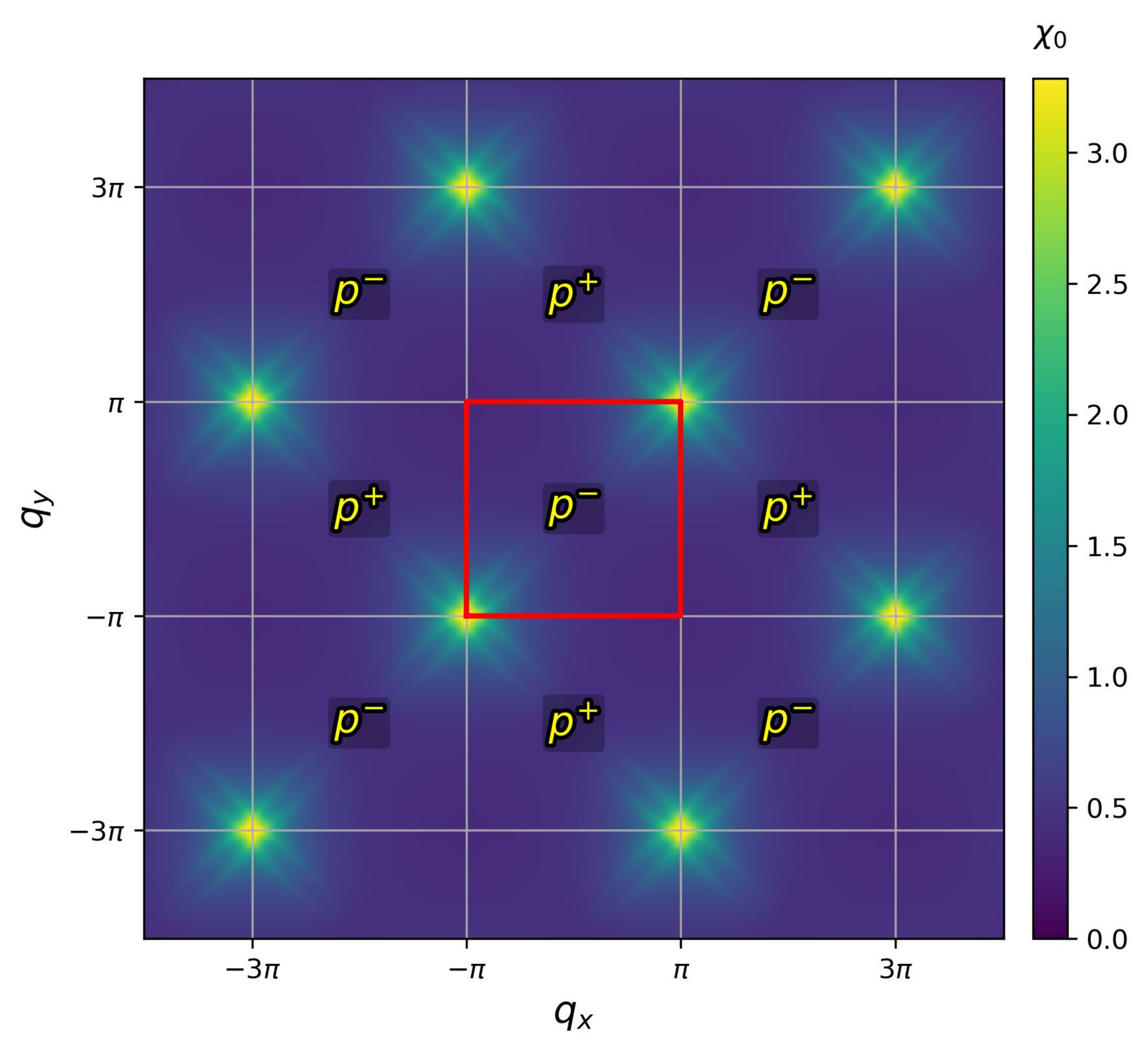}
    \caption{Extended Brillouin-zone representation of the bond-current susceptibility for the square-lattice $p^-$
 form factor, calculated using the same parameters as in Fig.~\ref{fig:Sussquare}. White lines denote neighboring Brillouin-zone replicas. Reciprocal-lattice translations interchange the symmetry character of the form factor, causing adjacent zones to alternate between the $p^-$ and $p^+$ channels.}
    \label{fig:EXTENDSUSSQUARE}
\end{figure}

\begin{figure}[t]
    \centering
    \includegraphics[width=0.75\linewidth]{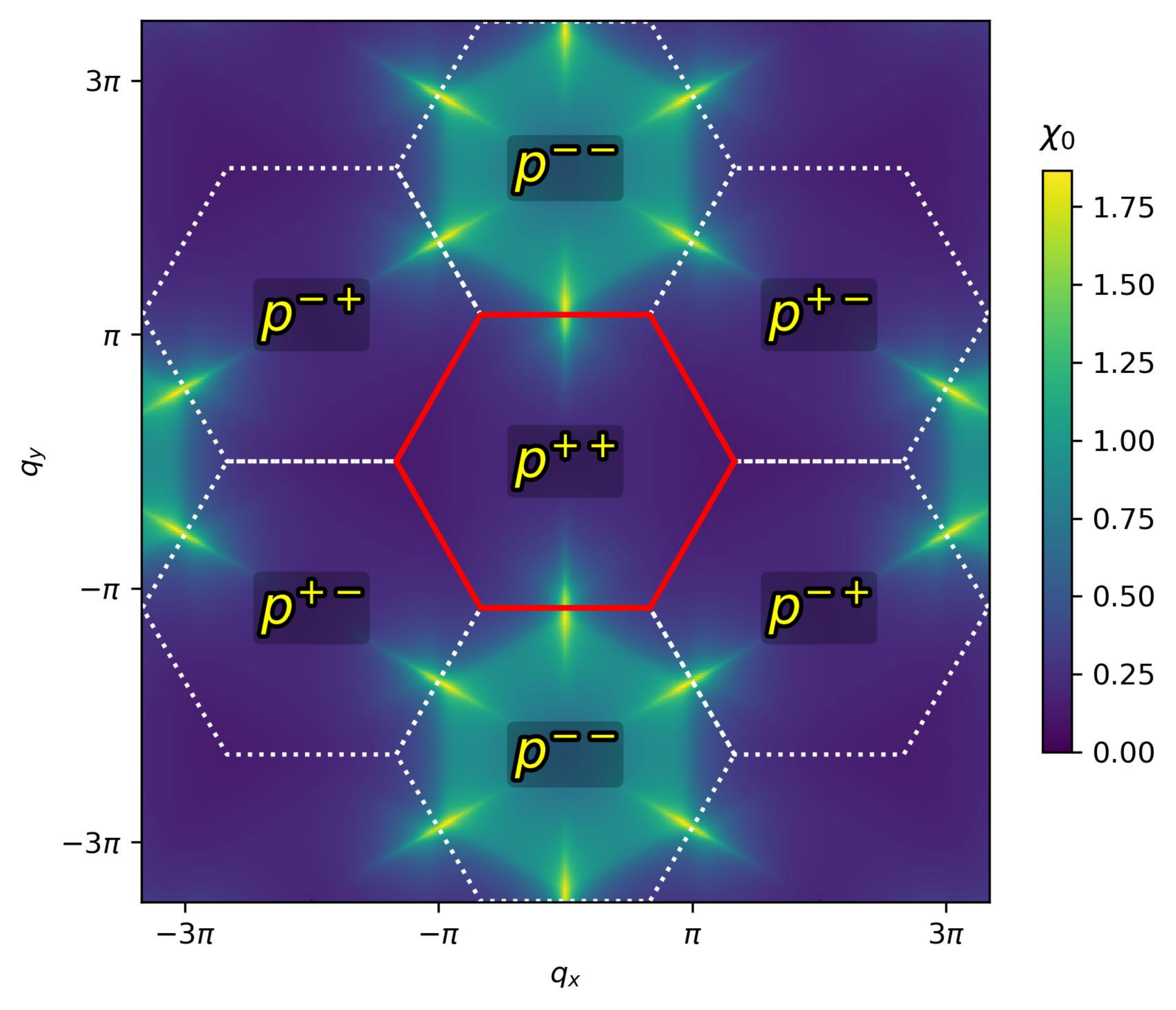}
    \caption{Extended Brillouin-zone representation of the bond-current susceptibility for the triangular-lattice $p^{++}$ form factor, calculated using the same parameters as in Fig.~\ref{fig:SusTri}. White-dotted lines denote neighboring Brillouin-zone replicas. Under reciprocal-lattice translations, the form factor transforms among the $p^{++}, p^{+-}, p^{+-}$, and $p^{-+}$ channels, producing a symmetry-dependent susceptibility pattern across the extended zone.}
    \label{fig:EXTENDSUSTRI}
\end{figure}

\begin{figure}[t]
    \centering
    \includegraphics[width=0.75\linewidth]{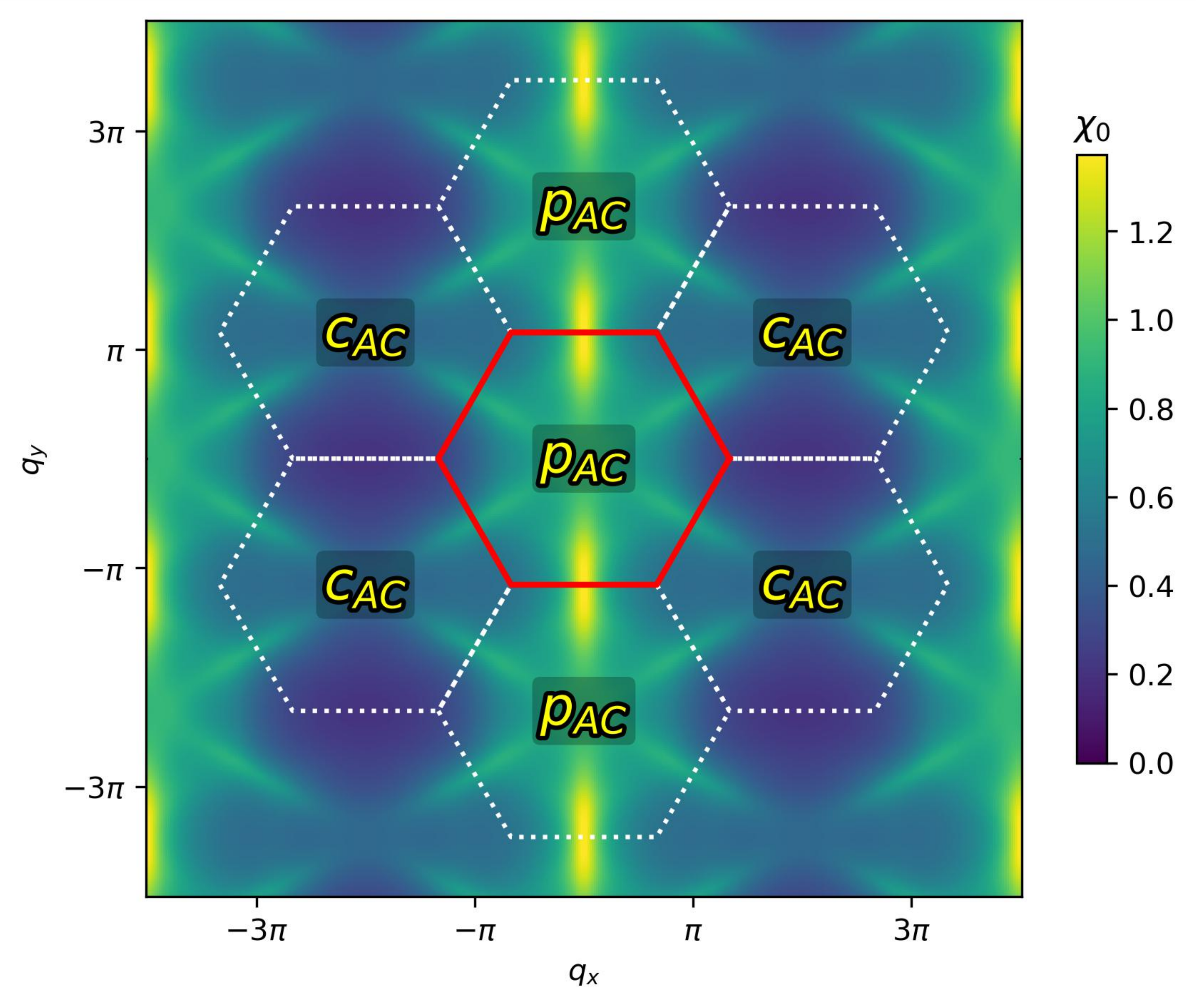}
    \caption{Extended Brillouin-zone structure of the bond-current susceptibility for the kagome-lattice $p_{AC}$ form factor, using the same parameters as in Fig.~\ref{fig:SusKagome}. White-dotted lines denote neighboring Brillouin-zone replicas. Reciprocal-lattice translations interchange the $p_{AC},c_{AC}$ channels, producing alternating bond-current and bond-charge character in adjacent Brillouin-zone blocks.}
    \label{fig:EXTENDSUSKagome}
\end{figure}

The form-factor-resolved susceptibility considered in this work exhibits a nontrivial structure in the extended Brillouin zone. Unlike conventional charge or spin susceptibilities, the bond-current susceptibility contains an additional momentum-dependent form factor. As a result, its symmetry properties under reciprocal-lattice translations can differ from those of the ordinary Lindhard susceptibility.

For a conventional susceptibility,
\begin{equation}
    \chi(\mathbf{Q})
    =
    \chi(\mathbf{Q}+\mathbf{G}),\label{eq.periodicity}
\end{equation}
where $\mathbf{G}$ is a reciprocal lattice vector. This periodicity follows directly from the translational invariance of the electronic structure and is recovered from Eq.~\eqref{eq.x0} when the form factor is replaced by a constant, $\gamma_{\mathbf{k}}=1$.

In the present case, however, the susceptibility depends explicitly on the form factor through the bond-current operator, Eq.~\eqref{eq.BondOperator}, or equivalently through the weight $\gamma_{\mathbf{k}+\mathbf{Q}/2}$ appearing in Eq.~\eqref{eq.x0}. Under a reciprocal-lattice translation,
\begin{equation}
    \mathbf{Q}\rightarrow\mathbf{Q}+\mathbf{G},
\end{equation}
the electronic dispersion remains invariant, whereas the form factor may transform into a different symmetry channel. Consequently, neighboring Brillouin-zone blocks need not represent the same bond-current mode.

For the square lattice, the reciprocal lattice vectors are
\begin{equation}
\mathbf{G}_x=(2\pi,0),
\qquad
\mathbf{G}_y=(0,2\pi).
\end{equation}
Using Eq.~\eqref{eq.pFormfactors}, one obtains
\begin{align}
    p^{\pm}\left(\mathbf{k}+ \frac{\mathbf{Q}+\mathbf{G}_{x}}{2}\right) &= - p^{\mp} \left(\mathbf{k}+\frac{\mathbf{Q}}{2}\right), \\
    p^{\pm}\left(\mathbf{k}+ \frac{\mathbf{Q}+\mathbf{G}_{y}}{2}\right) &=  p^{\mp} \left(\mathbf{k}+\frac{\mathbf{Q}}{2}\right),
\end{align}
showing that a reciprocal-lattice translation maps the $p^{-}$ channel onto the $p^{+}$ channel, producing the extended-zone pattern shown in Fig.~\ref{fig:EXTENDSUSSQUARE}. Similar transformations occur for the triangular- and kagome-lattice form factors, as illustrated in Figs.~\ref{fig:EXTENDSUSTRI} and \ref{fig:EXTENDSUSKagome}.

Although reciprocal-lattice translations change the form-factor representation, they do not modify the susceptibility anisotropy or the associated real-space current configurations discussed in the main text.

\section{Bond-current susceptibilities at different electronic filling}\label{app.ChangeOfSUSonMu}
The main text focuses on the vicinity of van Hove filling. Here, we examine the evolution of the susceptibility over a broader range of electron fillings and discuss the corresponding changes in the preferred ordering vectors and real-space current configurations.

For the square lattice, the evolution of the dominant ordering vector with filling is comparatively weak and has been discussed previously in Ref.~\cite{bejas12,bejas14}. We therefore focus on the triangular and kagome lattices, where the filling dependence is more pronounced.

\subsection{Triangular lattice}
\begin{figure*}
    \centering
     \includegraphics[width = .8\linewidth]{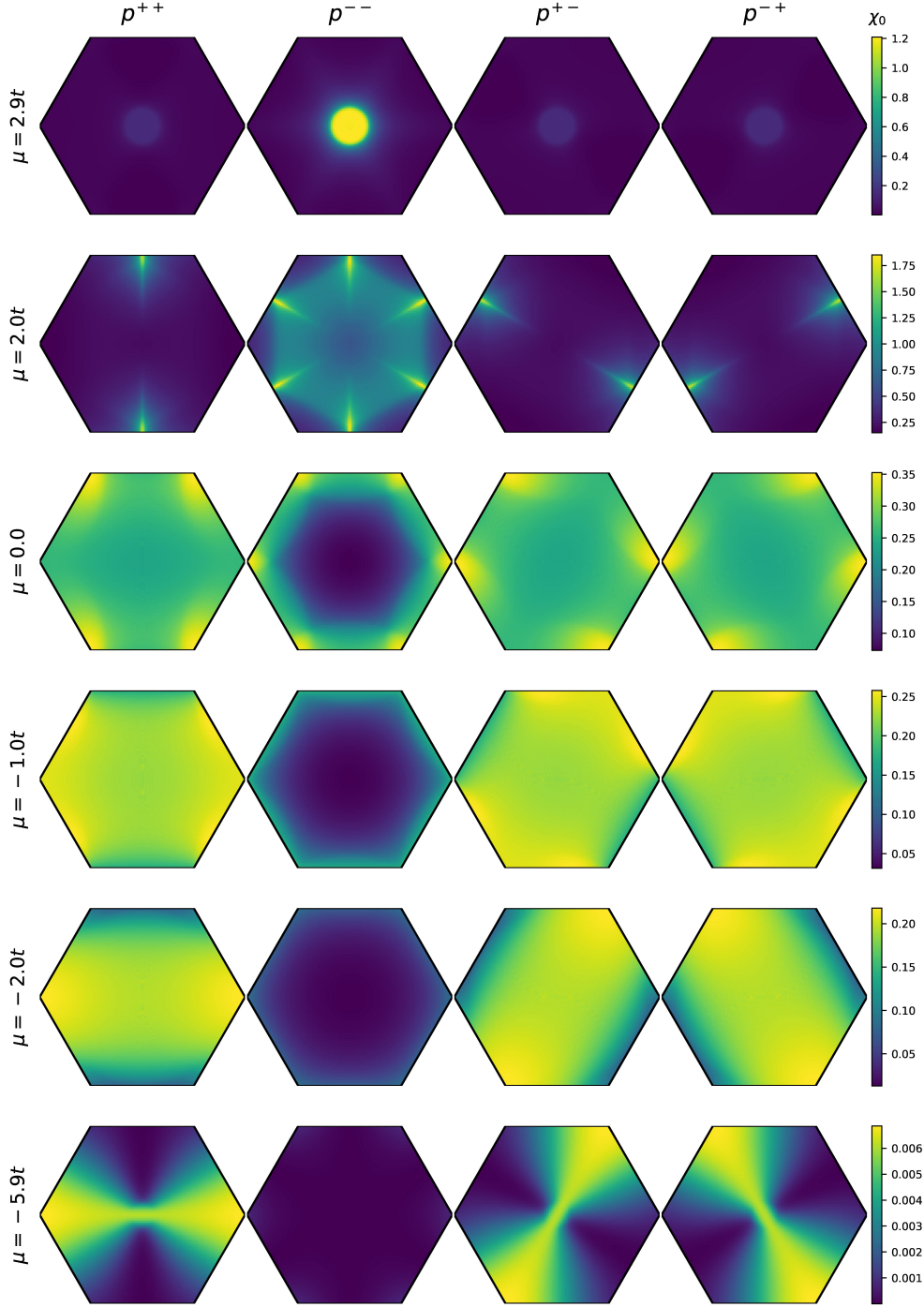}
\caption{
Momentum-space bond-current susceptibilities on the triangular lattice for representative chemical potentials at $T=0.01t$. Columns show the four bond-current channels $p^{++}$, $p^{--}$, $p^{+-}$, and $p^{-+}$, and rows correspond to different fillings. As the chemical potential varies, the dominant susceptibility maxima shift between different high-symmetry points of the Brillouin zone, revealing a strong filling dependence of the preferred bond-current ordering vectors.
}
    \label{fig:px_sus}
\end{figure*}

The momentum-space structure of the bond-current susceptibility evolves substantially with electron filling. For the isotropic triangular lattice, the electronic band shown in Fig.~\ref{fig:EnergyDispersion} spans the range $-6t\le\mu\le 3t$, with van Hove singularities located at the six $M$ points at $\mu=2t$. To illustrate the filling dependence, we consider six representative chemical potentials,
\[
\mu/t=-5.9,\,-2.0,\,-1.0,\,0.0,\,2.0,\ \text{and}\ 2.9,
\]
spanning the lower and upper band edges, intermediate fillings, and the van Hove regime. The corresponding susceptibilities for the four bond-current form factors are shown in Fig.~\ref{fig:px_sus}.

The susceptibility is maximized at van Hove filling, where the dominant ordering vectors are the six symmetry-related $M_i$ points discussed in the main text.  Away from $\mu=2t$, however, the susceptibility landscape changes qualitatively. Around $\mu=0$, the dominant spectral weight shifts toward the six $K_i$ points at the corners of the Brillouin zone, indicating a change in the preferred ordering vector. This evolution reflects the strong dependence of the bond-current order on the Fermi-surface geometry.

\begin{figure*}
    \centering
    \includegraphics[width=0.9\linewidth]{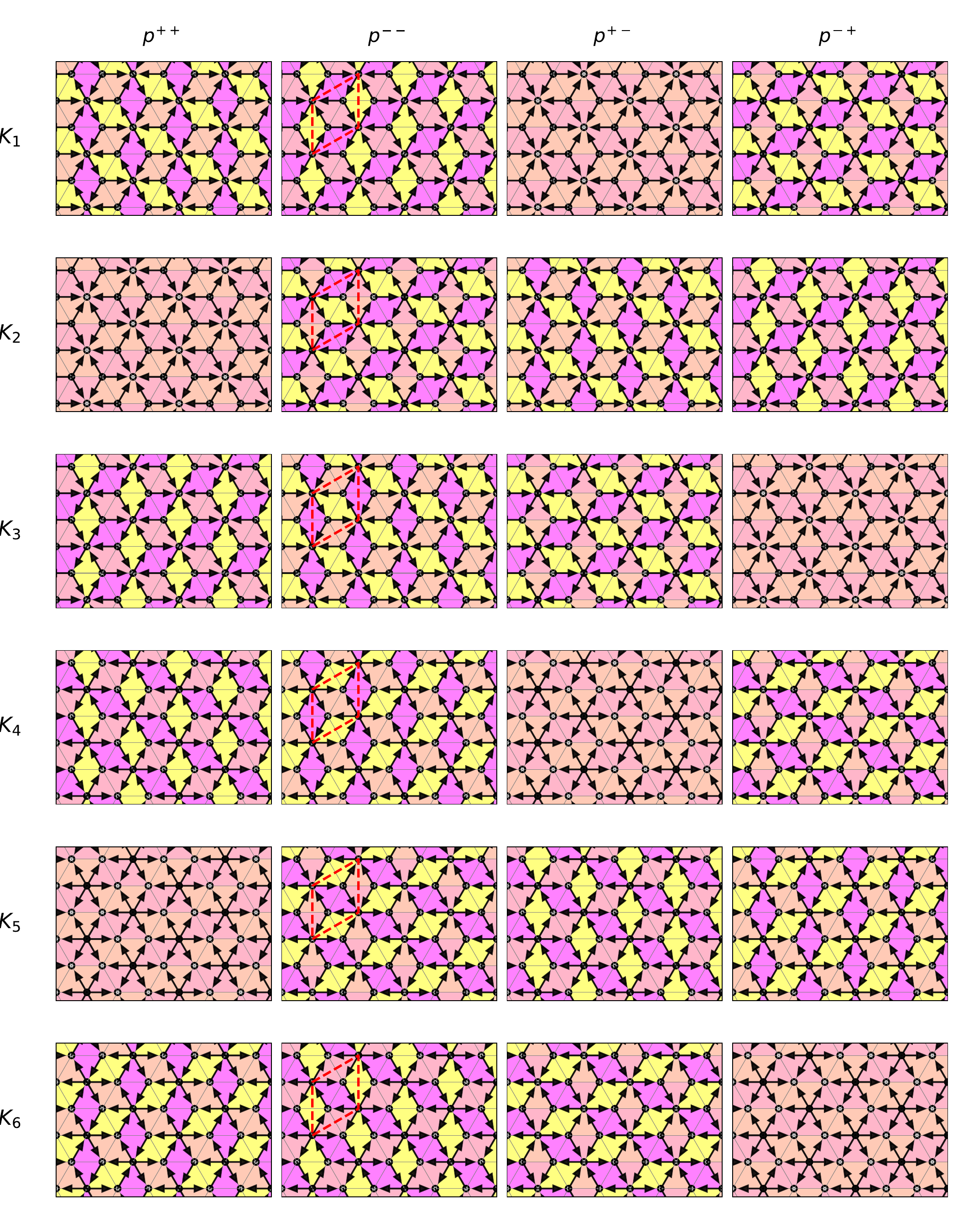}
\caption{
Real-space bond-current configurations on the triangular lattice associated with the six $K$ ordering vectors. The four columns correspond to the form factors $p^{++}$, $p^{--}$, $p^{+-}$, and $p^{-+}$, respectively. Arrows indicate the direction of the bond-currents, while pink and yellow plaquettes denote clockwise and counterclockwise circulating currents.  The dashed-red lines describe the enlarged $\sqrt{3}\times \sqrt{3}$ unit cells.
} 
\label{fig:K1}
\end{figure*}

%The real-space current patterns associated with the $M_i$ ordering vectors were presented in Fig.~\ref{fig:M1}. 
Figure~\ref{fig:K1} shows representative current configurations generated by the $K_i$ ordering vectors, which become dominant near $\mu=0$. Although these states differ from the $M_i$-point orders, they likewise exhibit diamond-shaped loop-current patterns that break translational and time-reversal symmetries. The ordered state forms a $\sqrt{3}\times \sqrt{3}$ superstructure. Thus, the correspondence between enhanced bond-current susceptibility and circulating-current order persists over a substantial range of fillings around the van Hove regime.

The correspondence is nevertheless not universal. As the filling moves further away from the van Hove point, susceptibility maxima may occur at ordering vectors that do not support closed current loops. For example, at $\mu=-1.0t$, the $p^{++}$ susceptibility develops pronounced maxima near $M_2$, $M_3$, $M_5$, and $M_6$. As shown in Fig.~\ref{fig:M1}, the corresponding current configurations do not form closed-loop structures. Similarly, at $\mu=-2t$, the dominant $p^{++}$ susceptibility occurs near the $K_2$ and $K_5$ points, where the resulting current pattern does not exhibit circulating currents. At the opposite end of the band, near $\mu=2.9t$, the $p^{--}$ susceptibility is maximal at the $\Gamma$ point; however, the associated bond-current form factor vanishes identically, and no loop-current state is realized.
These results indicate that the position of a susceptibility maximum alone is insufficient to determine the nature of the ordered state. The corresponding real-space current topology must also be examined.

\subsection{Kagome lattice}
\begin{figure}[t]
    \centering
\begin{minipage}{0.48\linewidth}
        \centering
        (a) $\mu = 1.9t$\\
        \includegraphics[width=0.95\linewidth]{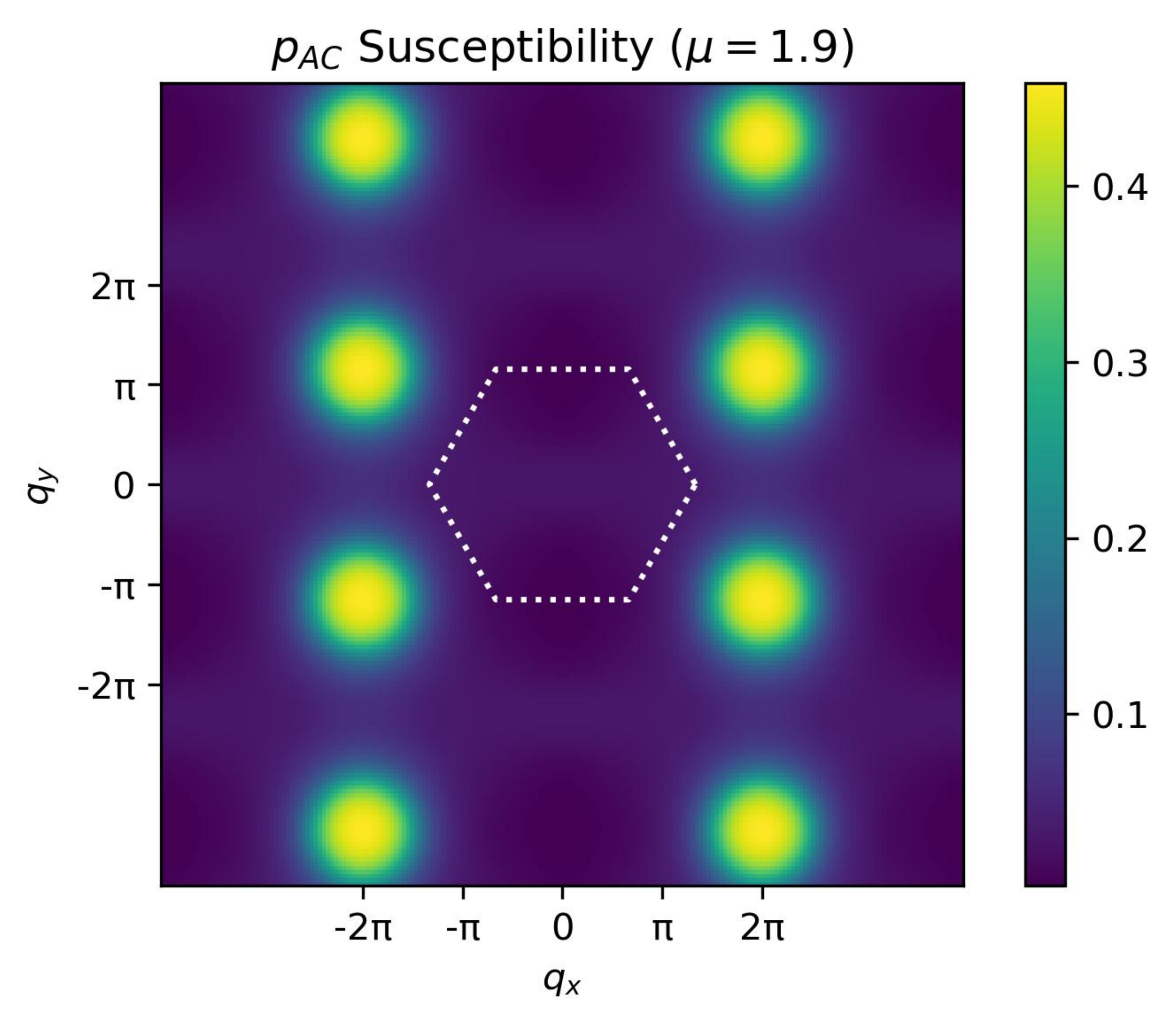}
    \end{minipage}
    \begin{minipage}{0.48\linewidth}
        \centering
        (b) $\mu = 1.0t$\\
        \includegraphics[width=0.95\linewidth]{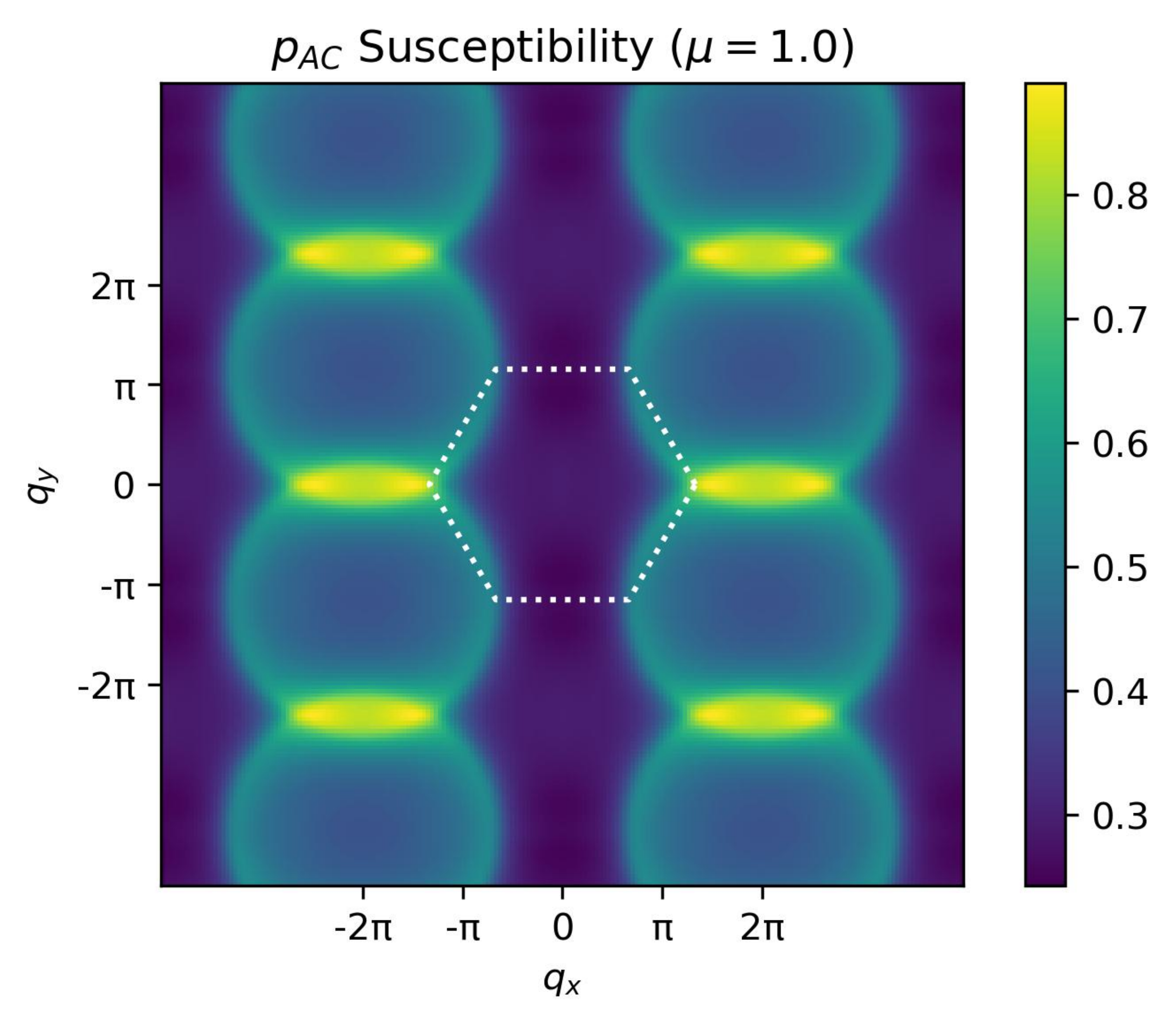}
    \end{minipage}

            \begin{minipage}{0.48\linewidth}
        \centering
        (c) $\mu = 0.7t$\\
        \includegraphics[width=0.95\linewidth]{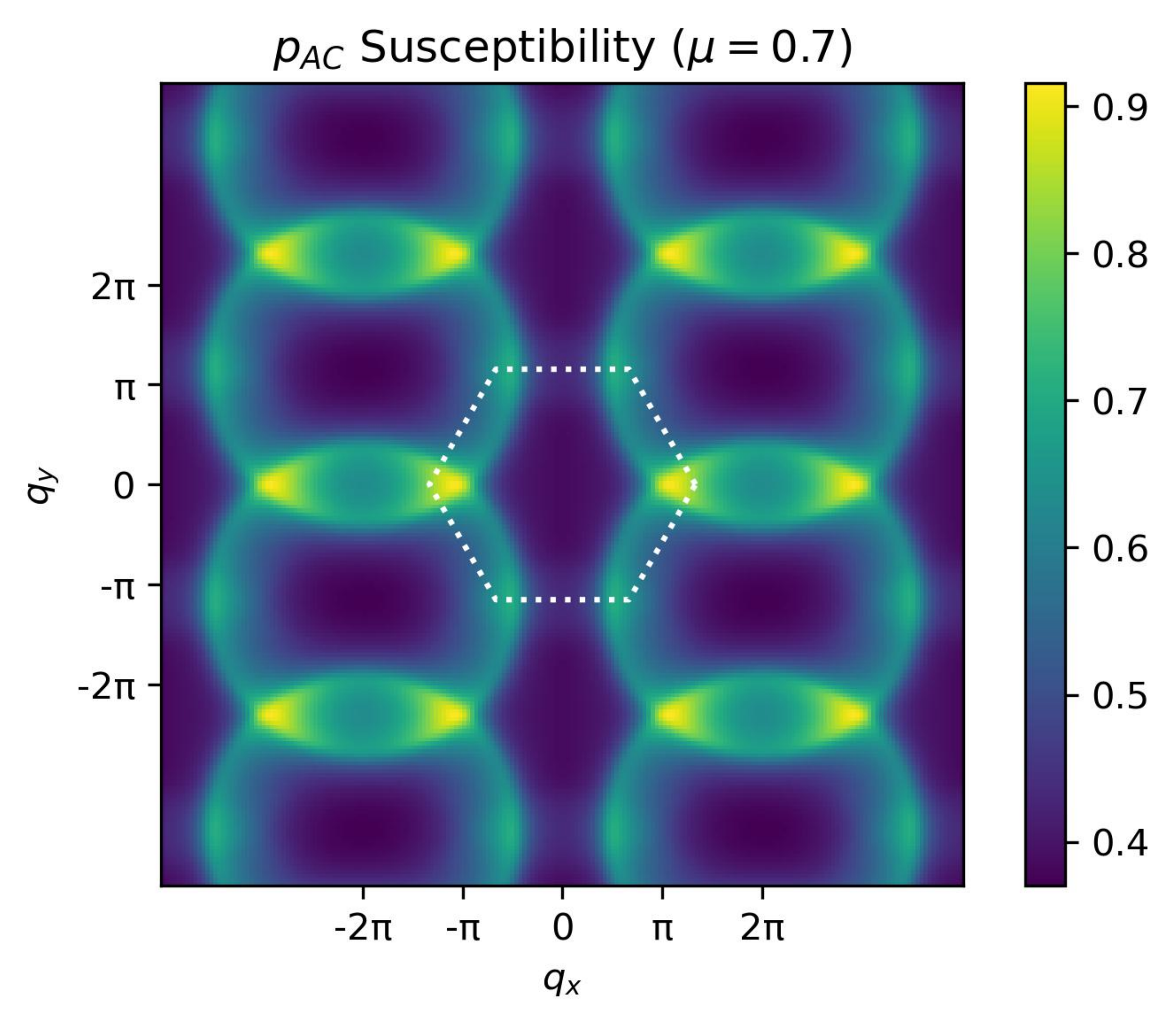}
    \end{minipage}
    \begin{minipage}{0.48\linewidth}
        \centering
        (d) $\mu = 0.5t$\\
        \includegraphics[width=0.95\linewidth]{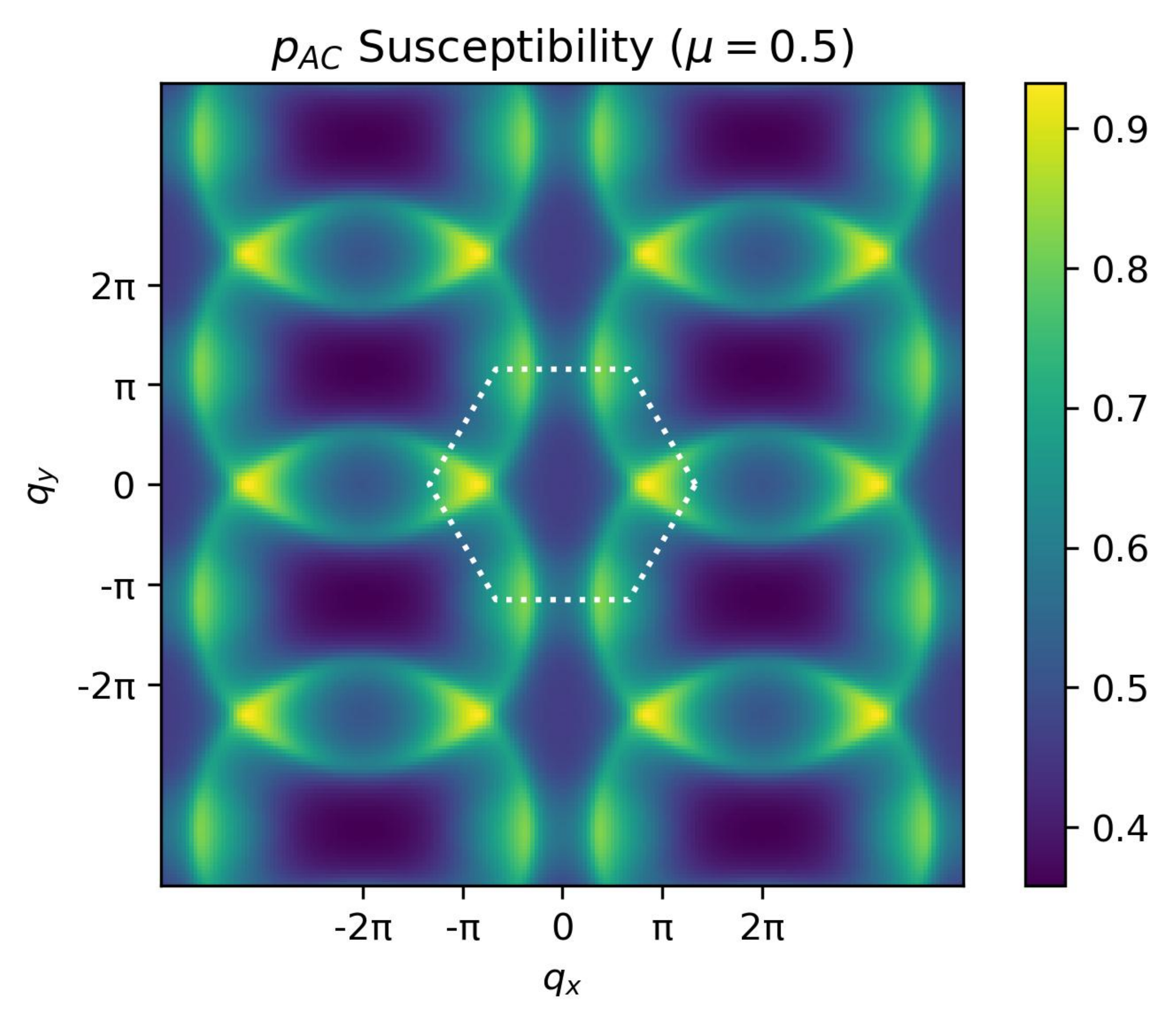}
    \end{minipage}

    \begin{minipage}{0.48\linewidth}
        (e) $\mu = 0.0$\\
        \centering
        \includegraphics[width=0.95\linewidth]{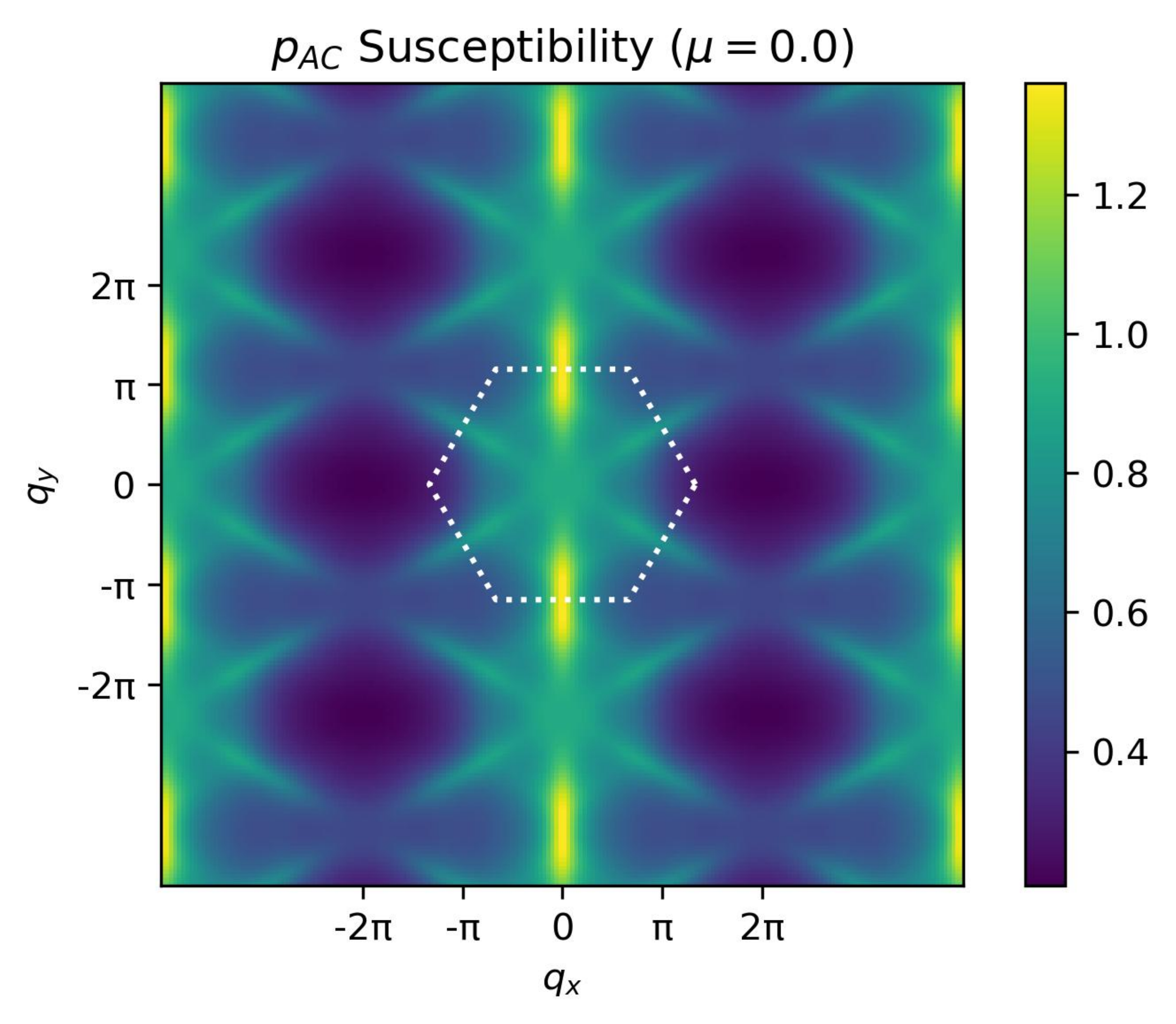}
    \end{minipage}
 \begin{minipage}{0.48\linewidth}
        \centering
        (f) $\mu = -0.9t$ \\
        \includegraphics[width=0.95\linewidth]{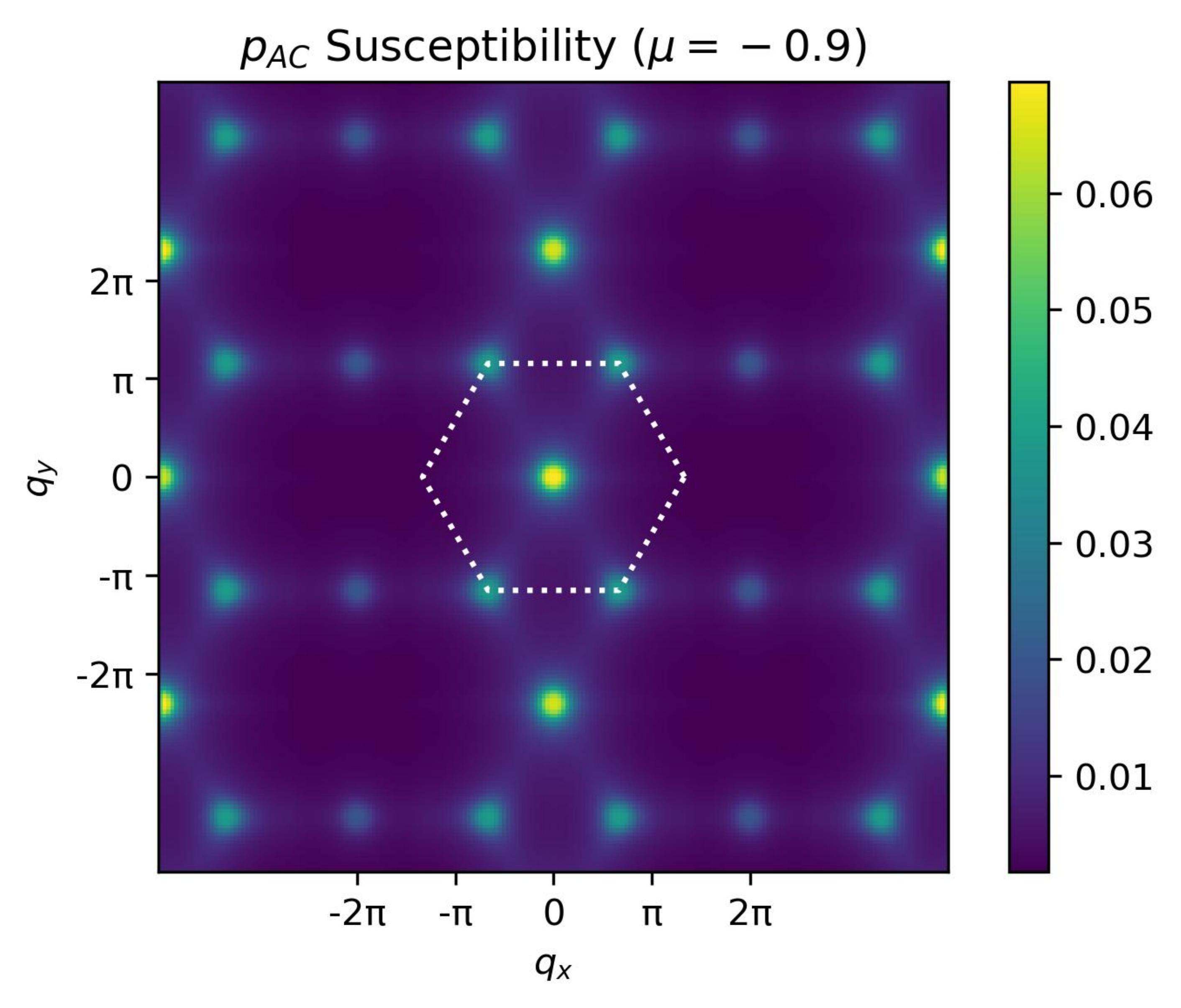}
    \end{minipage}
   \caption{Momentum-space bond-current susceptibility for the kagome-lattice $p_{AC}$ form factor at representative chemical potentials. Panels (a)--(f) correspond to $\mu/t=1.9$, $1.0$, $0.7$, $0.5$, $0.0$, and $-0.9$, respectively. The white hexagon denotes the first Brillouin zone. The susceptibility landscape evolves strongly with filling, leading to a shift of the dominant ordering wave vectors away from the van Hove regime.}
    \label{fig:pAC_sus}
\end{figure}

We next examine the filling dependence of the bond-current susceptibility in the kagome lattice. While the main text focuses on van Hove filling of the dispersive $p$ band ($\mu=0$), the preferred bond-current order evolves significantly as the chemical potential is varied. To illustrate this evolution, we evaluate the susceptibility for representative chemical potentials
\[
\mu/t=-0.9, \ 0.0,\ 0.5,\ 0.7,\ 1.0,\ \text{and} \ 1.9.
\]
The corresponding susceptibility for the $p_{AC}$ form factor is presented in Fig.~\ref{fig:pAC_sus}; those for $p_{AB}$ and $p_{BC}$ exhibit analogous structures related by the rotational symmetries of the kagome lattice.

\begin{figure}[t]
    \centering
\begin{minipage}{0.48\linewidth}
        \centering
        (a)  $p_{AC}$, $\mathbf{Q}=K_2$\\
        \includegraphics[width=0.95\linewidth]{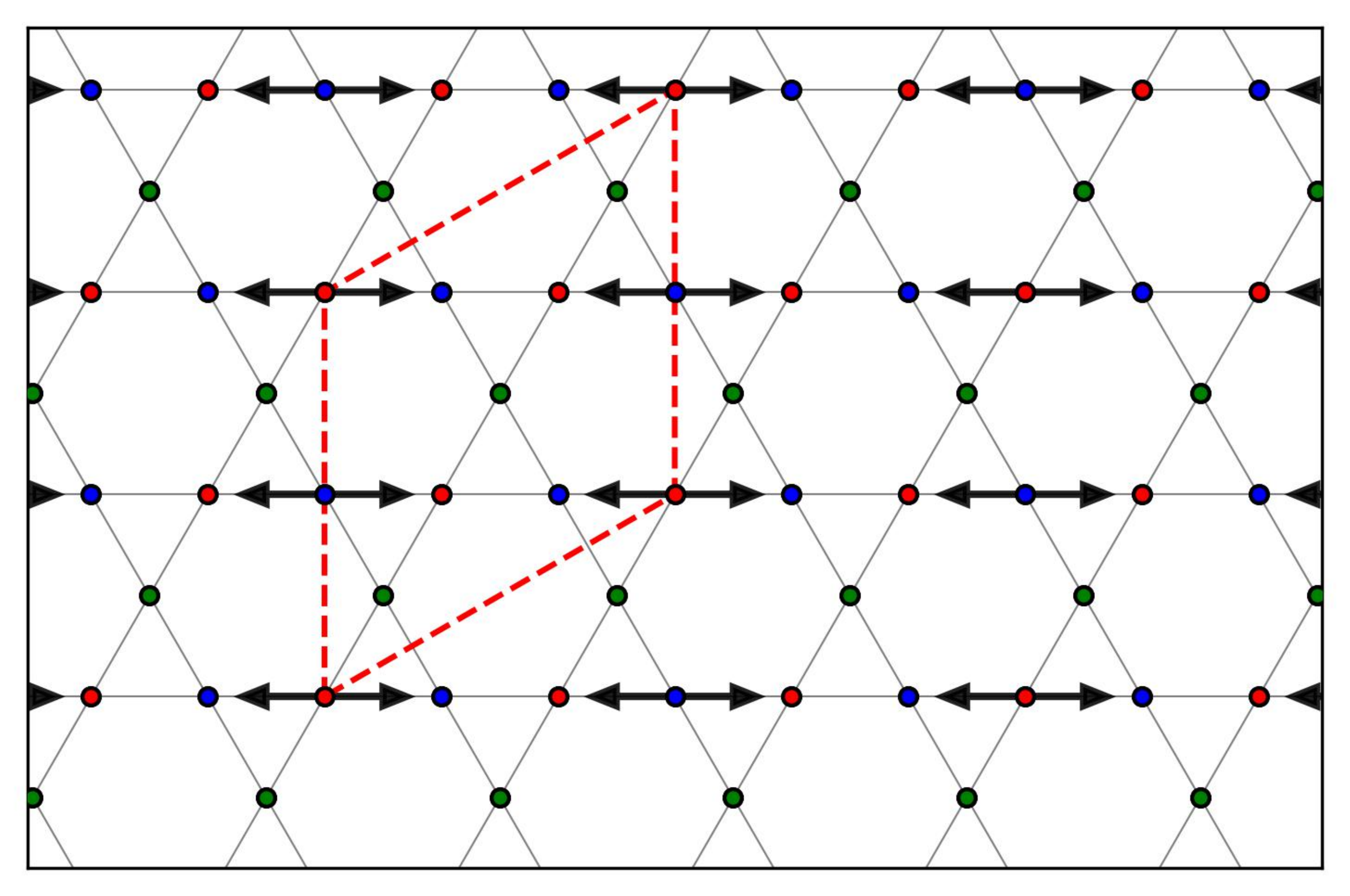}
    \end{minipage}
    \begin{minipage}{0.48\linewidth}
        \centering
        (b) $p_{AC}$, $\mathbf{Q}=K_5$\\
        \includegraphics[width=0.95\linewidth]{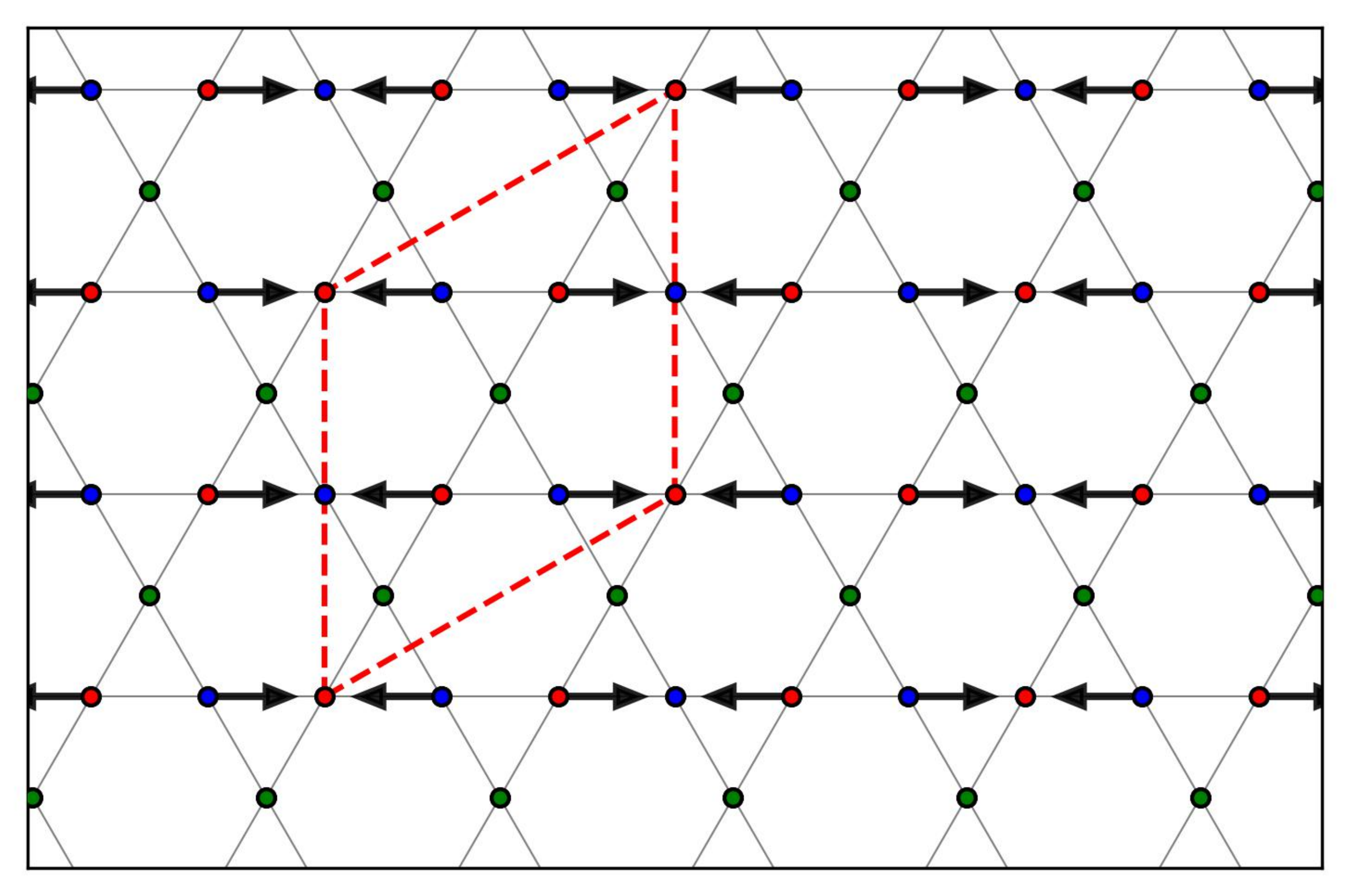}
    \end{minipage}

            \begin{minipage}{0.48\linewidth}
        \centering
        (c) $3\mathbf{Q}$: $(K_2,K_1,K_3)$\\
        \includegraphics[width=0.95\linewidth]{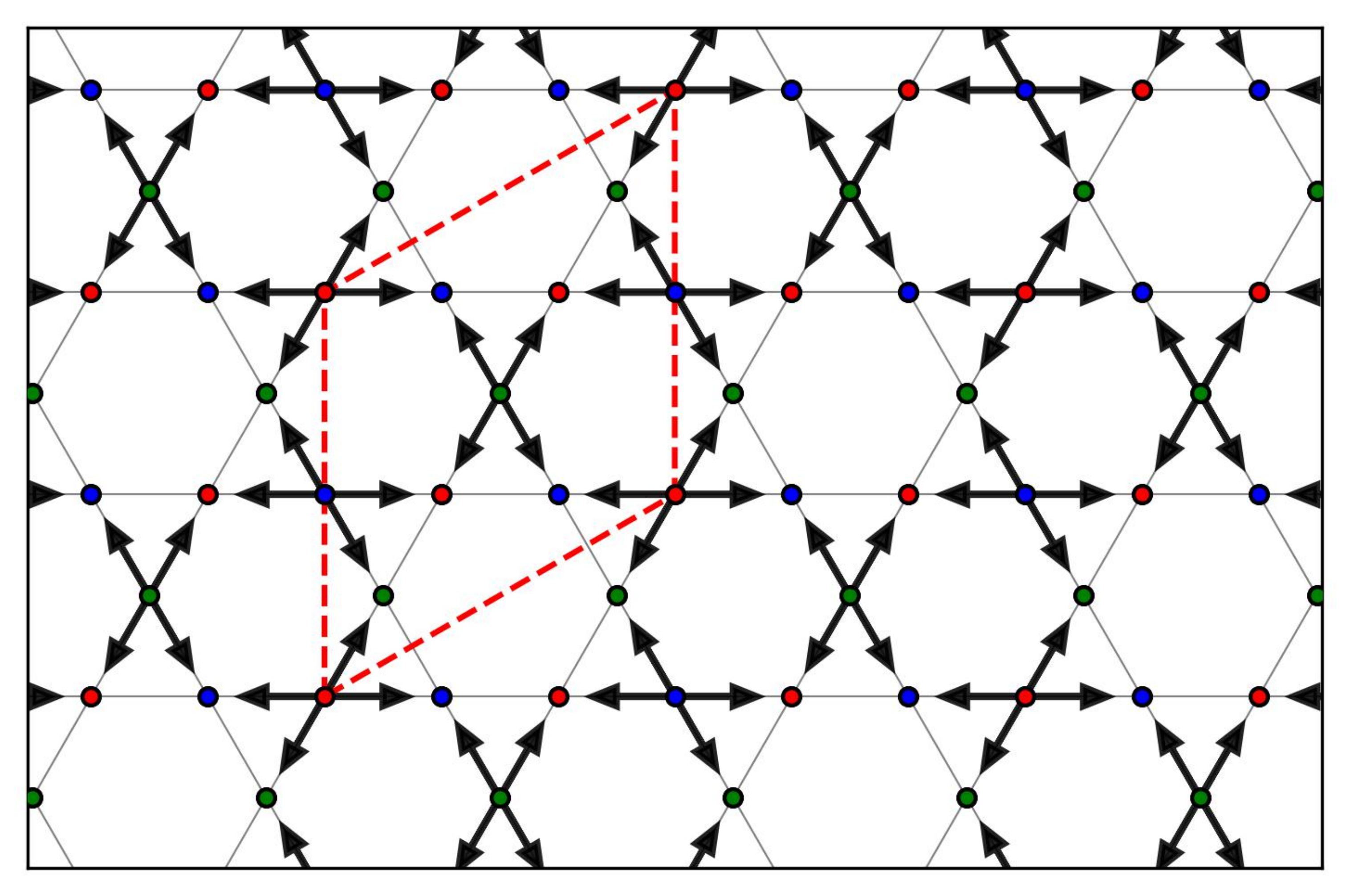}
    \end{minipage}
    \begin{minipage}{0.48\linewidth}
        \centering
        (d) $3\mathbf{Q}$: $(K_2,K_4,K_6)$\\
        \includegraphics[width=0.95\linewidth]{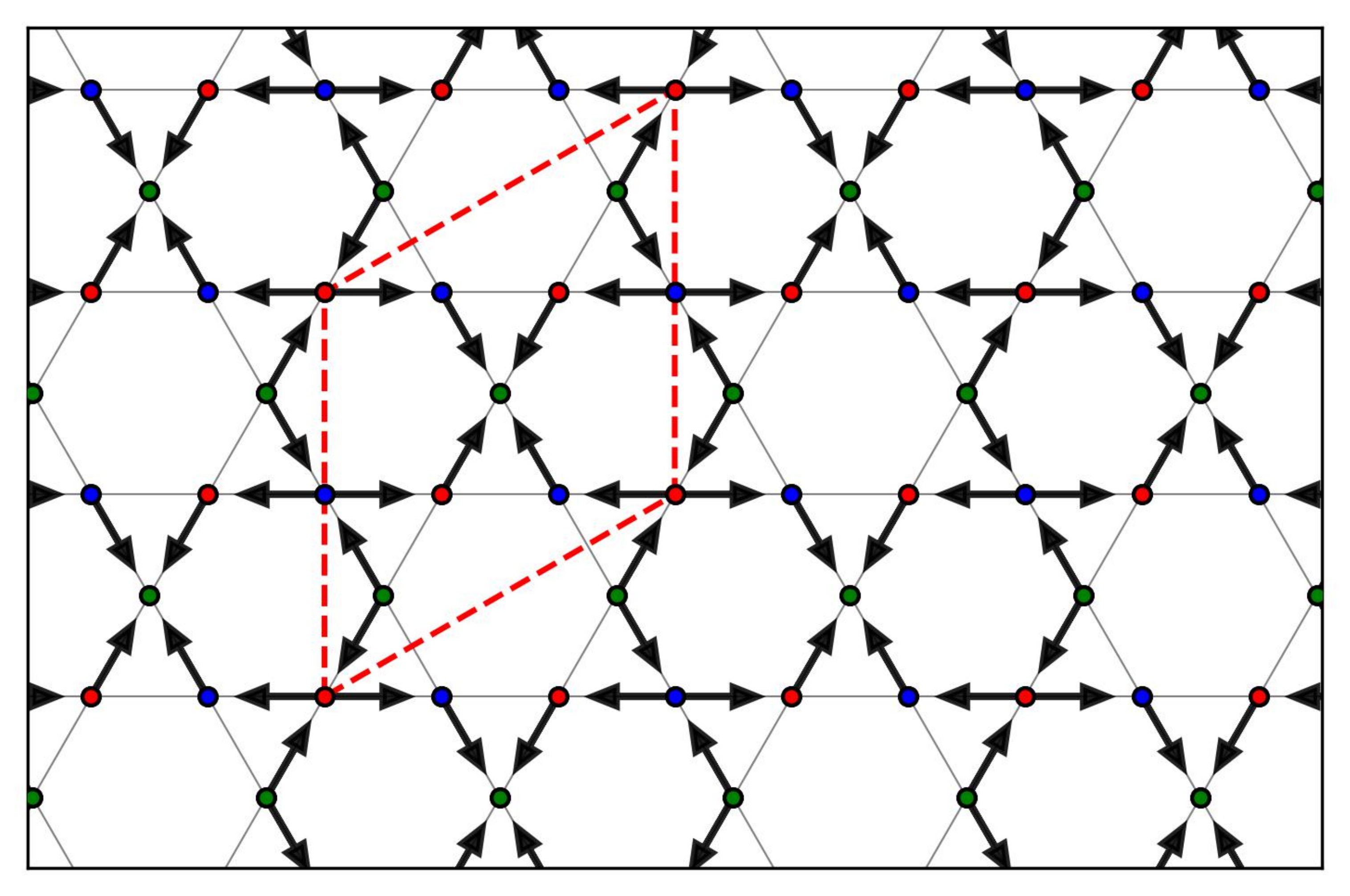}
    \end{minipage}

   \caption{
Real-space bond-current configurations on the kagome lattice associated with $K$-point ordering vectors. (a) $p_{AC}$ at $K_2$. (b) $p_{AC}$ at $K_5$. (c) and (d) Representative $3\mathbf{Q}$ states formed from the combinations $(p_{AC},p_{AB},p_{BC})=(K_2,K_1,K_3)$ and $(K_2,K_4,K_6)$, respectively. Arrows indicate the bond-current direction. The dashed-red lines describe the enlarged $\sqrt{3}\times \sqrt{3}$ unit cells. Unlike the loop-current states stabilized near van Hove filling, these configurations do not form closed circulating-current loops and instead exhibit source--sink current textures. 
}
    \label{fig:Kagome_node}
\end{figure}

Near van Hove filling, the susceptibility is maximized at the $M$ points, consistent with the results discussed in the main text. As the chemical potential moves away from $\mu=0$, however, the susceptibility maxima gradually shift from the $M$ points and eventually appear in neighboring Brillouin-zone corners of the extended-zone representation.  A particularly instructive regime occurs near $\mu=0.7t$, where the susceptibility develops pronounced maxima at the $K_i$ points:  $K_2$ and $K_5$ for $p_{AC}$, $K_1$ and $K_4$ for $p_{AB}$, and $K_3$ and $K_6$ for $p_{BC}$. 
The corresponding real-space current configurations for $p_{AC}$ are shown in Figs.~\ref{fig:Kagome_node}(a) and \ref{fig:Kagome_node}(b). Unlike the loop-current states stabilized near van Hove filling, these configurations do not form closed circulating-current loops. Instead, the currents converge toward and diverge from localized regions, giving rise to source--sink textures.

The absence of loop-current order persists when the dominant $K_i$-point orders are combined into a $3\mathbf{Q}$ state. Representative examples are shown in Figs.~\ref{fig:Kagome_node}(c) and \ref{fig:Kagome_node}(d), constructed from the ordering-vector combinations $(K_2,K_1,K_3)$ and $(K_2,K_4,K_6)$, respectively. Although these states preserve the three-wave-vector structure characteristic of the kagome lattice, they do not generate finite circulating currents. Instead, the resulting current patterns contain sources and sinks, violating the charge conservation. They therefore cannot represent physical states.

At higher fillings, the susceptibility continues to evolve. Near the upper edge of the dispersive $p$-band, $\mu=1.9t$, the dominant ordering vectors move into neighboring Brillouin-zone sectors of the extended-zone representation,
\begin{align}
    \mathbf{Q}'_{AC} &= \pm\left(2\pi,\frac{2\pi}{\sqrt{3}} \right), \left(\pm2\pi,\mp\frac{2\pi}{\sqrt{3}} \right) , \\
    \mathbf{Q}'_{AB} &= \left(0,\pm\frac{4\pi}{\sqrt{3}}\right), \pm\left(2\pi,\frac{2\pi}{\sqrt{3}} \right) , \\
    \mathbf{Q}'_{BC} &= \left(0,\pm\frac{4\pi}{\sqrt{3}}\right), \left(\pm 2\pi,\mp \frac{2\pi}{\sqrt{3}}\right).
\end{align}
As discussed in Appendix~\ref{app.ExtendedBZ}, reciprocal-lattice translations modify the symmetry character of the form factors in the extended-zone representation. Consequently, these ordering vectors are no longer associated with purely sine-type bond-current channels and acquire admixtures of bond-charge components.

\begin{figure}[t]
    \centering
        \begin{minipage}{0.48\linewidth}
        \centering
        (a)\\
        \includegraphics[width=\linewidth]{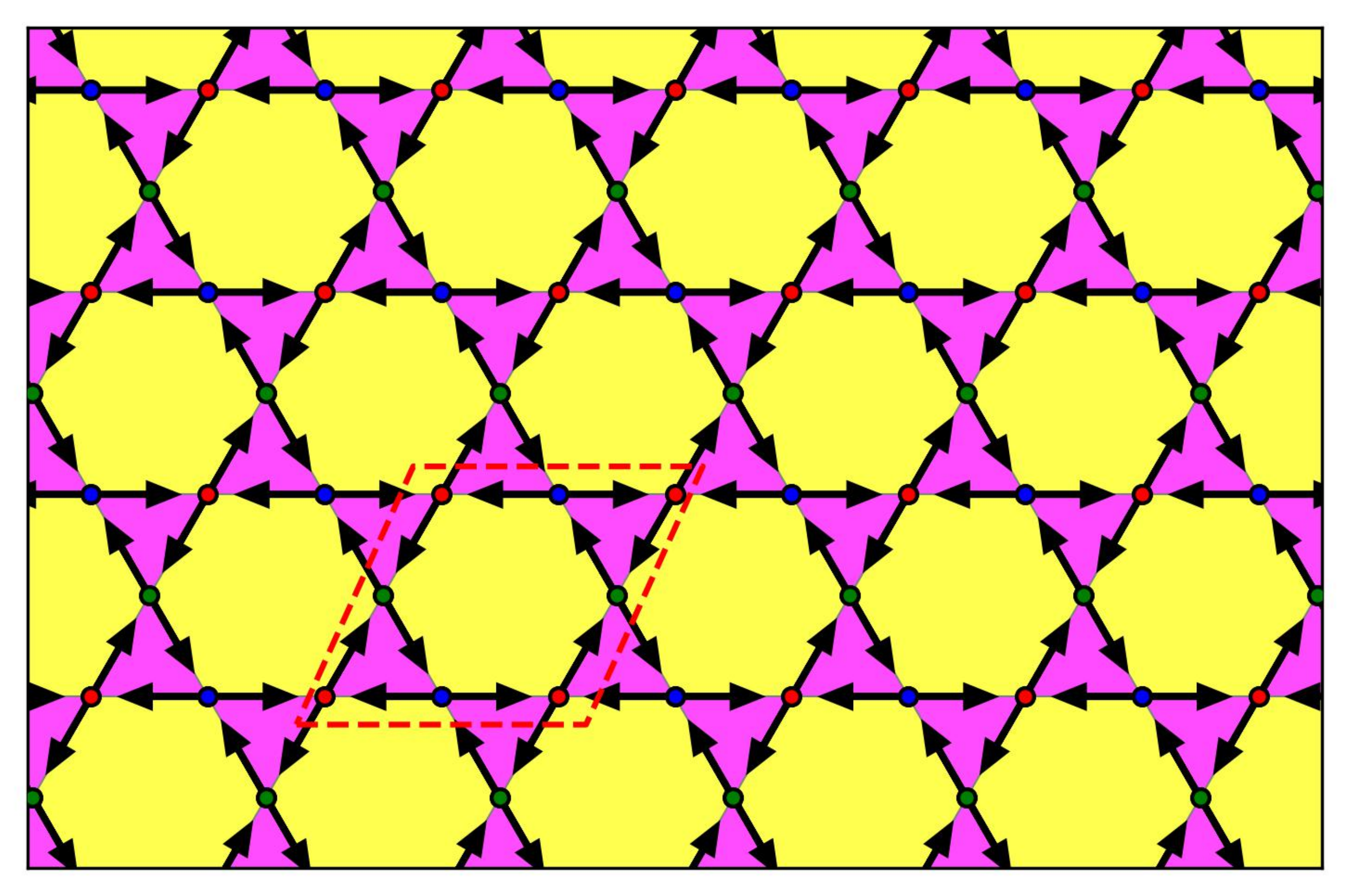}
    \end{minipage}
            \begin{minipage}{0.48\linewidth}
        \centering
        (b) \\
    \includegraphics[width=\linewidth]{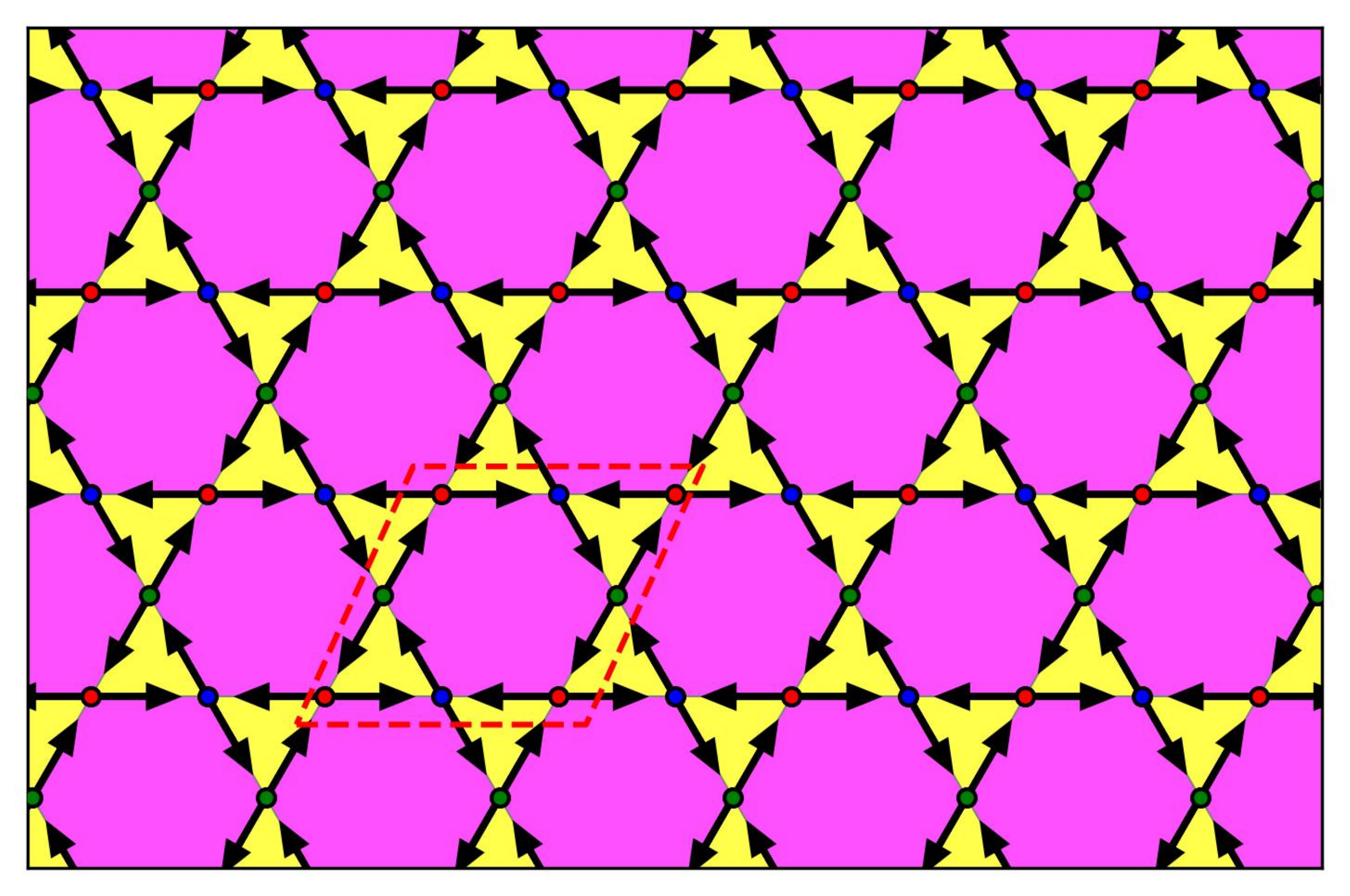}
    \end{minipage}
\caption{
Real-space bond-current configuration of the kagome lattice in the Nagaosa flux phase~\cite{ohgushi00}. Arrows indicate the bond-current direction, while pink (yellow) plaquettes denote clockwise (counterclockwise) circulating currents. The dashed red lines describe the primitive $1\times1$ unit cell.
}
    \label{fig:Nagaosa}
\end{figure}

Finally, as the filling approaches the lower edge of the $p$-band ($\mu = -0.9t$), the susceptibility develops maxima at the $\Gamma$ point, $\mathbf{Q} = (0,0)$, for all three odd-parity bond-current channels. However, according to the real-space current reconstruction in Eq.~\eqref{eq.realCurrentKagome}, the bond currents associated with the sine-form-factor representation vanish identically at $\mathbf{Q}=(0,0)$. Consequently, despite the enhanced susceptibility, these fluctuations cannot generate a finite bond-current order. As discussed in Appendix~\ref{app.FormFac}, the same ordering vector can instead be described using the cosine-form-factor representation, Eqs.~\eqref{eq.DcAC}--\eqref{eq.DcBC}, which remains finite at $\mathbf{Q}=\Gamma$.  In this representation, a closed-loop current state, namely the Nagaosa flux phase~\cite{ohgushi00,feng21a}, shown in Fig.~\ref{fig:Nagaosa}, can be realized, consistent with the Supplemental Material of Ref.~\cite{fu25}. Although the sine- and cosine-form-factor representations are equivalent descriptions of the bond-current operator, the loop-current state at $\Gamma$ is naturally captured only in the cosine representation because the odd-parity basis vanishes identically at this wave vector. Since the present work focuses on the odd-parity bond-current channels, the susceptibility analysis and the corresponding real-space current configurations discussed throughout the main text are based exclusively on the sine-form-factor representation.

These results also demonstrate that the magnitude of the susceptibility alone is insufficient to determine the resulting current topology. Away from van Hove filling, changes in the Fermi-surface topology can favor ordering vectors that produce source--sink current textures or even vanishing bond-current order rather than closed circulating currents. 

%\bibliography{Reference}
\bibliography{main.bib}

@article{affleck88a,
  title = {{Large-n limit of the Heisenberg-Hubbard model: Implications for high-${T}_{c}$ superconductors}},
  author = {Affleck, Ian and Marston, J. Brad},
  journal = {Phys. Rev. B},
  volume = {37},
  issue = {7},
  pages = {3774--3777},
  numpages = {0},
  year = {1988},
  month = {Mar},
  publisher = {American Physical Society},
  optdoi = {10.1103/PhysRevB.37.3774},
  opturl = {https://link.aps.org/doi/10.1103/PhysRevB.37.3774}
}

@article{auvray19,
	Author = {Auvray, N. and Loret, B. and Benhabib, S. and Cazayous, M. and Zhong, R. D. and Schneeloch, J. and Gu, G. D. and Forget, A. and Colson, D. and Paul, I. and Sacuto, A. and Gallais, Y.},
	Da = {2019/11/15},
	Doi = {10.1038/s41467-019-12940-w},
	Id = {Auvray2019},
	Isbn = {2041-1723},
	Journal = {Nat. Commun.},
	Number = {1},
	Pages = {5209},
	Title = {Nematic fluctuations in the cuprate superconductor {Bi}$_2${Sr}$_2${CaCu}$_2${O}$_{8+\delta}$},
	Ty = {JOUR},
	optUrl = {https://doi.org/10.1038/s41467-019-12940-w},
	Volume = {10},
	Year = {2019},
	optBdsk-Url-1 = {https://doi.org/10.1038/s41467-019-12940-w}}

@article{bejas12,
  title = {Possible charge instabilities in two-dimensional doped {Mott} insulators},
  author = {Bejas, Mat\'{\i}as and Greco, Andr\'es and Yamase, Hiroyuki},
  journal = {Phys. Rev. B},
  volume = {86},
  issue = {22},
  pages = {224509},
  numpages = {12},
  year = {2012},
  month = {Dec},
  publisher = {American Physical Society},
  optdoi = {10.1103/PhysRevB.86.224509},
  opturl = {https://link.aps.org/doi/10.1103/PhysRevB.86.224509}}

@article{bejas14,
        author = "Bejas, Mat{\'i}as and Greco, Andr{\'e}s and Yamase, Hiroyuki",
        journal = "New J. Phys.",
        number = "12",
        pages = "123002",
        title = "{Strong particle-hole asymmetry of charge instabilities in doped Mott insulators}",
        ourl = "http://stacks.iop.org/1367-2630/16/i=12/a=123002",
        volume = "16",
        year = "2014"
}

@article{bejas17,
  title = {Dual structure in the charge excitation spectrum of electron-doped cuprates},
  author = {Bejas, Mat\'{\i}as and Yamase, Hiroyuki and Greco, Andr\'es},
  journal = {Phys. Rev. B},
  volume = {96},
  issue = {21},
  pages = {214513},
  numpages = {12},
  year = {2017},
  month = {Dec},
  publisher = {American Physical Society},
  doi = {10.1103/PhysRevB.96.214513},
  url = {https://link.aps.org/doi/10.1103/PhysRevB.96.214513}
}

@article{cappelluti99,
  title = {{Interplay between superconductivity and flux phase in the $t\ensuremath{-}J$ model}},
  author = {Cappelluti, E. and Zeyher, R.},
  journal = {Phys. Rev. B},
  volume = {59},
  issue = {9},
  pages = {6475--6486},
  numpages = {0},
  year = {1999},
  month = {Mar},
  publisher = {American Physical Society},
  optdoi = {10.1103/PhysRevB.59.6475},
  opturl = {https://link.aps.org/doi/10.1103/PhysRevB.59.6475}
}

@article{chakravarty01,
  title = {Hidden order in the cuprates},
  author = {Chakravarty, Sudip and Laughlin, R. B. and Morr, Dirk K. and Nayak, Chetan},
  journal = {Phys. Rev. B},
  volume = {63},
  issue = {9},
  pages = {094503},
  numpages = {10},
  year = {2001},
  month = {Jan},
  publisher = {American Physical Society},
  optdoi = {10.1103/PhysRevB.63.094503},
  opturl = {https://link.aps.org/doi/10.1103/PhysRevB.63.094503}
}

@article{chen25,
  title = {Emergence of Chiral Phonons in Two-Dimensional Kagome Lattices Harboring Electronic Chirality},
  author = {Chen, Yanru and Qin, Wei and Zhang, Shunhong and Cui, Ping and Niu, Qian and Zhang, Zhenyu},
  journal = {Phys. Rev. Lett.},
  volume = {135},
  issue = {12},
  pages = {126608},
  numpages = {9},
  year = {2025},
  month = {Sep},
  publisher = {American Physical Society},
  doi = {10.1103/bfll-sdrb},
  url = {https://link.aps.org/doi/10.1103/bfll-sdrb}
}

@article{christensen22,
  title = {Loop currents in {AV}$_{3}${Sb}$_{5}$ kagome metals: Multipolar and toroidal magnetic orders},
  author = {Christensen, Morten H. and Birol, Turan and Andersen, Brian M. and Fernandes, Rafael M.},
  journal = {Phys. Rev. B},
  volume = {106},
  issue = {14},
  pages = {144504},
  numpages = {18},
  year = {2022},
  month = {Oct},
  publisher = {American Physical Society},
  doi = {10.1103/PhysRevB.106.144504},
  url = {https://link.aps.org/doi/10.1103/PhysRevB.106.144504}
}

@article{denner21,
  title = {Analysis of Charge Order in the Kagome Metal $\mathrm{AV}_{3}\mathrm{Sb}_{5}$ ($\mathrm{A}=\mathrm{K},\mathrm{Rb},\mathrm{Cs}$)},
  author = {Denner, M. Michael and Thomale, Ronny and Neupert, Titus},
  journal = {Phys. Rev. Lett.},
  volume = {127},
  issue = {21},
  pages = {217601},
  numpages = {6},
  year = {2021},
  month = {Nov},
  publisher = {American Physical Society},
  doi = {10.1103/PhysRevLett.127.217601},
  url = {https://link.aps.org/doi/10.1103/PhysRevLett.127.217601}
}

@article{dong23,
  title = {Loop-current charge density wave driven by long-range {C}oulomb repulsion on the kagom\'e lattice},
  author = {Dong, Jin-Wei and Wang, Ziqiang and Zhou, Sen},
  journal = {Phys. Rev. B},
  volume = {107},
  issue = {4},
  pages = {045127},
  numpages = {12},
  year = {2023},
  month = {Jan},
  publisher = {American Physical Society},
  doi = {10.1103/PhysRevB.107.045127},
  url = {https://link.aps.org/doi/10.1103/PhysRevB.107.045127}
}

@ARTICLE{feng21,
title = {Chiral flux phase in the Kagome superconductor $\mathrm{AV}_3\mathrm{Sb}_5$},
  author    = "Feng, Xilin and Jiang, Kun and Wang, Ziqiang and Hu, Jiangping",
  journal   = "Sci. Bull. ",
  publisher = "Elsevier BV",
  volume    =  66,
  number    =  14,
  pages     = "1384--1388",
  month     =  jul,
  year      =  2021,
doi = {https://doi.org/10.1016/j.scib.2021.04.043},
url = {https://www.sciencedirect.com/science/article/pii/S2095927321003224},
}

@article{feng21a,
  title = {Low-energy effective theory and symmetry classification of flux phases on the kagome lattice},
  author = {Feng, Xilin and Zhang, Yi and Jiang, Kun and Hu, Jiangping},
  journal = {Phys. Rev. B},
  volume = {104},
  issue = {16},
  pages = {165136},
  numpages = {10},
  year = {2021},
  month = {Oct},
  publisher = {American Physical Society},
  doi = {10.1103/PhysRevB.104.165136},
  url = {https://link.aps.org/doi/10.1103/PhysRevB.104.165136}
}

@article{fernandes14,
Author = {R. M. Fernandes and A. V. Chubukov and J. Schmalian},
Journal = {Nat. Phys.},
Pages = {97},
Volume = {10},
Year = {2014}}

@article{friedlan26,
  title = {Emergence of nematic loop-current bond order in kagome metals near van Hove singularities},
  author = {Friedlan, Alex and Kee, Hae-Young},
  journal = {Phys. Rev. B},
  volume = {113},
  issue = {12},
  pages = {125118},
  numpages = {11},
  year = {2026},
  month = {Mar},
  publisher = {American Physical Society},
  doi = {10.1103/dz1w-s9t2},
  url = {https://link.aps.org/doi/10.1103/dz1w-s9t2}
}

@article{fu25,
    author = {Fu, Ruiqing and Zhan, Jun and D\"urrnagel, Matteo and Hohmann, Hendrik and Thomale, Ronny and Hu, Jiangping and Wang, Ziqiang and Zhou, Sen and Wu, Xianxin},
    title = {Exotic charge-density waves and superconductivity on the kagome lattice},
  journal   = "Natl. Sci. Rev.",
  publisher = "Oxford University Press (OUP)",
    volume = {12},
    number = {11},
    pages = {nwaf414},
    year = {2025},
    month = {11},
    issn = {2095-5138},
    doi = {10.1093/nsr/nwaf414},
    url = {https://doi.org/10.1093/nsr/nwaf414}
}

@article{gallais13,
	Author = {Y. Gallais and R. M. Fernandes and I. Paul and L. Chauvi\`{e}re and Y. -X. Yang and M. -A. M\'{e}asson and M. Cazayous and A. Sacuto and D. Colson and A. Forget},
	Journal = {Phys. Rev. Lett.},
	Optauthor = {{M. Fujita \it et al.}},
	Pages = {267001},
	Volume = {111},
	Year = {2013}}

@ARTICLE{gneist22,
  title     = "Functional renormalization of spinless triangular-lattice
               fermions: N-patch vs. truncated-unity scheme",
  author    = "Gneist, Nico and Kiese, Dominik and Henkel, Ravn and Thomale,
               Ronny and Classen, Laura and Scherer, Michael M",
  journal   = "Eur. Phys. J. B",
  publisher = "Springer Science and Business Media LLC",
  volume    =  95,
  number    =  9,
  pages={157},
  month     =  sep,
  year      =  2022,
doi = "https://doi.org/10.1140/epjb/s10051-022-00395-w"
}

@ARTICLE{guguchia23,
  title     = "Tunable unconventional kagome superconductivity in charge ordered $\mathrm{RbV}_3\mathrm{Sb}_5$ and $\mathrm{KV}_3\mathrm{Sb}_5$",
 author={Guguchia, Z.
and Mielke, C.
and Das, D.
and Gupta, R.
and Yin, J.-X.
and Liu, H.
and Yin, Q.
and Christensen, M. H.
and Tu, Z.
and Gong, C.
and Shumiya, N.
and Hossain, Md Shafayat
and Gamsakhurdashvili, Ts.
and Elender, M.
and Dai, Pengcheng
and Amato, A.
and Shi, Y.
and Lei, H. C.
and Fernandes, R. M.
and Hasan, M. Z.
and Luetkens, H.
and Khasanov, R.},
  journal   = "Nat. Commun.",
  publisher = "Springer Science and Business Media LLC",
  volume    =  14,
  number    =  1,
  pages     = "153",
  month     =  jan,
  year      =  2023,
  doi = "https://doi.org/10.1038/s41467-022-35718-z"
}

@article{guo09,
  title = {Topological insulator on the kagome lattice},
  author = {Guo, H.-M. and Franz, M.},
  journal = {Phys. Rev. B},
  volume = {80},
  issue = {11},
  pages = {113102},
  numpages = {4},
  year = {2009},
  month = {Sep},
  publisher = {American Physical Society},
  doi = {10.1103/PhysRevB.80.113102},
  url = {https://link.aps.org/doi/10.1103/PhysRevB.80.113102}
}

@ARTICLE{guo22,
  title     = "Switchable chiral transport in charge-ordered kagome metal
               $\mathrm{CsV}_3\mathrm{Sb}_5$",
  author    = "Guo, Chunyu and Putzke, Carsten and Konyzheva, Sofia and Huang,
               Xiangwei and Gutierrez-Amigo, Martin and Errea, Ion and Chen,
               Dong and Vergniory, Maia G and Felser, Claudia and Fischer, Mark
               H and Neupert, Titus and Moll, Philip J W",
  journal   = "Nature",
  publisher = "Springer Science and Business Media LLC",
  volume    =  611,
  number    =  7936,
  pages     = "461--466",
  month     =  nov,
  year      =  2022,
  doi = "https://doi.org/10.1038/s41586-022-05127-9"
}

@ARTICLE{jiang21,
  title = {Unconventional chiral charge order in kagome superconductor $\mathrm{KV}_3\mathrm{Sb}_5$},
  author    = "Jiang, Yu-Xiao and Yin, Jia-Xin and Denner, M Michael and
               Shumiya, Nana and Ortiz, Brenden R and Xu, Gang and Guguchia,
               Zurab and He, Junyi and Hossain, Md Shafayat and Liu, Xiaoxiong
               and Ruff, Jacob and Kautzsch, Linus and Zhang, Songtian S and
               Chang, Guoqing and Belopolski, Ilya and Zhang, Qi and Cochran,
               Tyler A and Multer, Daniel and Litskevich, Maksim and Cheng,
               Zi-Jia and Yang, Xian P and Wang, Ziqiang and Thomale, Ronny and
               Neupert, Titus and Wilson, Stephen D and Hasan, M Zahid",
  journal   = "Nat. Mater.",
  publisher = "Springer Science and Business Media LLC",
  volume    =  20,
  number    =  10,
  pages     = "1353--1357",
  month     =  oct,
  year      =  2021,
    url = {http://dx.doi.org/10.1038/s41563-021-01034-y},
  DOI = {10.1038/s41563-021-01034-y},
}

@article{keimer15,
author={Keimer, B.
and Kivelson, S. A.
and Norman, M. R.
and Uchida, S.
and Zaanen, J.},
title={From quantum matter to high-temperature superconductivity in copper oxides},
journal={Nature},
year={2015},
month={Feb},
day={01},
volume={518},
number={7538},
pages={179-186},
issn={1476-4687},
doi={10.1038/nature14165},
url={https://doi.org/10.1038/nature14165}
}

@article{khasanov22,
  title = {Time-reversal symmetry broken by charge order in $\mathrm{CsV}_{3}\mathrm{Sb}_{5}$},
  author = {Khasanov, Rustem and Das, Debarchan and Gupta, Ritu and Mielke, Charles and Elender, Matthias and Yin, Qiangwei and Tu, Zhijun and Gong, Chunsheng and Lei, Hechang and Ritz, Ethan T. and Fernandes, Rafael M. and Birol, Turan and Guguchia, Zurab and Luetkens, Hubertus},
  journal = {Phys. Rev. Res.},
  volume = {4},
  issue = {2},
  pages = {023244},
  numpages = {13},
  year = {2022},
  month = {Jun},
  publisher = {American Physical Society},
  doi = {10.1103/PhysRevResearch.4.023244},
  url = {https://link.aps.org/doi/10.1103/PhysRevResearch.4.023244}
}

@article{kiesel13,
  title = {Unconventional {F}ermi Surface Instabilities in the Kagome {H}ubbard Model},
  author = {Kiesel, Maximilian L. and Platt, Christian and Thomale, Ronny},
  journal = {Phys. Rev. Lett.},
  volume = {110},
  issue = {12},
  pages = {126405},
  numpages = {5},
  year = {2013},
  month = {Mar},
  publisher = {American Physical Society},
  doi = {10.1103/PhysRevLett.110.126405},
  url = {https://link.aps.org/doi/10.1103/PhysRevLett.110.126405}
}

@article{kivelson98,
author={Kivelson, S. A. and Fradkin, E. and Emery, V. J.},
title={Electronic liquid-crystal phases of a doped Mott insulator},
journal={Nature},
year={1998},
month={Jun},
day={11},
publisher={Macmillan Magazines Ltd.},
volume={393},
pages={550},
url={http://dx.doi.org/10.1038/31177}
}

@article{klein18,
  title = {Superconductivity near a nematic quantum critical point: Interplay between hot and lukewarm regions},
  author = {Klein, Avraham and Chubukov, Andrey},
  journal = {Phys. Rev. B},
  volume = {98},
  issue = {22},
  pages = {220501},
  numpages = {5},
  year = {2018},
  month = {Dec},
  publisher = {American Physical Society},
  doi = {10.1103/PhysRevB.98.220501},
  url = {https://link.aps.org/doi/10.1103/PhysRevB.98.220501}
}

@article{lee06,
  title = {Doping a {Mott} insulator: Physics of high-temperature superconductivity},
  author = {Lee, Patrick A. and Nagaosa, Naoto and Wen, Xiao-Gang},
  journal = {Rev. Mod. Phys.},
  volume = {78},
  issue = {1},
  pages = {17--85},
  numpages = {0},
  year = {2006},
  month = {Jan},
  publisher = {American Physical Society},
  optdoi = {10.1103/RevModPhys.78.17},
  opturl = {https://link.aps.org/doi/10.1103/RevModPhys.78.17}
}

@article{li17,
  title = {{Nature of the effective interaction in electron-doped cuprate superconductors: A sign-problem-free quantum Monte Carlo study}},
  author = {Li, Zi-Xiang and Wang, Fa and Yao, Hong and Lee, Dung-Hai},
  journal = {Phys. Rev. B},
  volume = {95},
  issue = {21},
  pages = {214505},
  numpages = {7},
  year = {2017},
  month = {Jun},
  publisher = {American Physical Society},
  doi = {10.1103/PhysRevB.95.214505},
  url = {https://link.aps.org/doi/10.1103/PhysRevB.95.214505}
}

@ARTICLE{li22,
  title     = "Rotation symmetry breaking in the normal state of a kagome
               superconductor $\mathrm{KV}_3\mathrm{Sb}_5$",
  author    = "Li, Hong and Zhao, He and Ortiz, Brenden R and Park, Takamori
               and Ye, Mengxing and Balents, Leon and Wang, Ziqiang and Wilson,
               Stephen D and Zeljkovic, Ilija",
  journal   = "Nat. Phys.",
  publisher = "Springer Science and Business Media LLC",
  volume    =  18,
  number    =  3,
  pages     = "265--270",
  month     =  mar,
  year      =  2022,
  doi = "https://doi.org/10.1038/s41567-021-01479-7"
}

@article{metzner00,
  title = {{$\mathit{d}$-Wave Superconductivity and Pomeranchuk Instability in the Two-Dimensional Hubbard Model}},
  author = {Halboth, Christoph J. and Metzner, Walter},
  journal = {Phys. Rev. Lett.},
  volume = {85},
  issue = {24},
  pages = {5162--5165},
  numpages = {0},
  year = {2000},
  month = {Dec},
  publisher = {American Physical Society},
  doi = {10.1103/PhysRevLett.85.5162},
  opturl = {https://link.aps.org/doi/10.1103/PhysRevLett.85.5162}
}

@article{mielke22,
  title = {Time-reversal symmetry-breaking charge order in a kagome superconductor},
  volume = {602},
  ISSN = {1476-4687},
  url = {http://dx.doi.org/10.1038/s41586-021-04327-z},
  DOI = {10.1038/s41586-021-04327-z},
  number = {7896},
  journal = {Nature},
  publisher = {Springer Science and Business Media LLC},
  author = {Mielke,  C. and Das,  D. and Yin,  J.-X. and Liu,  H. and Gupta,  R. and Jiang,  Y.-X. and Medarde,  M. and Wu,  X. and Lei,  H. C. and Chang,  J. and Dai,  Pengcheng and Si,  Q. and Miao,  H. and Thomale,  R. and Neupert,  T. and Shi,  Y. and Khasanov,  R. and Hasan,  M. Z. and Luetkens,  H. and Guguchia,  Z.},
  year = {2022},
  month = Feb,
  pages = {245--250}
}

@article{nakata21,
	author = {Nakata, S. and Horio, M. and Koshiishi, K. and Hagiwara, K. and Lin, C. and Suzuki, M. and Ideta, S. and Tanaka, K. and Song, D. and Yoshida, Y. and Eisaki, H. and Fujimori, A.},
	date = {2021/10/07},
	doi = {10.1038/s41535-021-00390-x},
	id = {Nakata2021},
	isbn = {2397-4648},
	journal = {npj Quantum Materials},
	number = {1},
	pages = {86},
	title = {Nematicity in a cuprate superconductor revealed by angle-resolved photoemission spectroscopy under uniaxial strain},
	url = {https://doi.org/10.1038/s41535-021-00390-x},
	volume = {6},
	year = {2021},
	optbdsk-url-1 = {https://doi.org/10.1038/s41535-021-00390-x}}

@article{nie22,
  title = {Charge-density-wave-driven electronic nematicity in a kagome superconductor},
  volume = {604},
  ISSN = {1476-4687},
  url = {http://dx.doi.org/10.1038/s41586-022-04493-8},
  DOI = {10.1038/s41586-022-04493-8},
  number = {7904},
  journal = {Nature},
  publisher = {Springer Science and Business Media LLC},
  author = {Nie,  Linpeng and Sun,  Kuanglv and Ma,  Wanru and Song,  Dianwu and Zheng,  Lixuan and Liang,  Zuowei and Wu,  Ping and Yu,  Fanghang and Li,  Jian and Shan,  Min and Zhao,  Dan and Li,  Shunjiao and Kang,  Baolei and Wu,  Zhimian and Zhou,  Yanbing and Liu,  Kai and Xiang,  Ziji and Ying,  Jianjun and Wang,  Zhenyu and Wu,  Tao and Chen,  Xianhui},
  year = {2022},
  month = Feb,
  pages = {59--64}
}

@article{ohgushi00,
  title = {Spin anisotropy and quantum Hall effect in the kagom\'e lattice: Chiral spin state based on a ferromagnet},
  author = {Ohgushi, Kenya and Murakami, Shuichi and Nagaosa, Naoto},
  journal = {Phys. Rev. B},
  volume = {62},
  issue = {10},
  pages = {R6065--R6068},
  numpages = {0},
  year = {2000},
  month = {Sep},
  publisher = {American Physical Society},
  doi = {10.1103/PhysRevB.62.R6065},
  url = {https://link.aps.org/doi/10.1103/PhysRevB.62.R6065}
}

@article{ortiz19,
  title = {New kagome prototype materials: discovery of $\mathrm{KV}_{3}\mathrm{Sb}_{5},\mathrm{RbV}_{3}\mathrm{Sb}_{5}$, and $\mathrm{CsV}_{3}\mathrm{Sb}_{5}$},
  author = {Ortiz, Brenden R. and Gomes, L\'{\i}dia C. and Morey, Jennifer R. and Winiarski, Michal and Bordelon, Mitchell and Mangum, John S. and Oswald, Iain W. H. and Rodriguez-Rivera, Jose A. and Neilson, James R. and Wilson, Stephen D. and Ertekin, Elif and McQueen, Tyrel M. and Toberer, Eric S.},
  journal = {Phys. Rev. Mater.},
  volume = {3},
  issue = {9},
  pages = {094407},
  numpages = {9},
  year = {2019},
  month = {Sep},
  publisher = {American Physical Society},
  doi = {10.1103/PhysRevMaterials.3.094407},
  url = {https://link.aps.org/doi/10.1103/PhysRevMaterials.3.094407}
}

@article{ortiz20,
  title = {$\mathrm{Cs}\mathrm{V}_{3}\mathrm{Sb}_{5}$: A $\mathbb{Z}_{2}$ Topological Kagome Metal with a Superconducting Ground State},
  author = {Ortiz, Brenden R. and Teicher, Samuel M. L. and Hu, Yong and Zuo, Julia L. and Sarte, Paul M. and Schueller, Emily C. and Abeykoon, A. M. Milinda and Krogstad, Matthew J. and Rosenkranz, Stephan and Osborn, Raymond and Seshadri, Ram and Balents, Leon and He, Junfeng and Wilson, Stephen D.},
  journal = {Phys. Rev. Lett.},
  volume = {125},
  issue = {24},
  pages = {247002},
  numpages = {6},
  year = {2020},
  month = {Dec},
  publisher = {American Physical Society},
  doi = {10.1103/PhysRevLett.125.247002},
  url = {https://link.aps.org/doi/10.1103/PhysRevLett.125.247002}
}

@article{park21,
  title = {Electronic instabilities of kagome metals: Saddle points and Landau theory},
  author = {Park, Takamori and Ye, Mengxing and Balents, Leon},
  journal = {Phys. Rev. B},
  volume = {104},
  issue = {3},
  pages = {035142},
  numpages = {20},
  year = {2021},
  month = {Jul},
  publisher = {American Physical Society},
  doi = {10.1103/PhysRevB.104.035142},
  url = {https://link.aps.org/doi/10.1103/PhysRevB.104.035142}
}

@article{sachdev13,
  title = {Bond Order in Two-Dimensional Metals with Antiferromagnetic Exchange Interactions},
  author = {Sachdev, Subir and La Placa, Rolando},
  journal = {Phys. Rev. Lett.},
  volume = {111},
  issue = {2},
  pages = {027202},
  numpages = {5},
  year = {2013},
  month = {Jul},
  publisher = {American Physical Society},
  optdoi = {10.1103/PhysRevLett.111.027202},
  opturl = {https://link.aps.org/doi/10.1103/PhysRevLett.111.027202}
}

@article{sato17,
	Author = {Sato, Y. and Kasahara, S. and Murayama, H. and Kasahara, Y. and Moon, E. -G. and Nishizaki, T. and Loew, T. and Porras, J. and Keimer, B. and Shibauchi, T. and Matsuda, Y.},
	Date = {2017/07/24/online},
	Day = {24},
	Journal = {Nat. Phys.},
	L3 = {10.1038/nphys4205; https://www.nature.com/articles/nphys4205#supplementary-information},
	Month = {07},
	Pages = {1074},
	Publisher = {Nature Publishing Group},
	Title = {{Thermodynamic evidence for a nematic phase transition at the onset of the pseudogap in YBa$_2$Cu$_3$O$_y$}},
	Ty = {JOUR},
	Url = {http://dx.doi.org/10.1038/nphys4205},
	Volume = {13},
	Year = {2017}
	}

@article{scalapino12,
	  title = {A common thread: The pairing interaction for unconventional superconductors},
	    author = {Scalapino, D. J.},
	      journal = {Rev. Mod. Phys.},
	        volume = {84},
		  issue = {4},
		    pages = {1383--1417},
		      numpages = {0},
		        year = {2012},
			  month = {Oct},
			    publisher = {American Physical Society},
}

@ARTICLE{schultz26,
  title     = "Superconductivity in kagome metals due to soft loop-current
               fluctuations",
  author    = "Schultz, Daniel J and Palle, Grgur and Mitra, Asimpunya and Kim,
               Yong Baek and Fernandes, Rafael M and Schmalian, J{\"o}rg",
  journal   = "Nat. Commun.",
  publisher = "Springer Science and Business Media LLC",
  volume    =  17,
  number    =  1,
pages={4557},
  month     =  may,
  year      =  2026,
  doi = "https://doi.org/10.1038/s41467-026-72806-w"
}

@article{shekhter09,
  title = {Considerations on the symmetry of loop order in cuprates},
  author = {Shekhter, A. and Varma, C. M.},
  journal = {Phys. Rev. B},
  volume = {80},
  issue = {21},
  pages = {214501},
  numpages = {7},
  year = {2009},
  month = {Dec},
  publisher = {American Physical Society},
  doi = {10.1103/PhysRevB.80.214501},
  url = {https://link.aps.org/doi/10.1103/PhysRevB.80.214501}
}

@Article{suetsugu26,
author={Suetsugu, S.
and Hori, F.
and Shibata, M.
and Kitagawa, S.
and Ishida, K.
and Asaba, T.
and Nakazawa, S.
and Li, Q.
and Wen, H.-H.
and Shibauchi, T.
and Kontani, H.
and Matsuda, Y.},
title={Microscopic signatures of an imaginary charge density wave in a kagome metal},
journal={Nature Physics},
year={2026},
month={Jun},
day={17},
issn={1745-2481},
doi={10.1038/s41567-026-03339-8},
url={https://doi.org/10.1038/s41567-026-03339-8}
}

@ARTICLE{tazai23,
  title    = "Charge-loop current order and $\mathrm{Z}_3$ nematicity mediated by bond
              order fluctuations in kagome metals",
  author   = "Tazai, Rina and Yamakawa, Youichi and Kontani, Hiroshi",
  journal  = "Nat. Commun.",
  volume   =  14,
  number   =  1,
  pages    = "7845",
  month    =  nov,
  year     =  2023,
  doi = "https://doi.org/10.1038/s41467-023-42952-6"
}

@article{tranquada95,
	author = {Tranquada, J. M. and Sternlieb, B. J. and Axe, J. D. and Nakamura, Y. and Uchida, S.},
	date = {1995/06/01},
	doi = {10.1038/375561a0},
	id = {Tranquada1995},
	isbn = {1476-4687},
	journal = {Nature},
	number = {6532},
	pages = {561--563},
	title = {Evidence for stripe correlations of spins and holes in copper oxide superconductors},
	url = {https://doi.org/10.1038/375561a0},
	volume = {375},
	year = {1995}}

@article{varma97,
  title = {Non-Fermi-liquid states and pairing instability of a general model of copper oxide metals},
  author = {Varma, C. M.},
  journal = {Phys. Rev. B},
  volume = {55},
  issue = {21},
  pages = {14554--14580},
  numpages = {0},
  year = {1997},
  month = {Jun},
  publisher = {American Physical Society},
  doi = {10.1103/PhysRevB.55.14554},
  url = {https://link.aps.org/doi/10.1103/PhysRevB.55.14554}
}

@article{varma06a,
  title = {Theory of the pseudogap state of the cuprates},
  author = {Varma, C. M.},
  journal = {Phys. Rev. B},
  volume = {73},
  issue = {15},
  pages = {155113},
  numpages = {17},
  year = {2006},
  month = {Apr},
  publisher = {American Physical Society},
  doi = {10.1103/PhysRevB.73.155113},
  url = {https://link.aps.org/doi/10.1103/PhysRevB.73.155113}
}

@ARTICLE{varma14,
  title     = "Pseudogap in cuprates in the loop-current ordered state",
  author    = "Varma, C M",
  journal   = "J. Phys. Condens. Matter",
  publisher = "IOP Publishing",
  volume    =  26,
  number    =  50,
  pages     = "505701",
  month     =  dec,
  year      =  2014,
  doi = "10.1088/0953-8984/26/50/505701"
}

@article{wang05,
doi = {10.1209/epl/i2004-10357-4},
url = {https://doi.org/10.1209/epl/i2004-10357-4},
year = {2005},
month = {jan},
publisher = {},
volume = {69},
number = {3},
pages = {404},
author = {Y. F. Wang and C. D. Gong and S. Y. Zhu},
title = {Field-induced gap, pseudogap and new 
Van Hove singularity in the triangular lattice},
journal = {Europhysics Letters}
}

@article{wang06,
  title = {Extended staggered-flux phases in two-dimensional lattices},
  author = {Wang, Yi-Fei and Gong, Chang-De and Zhu, Shi-Yao},
  journal = {Phys. Rev. B},
  volume = {73},
  issue = {19},
  pages = {193106},
  numpages = {4},
  year = {2006},
  month = {May},
  publisher = {American Physical Society},
  doi = {10.1103/PhysRevB.73.193106},
  url = {https://link.aps.org/doi/10.1103/PhysRevB.73.193106}
}

@article{wu21,
  title = {Nature of Unconventional Pairing in the Kagome Superconductors $\mathrm{A}\mathrm{V}_{3}\mathrm{Sb}_{5}$ ($\mathrm{A}=\mathrm{K},\mathrm{Rb},\mathrm{Cs}$)},
  author = {Wu, Xianxin and Schwemmer, Tilman and M\"uller, Tobias and Consiglio, Armando and Sangiovanni, Giorgio and Di Sante, Domenico and Iqbal, Yasir and Hanke, Werner and Schnyder, Andreas P. and Denner, M. Michael and Fischer, Mark H. and Neupert, Titus and Thomale, Ronny},
  journal = {Phys. Rev. Lett.},
  volume = {127},
  issue = {17},
  pages = {177001},
  numpages = {7},
  year = {2021},
  month = {Oct},
  publisher = {American Physical Society},
  doi = {10.1103/PhysRevLett.127.177001},
  url = {https://link.aps.org/doi/10.1103/PhysRevLett.127.177001}
}

@ARTICLE{xu22,
  title     = "Three-state nematicity and magneto-optical $\mathrm{K}$err effect in the
               charge density waves in kagome superconductors",
  author    = "Xu, Yishuai and Ni, Zhuoliang and Liu, Yizhou and Ortiz, Brenden
               R and Deng, Qinwen and Wilson, Stephen D and Yan, Binghai and
               Balents, Leon and Wu, Liang",
  journal   = "Nat. Phys.",
  publisher = "Springer Science and Business Media LLC",
  volume    =  18,
  number    =  12,
  pages     = "1470--1475",
  month     =  dec,
  year      =  2022,
  doi = "https://doi.org/10.1038/s41567-022-01805-7"
}

@article{yamase00a,
	author = {Yamase, Hiroyuki and Kohno, Hiroshi},
	doi = {10.1143/JPSJ.69.332},
	OPTeprint = {https://doi.org/10.1143/JPSJ.69.332},
	journal = {J. Phys. Soc. Jpn.},
	number = {2},
	pages = {332},
	title = {{Possible Quasi-One-Dimensional Fermi Surface in La$_{2-x}$Sr$_{x}$CuO$_{4}$}},
	opturl = {https://doi.org/10.1143/JPSJ.69.332},
	volume = {69},
	year = {2000}
}

@article{yamase00b,
	author = {Yamase, Hiroyuki and Kohno, Hiroshi},
	doi = {10.1143/JPSJ.69.2151},
	OPTeprint = {https://doi.org/10.1143/JPSJ.69.2151},
	journal = {J. Phys. Soc. Jpn.},
	number = {7},
	pages = {2151},
	title = {{Instability toward Formation of Quasi-One-Dimensional Fermi Surface in Two-Dimensional {t-J} Model}},
	opturl = {https://doi.org/10.1143/JPSJ.69.2151},
	volume = {69},
	year = {2000}
}

@article{yamase15b,
	author = {Yamase, Hiroyuki and Bejas, Mat\'{\i}as and Greco, Andr\'es},
	title = {d-wave bond-order charge excitations in electron-doped cuprates},
	optDOI= "10.1209/0295-5075/111/57005",
	opturl= "https://doi.org/10.1209/0295-5075/111/57005",
	journal = {Europhys. Lett.},
	year = {2015},
	volume = {111},
	number = {5},
	pages = {57005}}

@article{yamase21,
	author = {Yamase, Hiroyuki and Sakurai, Yoshiharu and Fujita, Masaki and Wakimoto, Shuichi and Yamada, Kazuyoshi},
	da = {2021/04/13},
	doi = {10.1038/s41467-021-22229-6},
	id = {Yamase2021},
	isbn = {2041-1723},
	journal = {Nat. Commun.},
	number = {1},
	pages = {2223},
	title = {{Fermi surface in La-based cuprate superconductors from Compton scattering imaging}},
	ty = {JOUR},
	url = {https://doi.org/10.1038/s41467-021-22229-6},
	volume = {12},
	year = {2021},
	optBdsk-Url-1 = {https://doi.org/10.1038/s41467-021-22229-6}}

@article{yamase21c,
author = {Yamase ,Hiroyuki},
title = {Theoretical Insights into Electronic Nematic Order, Bond-Charge Orders, and Plasmons in Cuprate Superconductors},
journal = {J. Phys. Soc. Jpn.},
volume = {90},
number = {11},
pages = {111011},
year = {2021},
optdoi = {10.7566/JPSJ.90.111011},
optURL = {https://doi.org/10.7566/JPSJ.90.111011},
opteprint = {https://doi.org/10.7566/JPSJ.90.111011}
}

@article{yamase26b,
doi = {10.1088/1361-648X/ae5142},
url = {https://doi.org/10.1088/1361-648X/ae5142},
year = {2026},
month = {mar},
publisher = {IOP Publishing},
volume = {38},
number = {12},
pages = {123001},
author = {Yamase, Hiroyuki},
title = {Theoretical perspectives on charge dynamics in high-temperature cuprate superconductors},
journal = {Journal of Physics: Condensed Matter}
}

@article{yu21,
  title = {Concurrence of anomalous Hall effect and charge density wave in a superconducting topological kagome metal},
  author = {Yu, F. H. and Wu, T. and Wang, Z. Y. and Lei, B. and Zhuo, W. Z. and Ying, J. J. and Chen, X. H.},
  journal = {Phys. Rev. B},
  volume = {104},
  issue = {4},
  pages = {L041103},
  numpages = {7},
  year = {2021},
  month = {Jul},
  publisher = {American Physical Society},
  doi = {10.1103/PhysRevB.104.L041103},
  url = {https://link.aps.org/doi/10.1103/PhysRevB.104.L041103}
}

@article{zafur24,
  title = {Spin and bond-charge excitation spectra in correlated electron systems near an antiferromagnetic phase},
  author = {Zafur, Muhammad and Yamase, Hiroyuki},
  journal = {Phys. Rev. B},
  volume = {109},
  issue = {24},
  pages = {245127},
  numpages = {26},
  year = {2024},
  month = {Jun},
  publisher = {American Physical Society},
  doi = {10.1103/PhysRevB.109.245127},
  url = {https://link.aps.org/doi/10.1103/PhysRevB.109.245127}
}

@article{zhan26,
  title = {Loop Current Order on the Kagome Lattice},
  author = {Zhan, Jun and Hohmann, Hendrik and D\"urrnagel, Matteo and Fu, Ruiqing and Zhou, Sen and Wang, Ziqiang and Thomale, Ronny and Wu, Xianxin and Hu, Jiangping},
  journal = {Phys. Rev. Lett.},
  volume = {136},
  issue = {12},
  pages = {126001},
  numpages = {7},
  year = {2026},
  month = {Mar},
  publisher = {American Physical Society},
  doi = {10.1103/5vyy-rj6v},
  url = {https://link.aps.org/doi/10.1103/5vyy-rj6v}
}

\end{document}